\documentclass[twocolumn,authoryear]{els-mrw} 

\usepackage{amsmath,amssymb,amsfonts,amsthm,makeidx,graphicx}
\usepackage{txfonts}
\usepackage{helvet}
\usepackage[colorlinks=true,citecolor=blue,filecolor=blue,linkcolor=black,urlcolor=blue,pdftex]{hyperref}
\usepackage{caption}
\usepackage{subcaption}

\def\araa{ARA\&A}%
\def\apj{ApJ}%
\def\apjl{ApJ}%
\def\apjs{ApJS}%
\def\apss{Ap\&SS}%
\def\aap{A\&A}%
\def\mnras{MNRAS}%
\def\prc{Phys.~Rev.~C}%
\def\prd{Phys.~Rev.~D}%
\def\prl{Phys.~Rev.~Lett.}%
\def\nat{Nature}%
\def\physrep{Phys.~Rep.}%
\newcommand{\pasa}{Publ.\ Astron.\ Soc.\ Aust.\ }
\newcommand{\nar}{New Astron.\ Rev.\ }
\newcommand{\jcap}{JCAP}

\newcommand{\msun}{\,\mathrm{M}_\odot}
 \newcommand{\pd}{\partial}
 \newcommand{\ud}{\mathrm{d}}

\begin{document}

\chapter{Core-collapse supernovae and supernova neutrinos}\label{chap1}

\author[1]{Bernhard Müller}
\author[1]{Bailey Sykes}

\address[1]{\orgname{Monash University}, \orgdiv{School of Physics and Astronomy}, \orgaddress{10 College Walk, Clayton, VIC 3800, Australia.}}

\articletag{Chapter Article tagline: update of previous edition,, reprint..}

\maketitle

\begin{glossary}[Nomenclature]
\begin{tabular}{@{}lp{34pc}@{}}
CCSN & Core-collapse supernova\\
ZAMS & Zero-Age Main Sequence\\
NS & Neutron star\\
PNS & Proto-neutron star\\
BH & Black hole\\
QCD & Quantum Chromodynamics\\
EoS & Equation of State\\
SASI & Standing Accretion Shock Instability \\
GW & Gravitational Wave
\end{tabular}
\end{glossary}

\begin{abstract}[Abstract]
Core-collapse supernovae are the terminal explosions of massive stars. After successive phases of nuclear fusion proceeding up to silicon burning, these stars form an iron core that is supported by electron degeneracy pressure. The core eventually collapses to a proto-neutron star, and in most cases the outer layers of the star are ejected by a shock wave, with a kinetic energy of order $10^{51}\,\mathrm{erg}$. Neutrinos and multi-dimensional fluid flow play a key role in extracting energy from the collapsed core to drive the explosion. After adumbrating the astrophysical context of stellar evolution and transient observations, this chapter sketches the modern theory of neutrino-driven supernova explosions, and discusses the key role of nuclear physics and neutrino interaction rates in the supernova problem. It also outlines the role of neutrinos and gravitational waves as probes into the supernova core.
\end{abstract}

\begin{BoxTypeA}[chap1:box1]{Key points}
\begin{itemize}
    \item Massive stars end their lives as core-collapse supernovae, which leave behind neutron stars or black holes as compact remnants.
    \item Neutrinos play a crucial role in the dynamics of supernova explosions.
    \item The properties of nuclear matter influence supernova dynamics both via the structure of the young
    proto-neutron star and via the neutrino interaction rates.
    \item Neutrinos and gravitational waves can directly probe the supernova core and shed light on supernova dynamics, proto-neutron star structure, and the properties of nuclear matter.
\end{itemize}
\end{BoxTypeA}

\section{Introduction}\label{chap1:sec1}
One of the most interesting astrophysical laboratories for nuclear physics is furnished by \emph{neutron stars} (NSs). These compact stars, whose densities exceed that of nuclei, are formed by the collapse of \emph{massive stars} at the end of their life. The formation of a NS is accompanied by the shedding of the outer layers in an explosion know as a
\emph{core-collapse supernova} (CCSN). Sometimes, the collapse of a massive star instead results in the formation of a black hole (BH).

Nuclear physics plays a key role in CCSNe from the initiation of the collapse, through the crucial phase that decides whether the star successfully explodes and whether a NS or a BH is formed, and on through the further evolution of the NS (Chapters 7--9). CCSNe also offer the opportunity to learn about nuclear physics through a number of direct and indirect probes of the core of the explosion. In particular, the neutrinos emitted from the hot young ``proto-neutron star'' formed during the explosion have significant diagnostic potential, and have already been observed once in the case of supernova SN~1987A in the Large Magellanic Cloud \citep{bionta_87,hirata_87,alekseev_87}, albeit in very small quantities. Gravitational waves (GWs), which are yet to be observed from a CCSN, are another potential probe of the physics in the supernova core. The elemental and isotopic composition of the material synthesized during the explosion, which is discussed in Chapter 13, also provides constraints on supernova physics.

In this chapter, we provide a brief overview of the dynamics and phenomenology of CCSNe. We then discuss in  the impact of nuclear physics on the outcomes of collapse and explosion in more detail. The role of nuclear physics in CCSNe is strongly tied to that of neutrinos, which are an important agent in the dynamics of supernova explosion. After discussing salient aspects of neutrino-matter interactions and neutrino transport in CCSNe, we segue to the role of neutrinos as a diagnostic of CCSN physics. In addition, we outline very briefly the potential of GW as another messenger from the interior of the supernova.

This chapter offers a basic introduction into these topics. There is an abundance of articles on each of those topics that can provide greater depth to the interested reader. 
A number of recent reviews extensively discuss the hydrodynamics and explosion outcomes of CCSNe \citep{mueller_20,burrows_21,janka_25}. The role of neutrinos in CCSNe is covered in greater detail by several other recent reviews
\citep{mirizzi_16,horiuchi_18,mueller_19d,mezzacappa_20}
and book chapters \citep{janka_17b,horiuchi_18,raffelt_26}. Among these, \citet{mezzacappa_20} provides a
very detailed technical discussion of the problem of neutrino transport. \citet{raffelt_26} contains an exhaustive
overview of all aspects of supernova neutrinos, ranging from the physical processes that govern their emission, through the problem of neutrino quantum kinetics, on to their potential as diagnostics for supernova physics, and also discusses CCSNe as a laboratory for particle physics more broadly, i.e., not limited to neutrinos. GWs as a probe of supernova and nuclear physics are treated in more depth in \citet{abdikamalov_22,mezzacappa_25,mueller_26}.

\begin{figure}
    \centering
    \includegraphics[width=\linewidth]{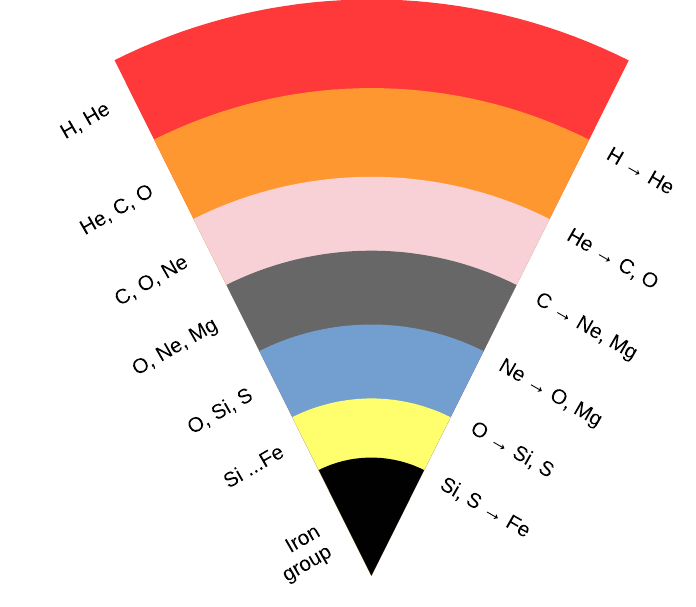}
    \caption{Sketch of the onion-shell structure of massive stars before the onset of core collapse. The elements dominating the composition of the shells are indicated on the left, and the major burning processes are indicated on the right. Burning takes place at the bottom of the shells and drives convection, which mixes the fuel and ashes in a shell. Note that silicon burning involves a complex quasi-equilibrium of 
    a whole range of intermediate mass nuclei
    mediated by light particle captures and dissociation
    \citep{bodansky_68,hix_99}.}
    \label{fig:prog}
\end{figure}

\section{The Astrophysical Context}
\subsection{From Massive Stars to Core-Collapse Supernovae}
\label{sec:sevol}
Classical CCSNe occur in stars that proceed through the whole range of hydrostatic nuclear burning stages from hydrogen burning up to silicon burning in their cores \citep{woosley_02}. Silicon  burning eventually leaves an iron core that is mostly supported by electron degeneracy pressure. Burning of lighter elements continues in the shell surrounding the core, leaving the star with a characteristic onion-shell structure (Figure~\ref{fig:prog}).

As silicon shell burning adds mass to the iron core and brings it close to its effective Chandrasekhar mass, the core contracts, reaching densities of $10^9\texttt{-}10^{10}\,\mathrm{g}\,\mathrm{cm}^{-3}$
and temperatures of ${\sim} 10^{10}\,\mathrm{K}$,
and eventually collapse to a NS as electron captures and photo-disintegration of heavy nuclei drain pressure support (Section~\ref{sec:sn_dyamics}). 

For single stars, evolution towards iron core formation and collapse occurs in those stars born with about $8\texttt{-}130\msun$ \citep{heger_03,heger_23} on the zero-age main sequence (ZAMS). The lower boundary of this mass range somewhat depends on the initial content of heavy elements in the star (metallicity)
\citep{ibeling_13,cinquegrana_23}, and is subject to uncertainties in the treatment of convection in stellar evolution
\citep{eldridge_04}. The evolutionary outcomes at the transition between low- and high-mass stars are particularly complex. Here, stars develop degenerate cores after carbon burning and before iron-core formation. One possible outcome of these conditions is a dynamical collapse already after the formation of a degenerate oxygen-neon-magnesium core by
electron captures on $^{20}\mathrm{Ne}$
and $^{24}\mathrm{Mg}$ (electron-capture supernova) as first proposed in the 1980s \citep{nomoto_84,nomoto_87}. Other fates are possible in this transition regime as well, with outcomes determined by an intricate interaction between nuclear burning, electron captures and convective mixing, as we shall discuss in more detail in Section~\ref{sec:nuc_collapse}.

The upper boundary of the CCSN mass range is determined by the transition to collapse due to electron-positron pair creation after carbon burning \citep{barkat_67,heger_02}. Stars with ZAMS masses of $\sim 130\texttt{-}260\msun$ are expected to be completely disrupted by oxygen and neon burning when they encounter the pair instability, giving rise to \emph{pair instability supernovae}, while stars of higher mass are expected to collapse directly to BHs \citep{heger_03,heger_23}. At ZAMS masses of $\sim 85\texttt{-}130\msun$, there is a transition regime where pair instability leads to partial mass ejection in pulses (pulsational pair instability supernovae), but the stars eventually evolve to iron core collapse \citep{heger_03,woosley_17}. The mass ranges for pair instability and pulsational pair instability supernovae are also subject to uncertainties in nuclear reaction rates and stellar physics \citep{woosley_21}. For solar or moderately sub-solar metallicity, stars of such very high ZAMS mass experience substantial wind mass loss during their lives and may not reach the pair instability regime at all, but evolve towards iron core collapse as normal massive stars
\citep{heger_03,belczynski_16}.

\subsection{Outline of Supernova Phenomenology}
The collapse of the core of a massive star to a NS with mass $M$ of order $1.4\msun$ and radius $R\approx12\,\mathrm{km}$ \citep{abbott_18,miller_21}
liberates a huge amount of gravitational potential energy of $E_\mathrm{grav}\sim GM^2/R \gtrsim 10^{53}\,\mathrm{erg}$. Transferring only a fraction of that energy into the envelope will result in an energetic explosion. 
At a time when NSs had not been observed, and the evolution of massive stars had not yet been worked out, the idea of NS formation as an energy source for stellar explosions was first anticipated in very crude form by \citet{baade_34b,baade_34c} as an explanation for a class of very bright astronomical transients in other galaxies that they had termed ``super-novae'' \citep{baade_34a}.

Today, the connection between the collapse of massive stars, supernova explosions, and (at least in many cases) NS formation is well established. Some milestones included the identification of gaseous remnants of ejected supernova material, starting with the association of the Crab Nebula with the historic supernova of 1054 \citep{hubble_28},  the discovery of a pulsar in the Crab Nebula \citep{staelin_68}, the observation of SN~1987A in the Large Magellanic Cloud from a known blue supergiant progenitor and with a coincident neutrino signal to prove the collapse to a NS
\citep{arnett_89}, and recent evidence for a ionizing radiation from the NS formed
in SN~1987A \citep{fransson_24}.

Observations of CCSNe and their compact and gaseous remnants have, however, furnished considerably more quantitative information on the physics of these explosions, the systematics and heterogeneities of explosion and remnant properties across the supernova population, and on the progenitors of CCSNe.

The most readily available source of information on CCSNe and their progenitors are their light curves and spectra. In the standard spectral classification of supernovae  based on the presence or absence of spectral lines \citep{filippenko_97,gal-yam_17}, CCSNe comprise the classes of Type~II SNe (with hydrogen lines), Type Ib (without hydrogen lines,
without a strong $6355\,\AA$ silicon line and with helium lines), and Type Ic (like Ib, but without helium lines). Among these, the Type~II SNe outnumber the Type Ib/c SNe by about 2:1 \citep{smith_11,jerkstrand_26}. The progenitors of Type Ib/c SNe must have lost their hydrogen envelope; the relatively high fraction of Type Ib/c events indicates that the mass loss is usually  due to binary interactions and not wind mass loss in massive progenitors \citep{smith_11,eldridge_13}.

The different envelope structure of Type~II and Type~Ib/c supernova progenitors has important implications on the physics that governs the light curves. The light curves of the most widespread subtype (Type~IIP) are characterized by a plateau of about $100\,\mathrm{d}$. During the plateau, the emission of light primarily feeds on the thermal energy of the shock-heated hydrogen envelope, which is slowly released as the photosphere recedes further into the ejecta due to hydrogen recombination \citep{kasen_09,zampieri_17}. This is followed by a tail phase, where the observable transient is primarily powered by the decay of the radioactive $^{56}\mathrm{Ni}$ made during the explosion and its daughter isotope  $^{56}\mathrm{Co}$. The dependence of the light curves on explosion and progenitor parameter is quite well understood for Type~IIP SNe, so that one can constrain explosion energies, progenitor masses and radii and the mass of ejected $^{56}\mathrm{Ni}$ based on simply scaling laws \citep{popov_93,kasen_09}, especially when supplementing the light curves with spectral information to constrain ejecta velocities. Well-observed samples of Type~IIP supernovae point to a range of explosion energies of $0.1\texttt{-}2\times10^{51}\,\mathrm{erg}$
and $^{56}\mathrm{Ni}$ masses of
$5\times10^{-3}\texttt{-}0.28\msun$ (with a mean value of $0.046\msun$), with indications of higher explosion energies for higher progenitor masses
\citep{pejcha_15c,mueller_t_17}.

Among the various supernova types, the direct identification of progenitor stars in archival images has also been most successful for Type~IIP SNe, and has clearly established them as originating from red supergiants \citep{smartt_09a,smartt_15}. 
The distribution of the progenitor masses suggests a paucity of explosions for progenitors above $15\texttt{-}18\msun$ \citep{smartt_15}, i.e., BH formation without any explosion.
There is also tentative evidence for the disappearance and hence, presumably, BH formation of two massive stars in M31
\citep{adams_17,de_26}.
Systematic uncertainties in the determination of red supergiant luminosities and masses remain a major challenge in determining the mass range of exploding massive stars, however \citep{beasor_25}.

Due to the availability of high-quality observations, considerable work has also been invested into the physical properties of SN~1987A, the closest supernova in recent history, which displayed a peculiar light curve shape (Type IIpec) due to its compact blue supergiant progenitor. Analyses point to an explosion energy of $\sim 1.5\times10^{51}\,\mathrm{erg}$, an ejecta
mass of ${\sim} 14\msun$ \citep{jerkstrand_20}
and $^{56}\mathrm{Ni}$ mass of $0.07\msun$ \citep{seitenzahl_14}.

Compared to Type~II supernovae, the quantitative determination of explosion energies and ejecta masses for Type Ib/c supernovae is more challenging and cannot be fully summarized here. One relevant aspect, however, consists of the presence of a small sub-class of Type~Ic supernovae that are characterized by very broad lines (Type~Ic-BL); these comprise about $1\%$ of all CCSNe in the local universe \citep{smith_11}. The properties of Type~Ic-BL supernovae indicate very high explosion energies of up to ${\sim} 10^{52}\,\mathrm{erg}$, about ten times higher than ordinary core-collapse events \citep{woosley_06b}. For some Type~Ic-BL supernovae, a coincident gamma-ray burst jet has been detected, starting with the prototypical event SN~1998bw \citep{iwamoto_98}. Based on corrections for the jet opening angle, a substantial fraction of Type~Ic-BL are thought to involve gamma-ray bursts \citep{woosley_06b}. The unusually high explosion energies suggest that there is not a single, universal explosion mechanism for all CCSNe. Chapter ``Exotic Transients: Pair Instability and Jet-Driven Explosions'' deals with  scenarios for exotic and extreme explosions in greater detail.

Observations show CCSNe to be inherently multi-dimensional. For example, SN~1987A provided evidence for strong mixing of iron and nickel far out into the hydrogen envelope \citep{mccray_93}, and polarization measurements indicate substantial global asymmetries especially in the inner ejecta of many explosions \citep{wang_08}. Supernova remnants
provide a very detailed late-stage view of explosion asymmetries. The connection of their structure to the initial explosion asymmetries is more involved, but increasingly better understood \citep[for current overviews, see][]{janka_25,jerkstrand_26}. The seeds for these asymmetries are set already during the first seconds of a supernova deep inside the stellar core.

Further observational constraints on supernova explosion physics come from NS and BH birth properties. Substantial birth velocities (kicks) of compact objects -- up to $1000\,\mathrm{km}\,\mathrm{s}^{-1}$ for NSs and several $100\,\mathrm{km}\,\mathrm{s}^{-1}$ for some BHs \citep{popov_25} -- are of particular interest. They are again suggestive of large-scale asymmetries deep in the core of the explosions. Furthermore, the kicks of some BHs indicate that BH formation is sometimes accompanied by partial, asymmetric mass ejection, which is required to reconcile the substantial kick with total momentum conservation.

For prospective multi-messenger observations of supernova neutrinos and GWs, the Galactic CCSN rate is a major hurdle, as these observations will be mostly limited to the Milky Way and its immediate vicinity. Complementary evidence from supernovae in nearby galaxies \citep{li_11} and from the radioactive $^{26}\mathrm{Al}$ ejected by massive stars \citep{diehl_06} indicate a galactic rate of about 1-3 events per century.

\section{Dynamics of Collapse and Explosion}
\label{sec:sn_dyamics}

The collapse and (in most cases) the subsequent explosion of massive stars takes place over several phases. Starting from the collapse of the iron core, the star contracts but halts after a short time when central densities become high enough for nuclear repulsive forces to come into play. This sudden stiffening of the fluid causes a so-called ``bounce'', which launches a shock wave out to a radius of a few hundred kilometers. The shock wave lingers here for a while, having expended its initial kinetic energy shortly after bounce. Various mechanisms have been proposed to ``revive'' the shock and, if successful, this again launches the shock outwards into the star. While this shock may take up to about a day to propagate from the core region to the stellar surface where an optical transient is produced, the final phase of collapse through to the revival of the shock only takes about a second.

\subsection{Final Collapse of the Core}
After numerous burning stages towards the end of its life, the star is left with the onion-shell structure depicted in Figure~\ref{fig:prog}. Key to this structure is the iron core, which grows in mass as silicon is burned in the core. At a certain point, the density has increased enough for the rates of electron captures (on nuclei and free protons; see Section \ref{sec:nuc_collapse}) to become significant. By reducing the abundance of free electrons, the electron degeneracy pressure which primarily supports the core against gravitational collapse, is weakened. Additionally, the contraction of the core and the associated increase in temperature is favorable for the production of high-energy photons which can cause endothermic photodisintegration of heavy nuclei, further sapping the support of the core against gravity. Eventually, the contraction transitions into a runaway collapse on a free-fall time scale of a few hundred milliseconds.

Initially during the collapse, neutrinos from electron capture are free to stream out of the core unimpeded, lowering the bulk lepton number. However, once the core density rises to $\gtrsim 10^{12} \, \mathrm{g \, cm^{-3}}$, neutrinos become trapped, i.e., their diffusion timescale becomes greater than the collapse timescale,  and the lepton number is locked to its present value. 

For non-rotating stars, the collapse remains almost spherically symmetric, while for rotating stars one can expect their shape to become slightly oblate during this phase, although not significantly so.

\begin{figure*}
    \centering
    \includegraphics[width=0.6\linewidth]{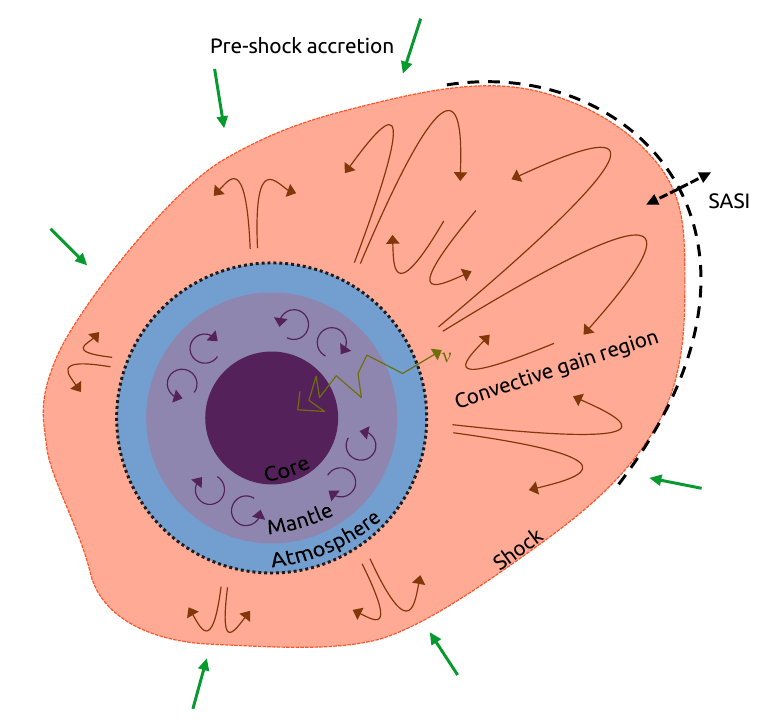}
    \includegraphics[width=0.8\linewidth]{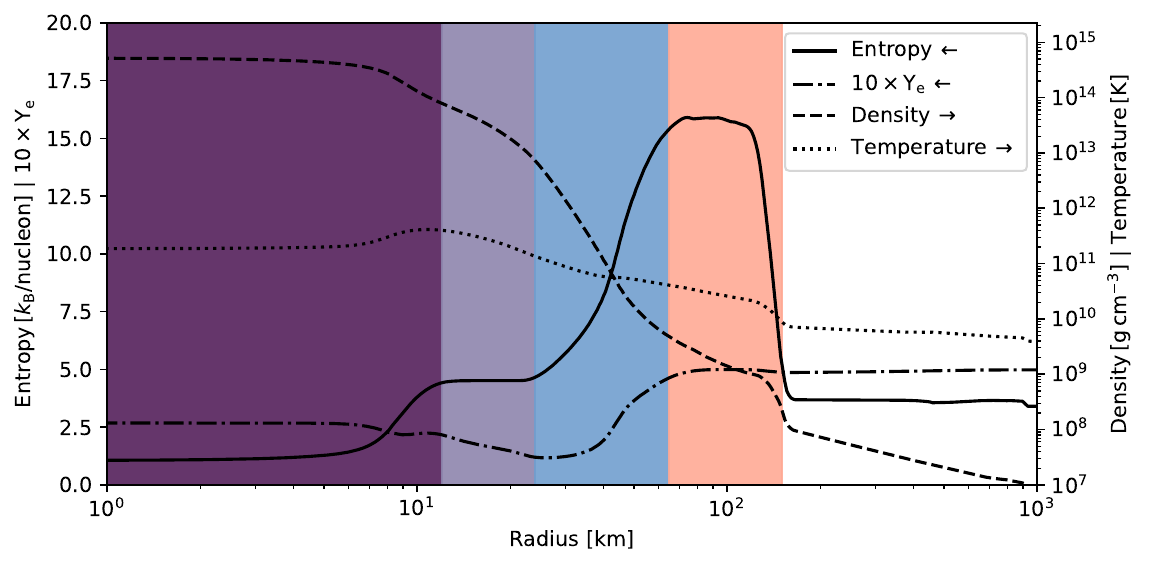}
    \caption{\textbf{Top}: Schematic diagram of the PNS and gain region out to the standing accretion shock during the pre-explosion phase. Convective fluid motions in the gain region and inside the PNS are indicated with curved arrows.  In addition to convective motions, the standing accretion shock instability (SASI) may lead to large-scale shock oscillations.
    A sample neutrino path is included for reference. \textbf{Bottom}: Sample profiles of entropy, electron fraction $Y_\mathrm{e}$ (multiplied by ten for visualization), density, and temperature. The plot area is colored according to the corresponding region in the above schematic.}
    \label{fig:ccsne_schematic}
\end{figure*}

\subsection{Bounce}
As the collapse proceed, the core eventually reaches nuclear densities. At this point repulsive nucleon-nucleon interactions become dominant and the matter strongly resists further compression. This process is often referred to as a  \emph{stiffening} of the equation of state (EoS). The collapsing material, having fairly significant radial momentum, slightly overshoots nuclear saturation density and consequently rebounds. This is called \emph{core bounce} (or just \textit{bounce}). The bounce launches a shock wave out into the star with a kinetic energy of order $10^{51} \, \mathrm{erg}$.

\begin{figure}
    \centering
    \includegraphics[width=\linewidth]{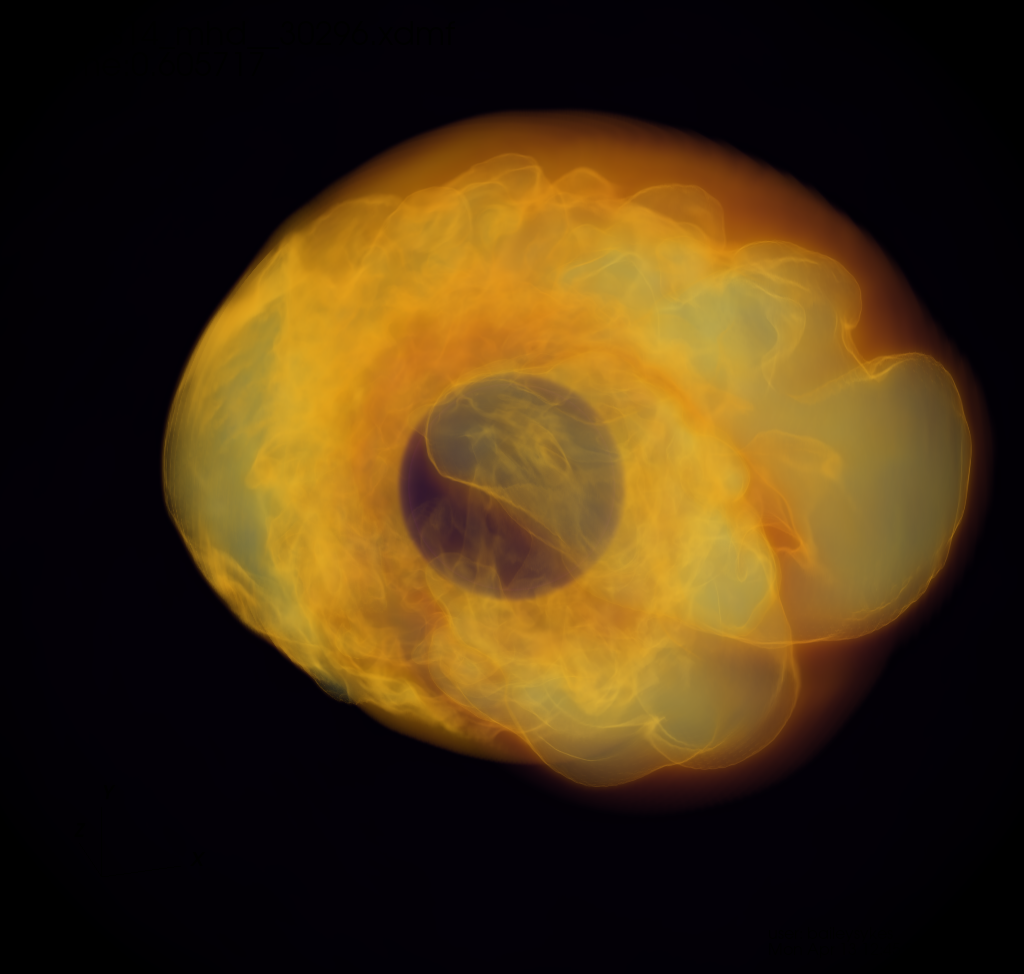}
    \caption{Rendering of the post-bounce entropy structure during the collapse of a $14 \msun$ star. The opaque central sphere is the low-entropy PNS, while the fluffy orange material surrounding it is the high-entropy gain region bounded on the exterior by the stalled shock.
    The deformation of the shock by the SASI is clearly visible in this snapshot, as are high-entropy bubbles behind the deformed shock.}
    \label{fig:sample_vol_render}
\end{figure}

\subsection{The Stalled Shock}
\label{sec:explosion}
The shock is formed deep inside the core, and the outer layers of the iron core are still collapsing at this stage. As the shock moves further out inside the iron core, its initial kinetic energy is quickly drained by nuclear dissociation and by neutrino energy losses, which happen rapidly once the post-shock density drops to $\sim 10^{11}\,\mathrm{g}\,\mathrm{cm}^{-3}$. The shock therefore \emph{stalls} and turns into an \emph{accretion shock}. Collapsing shells continue to fall through the shock and settle onto what is now a
\emph{proto-neutron star} at the center.
The shock is still pushed out to a radius of $100-200 \, \mathrm{km}$ over tens of milliseconds, and then hovers around this position.

At this point, the supernova core takes on a characteristic structure that it maintained until until an explosion develops (Figure~\ref{fig:ccsne_schematic}). The PNS at the center consists of several region. The core of about half a solar mass, has not been shock heated and retains a low entropy. It is surrounded by a
warm mantle of shock-heated matter, which quickly grows to about a solar mass by accretion. At the surface of the PNS, there is an atmosphere where the density drops precipitously. As accreted matters settles onto the PNS and is compressed to high densities while traversing the atmosphere, it cools efficiently by neutrino emission. This region is therefore also termed the \emph{cooling region} during the post-bounce accretion phase. Further outside in the \emph{gain region} between the PNS and the shock, heating by neutrinos from deeper layers dominates over neutrino cooling. The gain region develops a few tens of milliseconds after bounce.

Starting from these conditions, how is the shock \emph{revived} to eventually expel the outer layers of the star?
The prevailing theory, originally proposed by \citet{colgate_66} and later shaped into its modern form by \citet{bethe_85}, is that of the \textit{neutrino-driven explosion mechanism}, which is thought likely power the vast majority of CCSNe
with typical explosion energies. The idea runs as follows:
After the formation of the hot PNS at bounce, neutrinos are radiated from the PNS surface with high luminosities of several $10^{52} \, \mathrm{erg \, s^{-1}}$. A small fraction of these
neutrinos (more specifically, of the electron-flavor neutrinos) are reabsorbed behind the supernova shock, thus heating the \textit{gain region} behind the shock with a typical efficiency of ${\sim}10\%$. 

The conditions for successful shock revival can be formulated, in an idealized scenario, in terms of two parameters: the neutrino luminosity, and mass accretion rate through the shock \citep{burrows_93}. It has been shown that, above an accretion rate-dependent threshold of the neutrino luminosity, the shock becomes unstable to radial perturbations \citep{fernandez_12}. That is, for sufficiently strong neutrino emission, there is no equilibrium state for the shock and a runaway expansion -- i.e., an explosion -- will likely follow. It is also possible to formulate a similar argument as $t_\mathrm{adv} / t_\mathrm{heat} \gtrsim 1$, where $t_\mathrm{adv}$ and $t_\mathrm{heat}$ are the advection and heating timescales
\citep{janka_01}. The former is an approximation of the time spent by accreting material in the heating region, while the latter is roughly the time taken for neutrinos to deposit energy in the material equivalent to its gravitational binding energy.

It has been established for several decades, however, that neutrino heating is not sufficient to produce explosions in spherically symmetric models
\citep{liebendoerfer_00,rampp_00}, except for the lightest of supernova progenitors
\citep{kitaura_06}. In more massive progenitors, it is crucial that neutrino heating is aided by the effects of multi-dimensional flow.

Several fluid instabilities can operate in the supernova core to break spherical symmetry as illustrated by
Figure~\ref{fig:sample_vol_render}. The first is \emph{turbulent convection}, as has been recognized since the 1990s \citep{herant_94,burrows_95,janka_96}.
Neutrino heating produces a negative entropy gradient in the gain region. Consequently, it becomes convectively unstable, 
with high-entropy, neutrino-heated matter rising while cooler fluid elements sink towards the PNS. 
Turbulence aids revival of the shock through several mechanisms. 
The motion of the convective plumes hitting the shock effectively contributes ``turbulent''  pressure (or more precisely \emph{Reynolds stresses}; \citealp{murphy_11}) which expands the shock and increases the time available for neutrino heating. The transport of heat by convection from close to the PNS out the shock, and the dissipation of the turbulent flow also modify the structure of the gain region. The net effect is to substantially lower the neutrino luminosity required to trigger an explosion \citep{murphy_09,mueller_15a}.

In addition to convection in the gain region, the standing accretion shock instability (SASI) is another mechanism which breaks the symmetry of core-collapse and aids shock revival through the excitation of large-scale oscillation modes of the stalled shock \citep{blondin_03,foglizzo_07}. SASI oscillations are classified as sloshing modes if $\ell \gtrsim 1 \, \mathrm{and} \, m=0$, or spiral modes if $|m| > 0$, where $\ell$ and $m$ are, respectively, the degree and order of the corresponding spherical harmonic.

The SASI impacts shock revival in a similar manner as turbulent convection; i.e, the various sloshing and spiraling modes of the shock boost the post-shock Reynolds stresses, and also drive low-entropy downflows while raising high-entropy bubbles to larger radii, thus increasing the heating efficiency in the gain region.

The large-scale motions associated with the SASI have observational implications for both neutrinos and GWs. We will address this in more detail in Section~\ref{sec:neutrino_gw_signal}.

A more detailed understanding of the impact of multi-dimensional effects on CCSNe has emerged in recent years thanks to a growing collection of high-resolution 3D simulations by a number of different groups. Many of these have been able to obtain explosions and predictions for observable explosion and remnant properties and multi-messenger signals.
Due to the different treatment of neutrino transport, EoS, nuclear burning, magnetic fields, relativistic self-gravity, etc., the different simulations are often not in perfect quantitative agreement, but offer complementary insights into the CCSN mechanism. Code comparisons in 1D, 2D and 3D
\citep{liebendoerfer_05,mueller_10,cabezon_18,oconnor_18c,just_18,varma_21b}
have been and remain important for quantifying and controlling the associated uncertainties. Notable sets of 3D simulations with neutrino transport have been produced using
the codes \textsc{CoCoNuT-FMT} 
\citep{mueller_17,mueller_19a,chan_18,sykes_25b}, \textsc{Fornax}
\citet{vartanyan_19,burrows_20,burrows_24a},
\textsc{Chimera}
\citep{lentz_15,mezzacappa_20b},
\textsc{Prometheus-Vertex}
\citep{melson_15b,bollig_21,janka_24},
\textsc{3DnSNe-IDSA}
\citet{takiwaki_14,nakamura_22,matsumoto_24,nakamura_25}, \textsc{Flash}
\citet{oconnor_18b,kovalenko_26} and
\textsc{Fugra} \citep{kuroda_18}. The list of simulations is far from an exhaustive, and we refer the reader to recent reviews on neutrino-driven  \citep{mueller_20,janka_25} and magnetohydrodynamic explosions 
\citep{mueller_25b} for a more detailed discussion of the current state of simulations.

\subsection{Core Collapse and Black Hole Formation}
The shock is not always revived, however. In these situations, ongoing accretion onto the PNS increases its mass and it becomes more and more compact in terms of the relativistic compaction $GM/Rc ^2$. High temperatures may stabilize the PNS above the maximum mass for cold NSs, but even in this case,
further accretion or neutrino cooling will, at some point, cause
it to collapse. Eventually, the PNS becomes compact enough that an event horizon forms and a BH is born. The part of the PNS exterior to the horizon collapses in within a few microseconds. 

Depending on when the collapse of the PNS occurs, BH-forming CCSN candidates may or may not explode. Those which do not explode -- so-called \textit{failed supernovae} -- may explain the dearth of optical observations of higher mass CCSNe ($M_\mathrm{ZAMS}\gtrsim 20 \msun$): the `red supergiant problem' of \citet{smartt_09a}. The pathway of earliest BH formation, that is, immediate collapse of the stellar core, is disfavored by currently modeling, instead suggesting that all massive stellar cores must at least go through an intermediate hot PNS phase before cooling and collapsing as a BH; this pathway is still often referred to as \textit{direct collapse}. BH formation is also possible via two \emph{fallback} pathways: prompt and delayed \citep{wong_14}. Fallback here means that material previously (partially) ejected by a revived shock loses enough energy that it falls back onto the PNS, increasing its mass and triggering BH formation. Fallback is prompt if it occurs within a few seconds, or delayed if it occurs on time scales of hours to days. Such late fallback typically occurs as a result of reverse shocks that are formed when the blast wave traverses shell interfaces in the progenitor. The dynamics of explosions that undergo BH formation by fallback is complex, but multi-dimensional simulations have already yielded important insights into this pathway \citep{chan_18,chan_20b,rahman_22,sykes_25,burrows_25}.

While it is more likely that the CCSN remnant of more massive progenitors will be a BH, and less massive progenitors will leave behind a NS, there is no perfect indicator for how a given stellar collapse will proceed. Many authors argue that there is some predictive power in the \textit{compactness} of the progenitor core \citep{oconnor_11}, defined as,
\begin{equation}
    \xi_M = \frac{M}{\msun} \frac{1000 \, \mathrm{km}}{R(M_\mathrm{bary}=M)}.
\end{equation}
There is some freedom in choosing the reference mass, with common choices being $M_\mathrm{bary} = 1.75 \, \msun$ or $ 2.5\, \msun$. Stars with higher compactness appear more likely to produce a failed CCSN. Parameter studies of the collapse of stars in the core-collapse range suggest there may be islands of explodability in the progenitor parameter space
\citep{ugliano_12,mueller_16a,sukhbold_16,ebinger_19}, i.e., a non-monotonic dependence on properties of the star such as mass, metallicity, and rotation, to name a select few. The best predictive tool for the outcome of stellar collapse are high-resolution, multi-dimensional hydrodynamics simulations, but even these operate on idealized stellar models, retain modeling uncertainties, and are generally too expensive for systematic parameter and sensitivity studies.

Even for failed supernovae, it may still be possible to obtain a weak optical transient. The mechanism for this, described by \citet{Nadezhin:1980} and \citet{Lovegrove_Woosley:2013}, relies on the idea that neutrinos produced in the core and which propagate freely away, reduce the gravitational mass of the collapsing star. Thus, a formerly tenuously bound stellar envelop may become unbound and produce a very weak ${\sim}10^{47} \, \mathrm{erg}$ transient.

\subsection{Characteristics of the Explosion and Remnant}
\label{sec:explosion_phase}
Once shock revival occurs, neutrino heating powers up the explosion for several hundred milliseconds to seconds. SASI oscillation die off, while convection changes its character. The flow morphology is characterized by large-scale asymmetries with expanding neutrino-heated bubbles and fast downflows of colder material accreted through the shock.
Typically in more massive stars, the ejecta develop a unipolar or dipolar structure, while lower mass progenitors may develop bubbles on a smaller scale due to the imprint of prior convection \citep{mueller_19a,burrowS_19}.

The asymmetric ejecta (and anisotropically emitted neutrinos) carry with them a net linear and angular momentum. In return, the compact remnant acquires a net
linear and angular momentum in the opposite direction. 
Thus, the compact remnant usually receives a \emph{kick velocity} at birth which varies in magnitude between CCSNe but is typically on the order of several hundred kilometers per second. Asymmetric emission of neutrinos from the PNS may also transport away momentum, yielding a \textit{neutrino kick} on the order of a few tens of kilometers per second. The total kick imparted to the remnant is relevant for explaining the observed velocities of NSs and BHs.

Similarly, the angular momentum left on the compact remnant by asymmetrically accreted matter substantially influences the spin of the compact object, which can be very different from what one might expected from the rotation of the progenitor core assuming conservation of specific angular momentum \citep{wongwathanarat_13,mueller_19a}. Fallback at late times in particular could spin up NSs or BHs considerably, with NS spin periods down to the millisecond range \citep{chan_20b}.
Failed CCSNe, for which there is no explosion, do not produce sizable kicks or spins beyond that from transient asymmetric neutrino emission. 
\citet{janka_24,burrows_24a,popov_25} provide current overviews of the theory of compact object kicks and spins.

After the neutrino-driven engine has powered up the explosion for a few seconds, it takes about a day until the shock reaches the surface in red supergiant progenitors of Type~IIP supernovae, and somewhat less for more compact progenitors like blue supergiants for Type~IIpec explosions or stripped-envelope progenitors. As the shock propagates through the envelope, further mixing instabilities occur. This phase shapes the explosion asymmetries that can later be observed in supernova remnants.
We refer to dedicated texts on the late phases of the explosion up to and after shock breakout for details \citep{janka_25,jerkstrand_26}.

From the viewpoint of nuclear physics and nucleosynthesis, the long-term evolution of the PNS and its environment are of more immediate interest.
As accretion onto the PNS gradually dies away after the successful onset of an explosion, the PNS enters the so-called Kelvin-Helmholtz
\footnote{Note that this phenomenon has nothing to do with the hydrodynamics instability of the same name.} 
cooling phase. During this period, the PNS cools and contracts due to emission of neutrinos of all flavors on a diffusion timescale of several seconds 
\citep{pons_99,huedepohl_10,fischer_10}. A fraction of these neutrinos deposit energy in the layers just outside the neutrinosphere, resulting in a dilute overflow of material with an indicative initial rate of order ${\sim}10^{-3}\texttt{-}10^{-5} \msun \, \mathrm{s}^{-1}$. This outflow from the PNS surface is referred to as the \textit{neutrino-driven wind} \citep{duncan_86,qian_96,huedepohl_10,fischer_10}, which is potentially a significant site for nucleosynthesis beyond the iron-group \citep{cowan_21}.

\section{Nuclear Physics of the Collapse Phase}
\label{sec:nuc_collapse}
\subsection{Iron Core Collapse}
Let us now consider the role of nuclear physics in supernova explosions more closely.
Nuclear physics critically shapes the collapse phase, as it determines the rate of deleptonization in the collapsing iron core. Among the two possible channels, electron capture on nuclei,
\begin{equation}
    {}^A Z+\mathrm{e}^-\rightarrow {}^{A}(Z-1)+\nu_\mathrm{e}
\end{equation}
and on free protons,
\begin{equation}
    \mathrm{p}+\mathrm{e}^-\rightarrow \mathrm{n}+\nu_\mathrm{e},
\end{equation}
the former generally has a higher threshold energy ($Q$-value), but capture on protons is disfavored due to the low proton fraction due to low entropy and neutron-rich conditions. 
Whereas the composition of the core is initially dominated by familiar iron-group nuclei, deleptonization shifts the composition to highly neutron-rich and significantly more massive nuclei \citep[e.g.,][]{lattimer_91,janka_07,fischer_17}. Calculations of the weak transition rates in these nuclei have progressed over decades, culminating in modern shell
mode calculations \citep{langanke_03}. While seminal earlier calculations of the requisite Gamov-Teller transitions in the independent particle model \citep{fuller_82} had suggested that electron capture on heavy nuclei stops for neutron number $N\geq 40$ due to Pauli blocking, the modern calculations show that captures on nuclei remain possible throughout the collapse phase and consistently dominant over captures on free protons \citep{langanke_03,hix_03}.
This allows the lepton fraction to reach $\sim 0.3$
at trapping, and the final electron fraction at bounce to dip to $\sim0.25$.

The modern capture rates imply a small mass of the homologous inner core at bounce, since this mass scales as $M_\mathrm{core}\propto Y_\mathrm{e}^2$. With typical modern EoSs and modern rates, the core mass at bounce is about $0.5\msun$ \citep{hix_03}. There is some dependence on the EoS both via the electron capture rates and via its thermodynamic properties around nuclear saturation density, in particular the symmetry energy. The mass of the core at bounce remains imprinted onto the structure of the PNS on long time scales; it roughly corresponds to the mass of the low-entropy inner core of the PNS. While the impact of the core mass on supernova dynamics in the long term is less clear, it is of significance for a prospective GW signal from the core bounce, which is expected for progenitors with significant rotation \citep[e.g.,][]{dimmelmeier_08}. The mass of the core influences both the amplitude and the frequency of the GW signal \citep{dimmelmeier_08}.

\subsection{Electron-Capture Supernovae}
In the electron-capture supernova channel
at the transition between white-dwarf forming low-mass stars and high-mass stars (see Section~\ref{sec:sevol}), the detailed nuclear structure plays an intriguing role in determining the fate of the star after carbon burning. In the transition region, stars are left with a degenerate O-Ne-Mg core, whose further evolution
is shaped by electron captures and $\beta$-decays of Ne and Mg nuclei
\citep{nomoto_84,nomoto_87,miyaji_87,jones_14,kirsebom_19}: At densities of several $10^9\,\mathrm{g}\,\mathrm{cm}^{-3}$, two-stage electron captures 
$^{24}\mathrm{Mg}(\mathrm{e^-,\nu_\mathrm{e}})
^{24}\mathrm{Na}
(\mathrm{e^-,\nu_\mathrm{e}})
^{24}\mathrm{Ne}$
and, at higher densities, $^{20}\mathrm{Ne}(\mathrm{e^-,\nu_\mathrm{e}})
^{20}\mathrm{F}
(\mathrm{e^-,\nu_\mathrm{e}})
^{20}\mathrm{O}$ remove electron degeneracy pressure, but result in net heating of the core. At similar densities, URCA processes
involving an electron capture and a $\beta^-$-decay back to the original nuclei cool the core by neutrino emission, with
$^{25}\mathrm{Mg}\leftrightarrow ^{25}\mathrm{Na}$
and $^{23}\mathrm{Mg}\leftrightarrow ^{23}\mathrm{Na}$ being the predominant URCA pairs. The cycle of electron captures and $\beta$-decays is enabled by convection which shuffles matter between high densities (favoring electron captures) and low densities (favoring decays).

As the core contracts, sufficiently high temperatures to ignite oxygen burning off-center are eventually reached. If the density at ignition is still sufficiently low and the competition between nuclear energy generation and deleptonization favors the former, the star may undergo a thermonuclear explosion instead of collapsing \citep{jones_16}. The balance between core heating and URCA cooling, and the interaction of these processes with convection determine when ignition occurs and are therefore critical for the final fate of the stars in this transition range. Recent measurements of a strong forbidden $\beta$-decay transition of $^{20}\mathrm{F}$ have provided crucial information on the electron capture rate on $^{20}\mathrm{Ne}$, pointing to an onset of electron capture and core heating at lower densities and a higher likelihood for a thermonuclear explosion rather than for collapse to a NS
\citep{kirsebom_19}. While the fate of stars with degenerate O-Ne-Mg cores still hinges on other uncertainties in stellar evolution (in particular the treatment of convection), this presents a striking example where a small number of  states and transitions in a few nuclei can qualitatively impact the fates of stars.

\begin{figure*}
    \centering
    \includegraphics[width=0.8\linewidth]{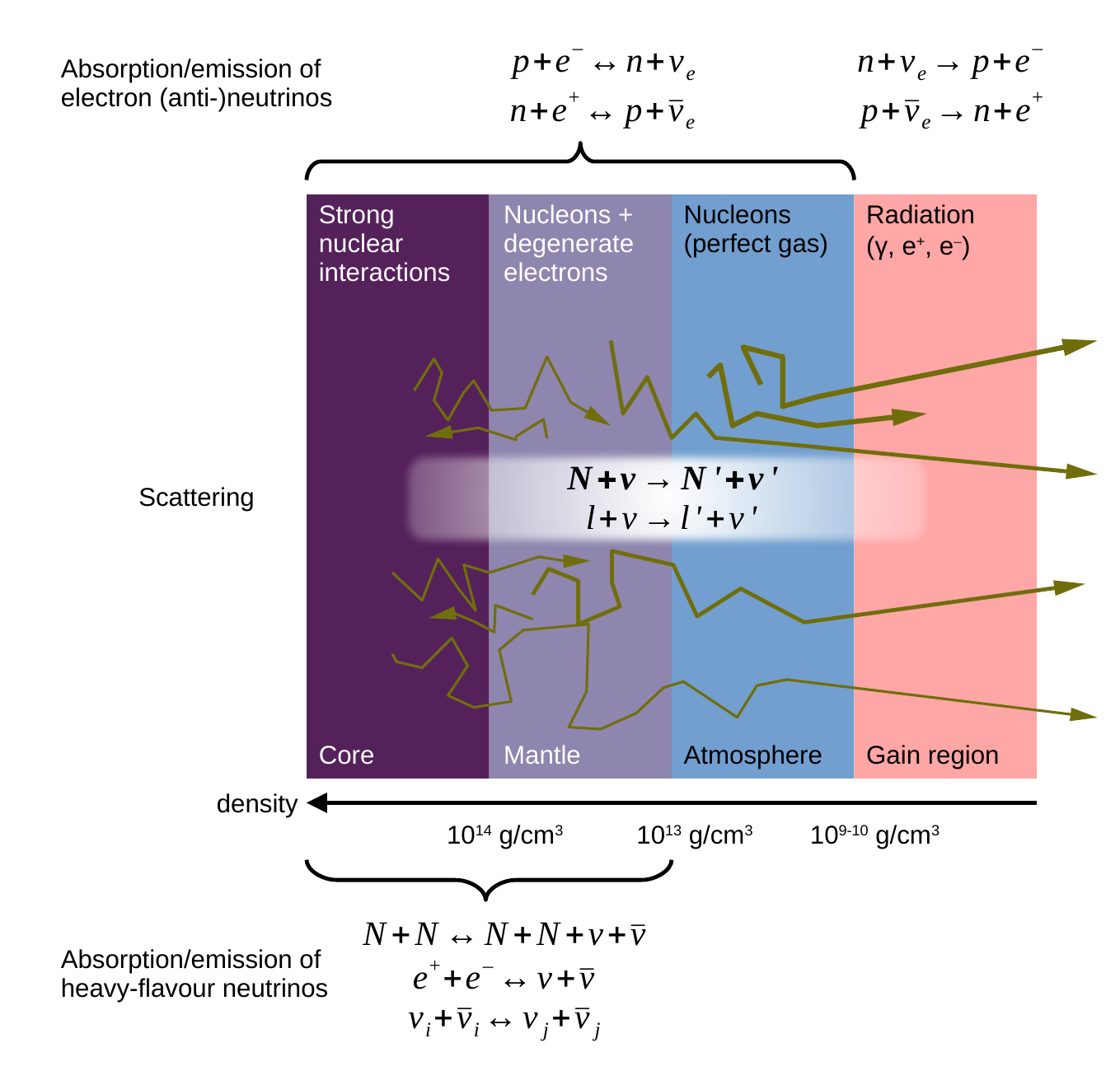}
    \caption{EoS and neutrino transport regimes in the supernova core, corresponding to the regions shown in Figure~\ref{fig:ccsne_schematic}. For each of the four regions from the core out to the gain region, the figure indicates the dominant contribution to the pressure. The key emission and absorption processes for electron-flavor and heavy-flavor neutrinos are listed at the top and at the bottom. Scattering processes are shown in the middle.     N denotes any type of nucleon, $\nu$ and $\bar{\nu}$ denote neutrinos and antineutrinos of any flavor, and l indicates leptons, i.e., electrons or neutrinos (or muons at sufficiently high temperatures). The figure also indicates the transition from the diffusion regime to free streaming, and illustrates that heavy-flavor neutrinos originate from deeper layers than electron-flavor neutrinos.}
    \label{fig:eos_neutrinos}
\end{figure*}

\section{The Nuclear Equation of State during the Post-bounce Evolution}
The nuclear EoS also has a major influence on the post-bounce evolution of CCSNe. Before we discuss this in more detail, it is useful to first review the different EoS regimes encountered in the supernova core as context. 

Figure~\ref{fig:eos_neutrinos} shows the EoS regimes corresponding to the different regions of the supernova interior sketched in Figure~\ref{fig:ccsne_schematic}. 
In the core we find densities well above or close to nuclear saturation density
$\rho_\mathrm{nuc}\approx 2.3\times 10^{14}\,\mathrm{g}\,\mathrm{cm}^{-3}$ and temperatures of a few ten MeV. Under these conditions, the pressure is determined by the nuclear interactions, and degeneracy effects are important. The maximum density is, however, still lower than in cold NSs. At high densities, heavier hadrons, mesons, or deconfined quark matter can eventually appear, but these are most relevant during later phases when the PNS has cooled down and contracted to its final radius over time scales of seconds. The appearance of muons is already important during the accretion and explosion phase, however \citep{bollig_17}.

Typical densities in the mantle are somewhat lower than nuclear saturation density. During the accretion phase and early explosion phase, the highest temperatures are reached in the inner part of the mantle. In very massive PNSs, the maximum temperature can reach about $100\,\mathrm{MeV}$. Non-degenerate neutrons and protons provide the bulk of the pressure in this region, with some additional contribution from degenerate electrons. Nuclear interactions still play a role and are, e.g., relevant for exact calculations of neutrino interactions.

In the PNS atmosphere, nucleons largely act as a perfect non-interacting gas and still provide the bulk of the pressure. Temperatures are of order of a few MeV. At the lower densities encountered in the gain region, photon, electron and positron radiation pressure become dominant. The temperatures in the heating region are still high enough for matter to remain dissociated into free neutrons and protons.
Recombination into $\alpha$-particle occurs once the shock expand beyond $\sim 300\,\mathrm{km}\,\mathrm{s}^{-1}$. Although neutrons and protons dominate the nuclear composition in the supernova core, light clusters ($^2\mathrm{H}$, $^3\mathrm{H}$, $^3\mathrm{He}$ and $^4\mathrm{He}$) do appear at subnuclear densities and can have a non-negligible impact on neutrino opacities \citep{fischer_17,fischer_20}.

\subsection{Impact on Proto-Neutron Star Structure and Heating Conditions}
The influence of the nuclear EoS on the conditions for shock revival is mostly an indirect one: The high-density EoS determines the radius of the (warm) PNS, this in turn determines the neutrino luminosity and mean energy and also the quasi-hydrostatic shock position by setting an ``effective'' boundary condition for the accretion flow onto the central object. Faster NS contraction for a ``soft'' EoS results in a higher PNS surface temperature and hence stronger neutrino emission and better heating conditions. However, this is balanced by stronger shock retraction and tighter binding of the material in the gain region in the gravitational potential. In spherical symmetry, the dependence of the shock radius $R_\mathrm{sh}$
on the electron-flavor luminosity $L$, PNS 
radius $R$ surface temperature $T_\nu$ and  mass $M$, and the mass accretion rate $\dot{M}$, has a steep power-law law dependence in $R$ \citep{janka_12,mueller_15a},
\begin{equation}
\label{eq:rsh}
    R_\mathrm{sh}
    \propto \frac{(LT ^2)^{4/9} R^{16/9}}{\dot{M}^{2/3} M^{1/3}}.
\end{equation}
In multi-D, the dependence of the accretion shock radius on the PNS radius is often modified because stronger neutrino heating for a softer EoS can also boost convection, which can then push the shock further out.

Thus, the overall influence of the EoS on shock revival is the result of several competing effects, which need to be studied in simulations. Empirically, early 2D simulations already established that soft EoS with faster PNS contraction generally \emph{favor} the development of an explosion \citep{janka_12b,suwa_13}. Since some of these early simulations still used EoS with widely varying parameters for nuclear matter and substantially different radii and maximum masses for cold NSs, different choices for the EoS often made the difference between a successful explosion and failure.

As the influence of the EoS on the conditions for shock revival is very much determined by the mass-radius relation for warm PNSs, it is not straightforward to relate ``softness'' or ``stiffness'' to nuclear matter parameters or cold NS properties (e.g., the maximum allowed mass). A systematic analysis of
the nuclear parameters that determine the contraction of warm PNSs (and hence regulate explodability) was conducted by \citet{yasin_20} using 1D simulations. They demonstrated that the most critical parameter for the PNS contraction is the \emph{nucleon effective mass}, with the nuclear symmetry energy and incompressibility playing a minor role.

In addition to the sensitivity of PNS contraction on the properties of nuclear matter, the leptonic contribution to the EoS is also relevant. As shown by \citet{bollig_17}, muon creation softens the EoS and accelerates PNS contraction. This also facilitates the development of neutrino-driven explosions.

While attempts have been made to investigate the repercussions of the EoS dependence on the landscape of supernova explosion and remnant properties in 
1D \citep{schneider_a_19,ghosh_22}, multi-D simulations are required to incorporate indirect effects of the EoS on neutrino-driven convection and the SASI, as well as the direct effect of the EoS on PNS convection.  Changes in the neutrino heating and the PNS contraction affect the entropy gradient and the advection time scale in the heating region, and thus
the conditions that determine the growth of SASI and convection \citep{mueller_20}. PNS convection is affected directly by the thermodynamic derivatives of nuclear matter \citep{roberts_12b,jakobus_22}, and is another factor that modifies the contraction of the PNS \citep{buras_06b}.

Recent astrophysical constraints on the maximum NS mass \citep{fonseca_21} and radii \citep{abbott_18,miller_21}, as well as heavy-ion experiments and lattice QCD \citep[see][]{motornenko_20}, have considerably narrowed down the allowed parameter space for the nuclear EoS, and ruled out many EoSs historically used for CCSN simulations. Even within the current constraints, multi-D simulations indicate that the remaining uncertainties in the EoS still limit predictions about supernova outcomes. In a large 2D study with the SFHo and SFHx EoS of \citet{steiner_10} and the CMF (chiral-mean field) EoS of \citet{motornenko_20} with a hadron-quark crossover, 
\citet{powell_26} found explosions for only two out of fifteen stellar progenitor models between $9.71\,\mathrm{M}_\odot$ and $36\,\mathrm{M}_\odot$ with the CMF EoS, and explosions for almost all progenitors with SFHo and SFHx. Systematic 3D studies of the EoS-dependence of supernova explosions for many progenitor models are yet to be conducted. Extant 3D simulations using different modern EoS show differences of order 50\% in explosion energy, substantially different nucleosynthesis in exploding low-mass stars \citep{rusakov_26}, and a notable impact on black-hole formation time and explosion energy in progenitors of very high mass \citep{powell_21}. 
Whether BH formation in high-mass progenitors can be accompanied by fallback explosions depends on the EoS. For some EoSs, the shock may be revived before BH collapse, but does not become supersonic quickly enough to escape the BH \citep{powell_26}. 

\subsection{The Phase-Transition Mechanism}
A particularly strong impact on the dynamics can result from a  transition to deconfined quark matter at high densities. 
Based on 1D simulations using an EoS with a first-order hadron-quark phase transition, \citet{sagert_09} and \citet{fischer_11} first proposed a mechanism for triggering explosions by such a transition. In their models, the PNS collapses to a more compact configuration when it encounters the phase transition, and avoids collapsing to a BH. The rebound after this second collapse launches a very powerful shock wave. Different from the initial rebound, this shock was able to launch a powerful explosion in their models.

The EoS assumed in early studies of the phase-transition mechanism was not compatible with constraints on the maximum NS mass, however. This was addressed in subsequent work with hadron-quark equations of state more in line
with observational and experimental constraints. These constraints imply that a phase transition could only happen after the PNS has become substantially more massive than the bulk of the NS population. Hence explosions triggered by a phase transition could at best explain a small fraction of all CCSNe. Accordingly, \citet{fischer_18} put this scenario forward as a possible explanation for energetic explosion of high-mass stars above $\gtrsim 50\,\mathrm{M}_\odot$. However, the viability of the phase-transition mechanism remains very uncertain. While \citet{zha_21} found that the second collapse leads to a transiently stable compact star that undergoes some ringdown oscillations for $0.24\lesssim\xi_{2.5}\lesssim0.51$, the rebound never triggers an explosion in their models. For higher compactness, the second collapse directly leads to black-hole formation. Similar results were obtained by \citet{jakobus_22}, who only found weak explosions triggered by the second collapse for some progenitors with low compactness, which would likely explode by the neutrino-driven mechanism earlier if multi-D effects were taken into account. Moreover, these explosions were not robust between different equations of state with a first-order phase transition.

Moreover, while a transition to deconfined quark matter is ultimately expected at sufficiently high densities,
lattice QCD calculations point to a smooth crossover and
not a first-order phase transition \citep{motornenko_20}.
In the case of such a smooth crossover, \citet{jakobus_22}
never found a second bounce after the phase-transition induced collapse.

Thus it remains unclear whether the hadron-quark phase transition is able to trigger an explosion in some stars at all. Nonetheless, the phase transition may still lead to interesting dynamics in the supernova core (Section~\ref{sec:pns_convection}) and produce observables signatures in the supernova neutrino and GW signal (Section~\ref{sec:neutrino_gw_signal}).

\subsection{Proto-Neutron Star Convection}
\label{sec:pns_convection}
Whereas the nuclear EoS only exerts an indirect influence on the multi-dimensional flow in the gain region, it impacts the convection inside the PNS much more directly. PNS convection initially develops in the region outside the entropy
step to about $5\,k_\mathrm{B}/\mathrm{nucleon}$ that is established immediately after bounce by shock heating (Figure~\ref{fig:ccsne_schematic}). The dynamics of PNS convection is then determined by the loss of energy and lepton number from the PNS surface and the transport of energy and lepton number in the PNS interior.
Instability to convection is determined by the 
Brunt-V\"ais\"al\"a frequency (or buoyancy frequency),
$\omega_\mathrm{BV}$. In the Newtonian approximation,
$\omega_\mathrm{BV}$ is determined by the gravitational acceleration
$g$, and profiles of the density $\rho$, pressure $P$ sound speed $c_\mathrm{s}$, entropy $s$, and lepton fraction $Y_\mathrm{L}$ (number of leptons per baryon),
\begin{equation}
    \omega_\mathrm{BV}^2 = 
     \frac{g}{\rho c_\mathrm{s}^2}\left[\left(\frac{\pd P}{\pd s}\right)_{\!\rho,Y_\mathrm{L}}\frac{\ud s}{\ud r}+\left(\frac{\pd P}{\pd Y_\mathrm{L}}\right)_{\!\rho,s}\frac{\ud Y_\mathrm{L}}{\ud r}\right],
\end{equation}
where instability occurs for $\omega_\mathrm{BV}^2<0$.
Under normal circumstances, the derivative
$(\pd P/\pd s)_{\!\rho,Y_\mathrm{L}}$ is positive, so the negative entropy gradient established by neutrino losses at the PNS surface drives convection in the mantle. The 
coefficient  $(\pd P/\pd Y_\mathrm{L})_{\!\rho,s}$
of the lepton number gradient can have either sign, i.e., 
negative lepton number gradients can be either destabilizing or
stabilizing \citep{bruenn_95,jakobus_25}. 
Under the conditions found in modern supernova simulations, negative lepton number gradients often stabilize the stratification against convection \citep{bruenn_95,powell_19,glas_19,jakobus_25}, such that there is a competition of stabilizing and destabilizing gradients. In principle, such a situation can lead to doubly-diffusive instabilities \citep{bruenn_95}, i.e., instability driven by diffusive energy or lepton number transport despite dynamical stability. However, such phenomena have not been seen in multi-D simulations so far.
These simulations instead show a complex interaction between destabilizing and stabilizing gradients that lead to a remarkable global organization of the convective flow in the PNS. The electron fraction often becomes globally asymmetric, which also results in asymmetric emission of lepton number by neutrinos (Lepton-number Emission Self-sustained Asymmetry or LESA; \citealp{tamborra_14a}).
By contrast, the convective velocity field is dominated by smaller scales and has a different turbulence spectrum \citep{powell_19,jakobus_25}.

\begin{figure*}
    \centering
    \includegraphics[width=0.7\linewidth]{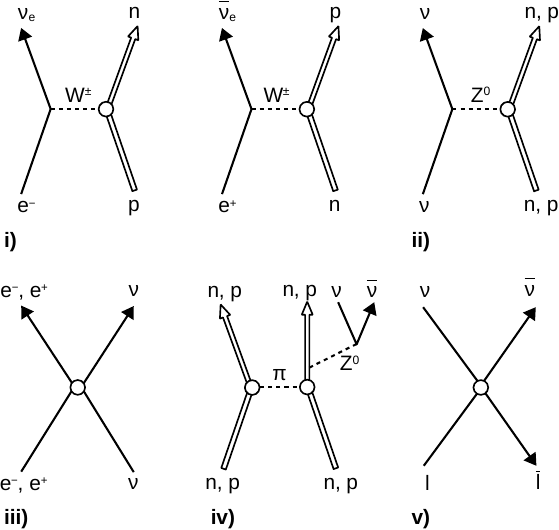}
    \caption{Simplified diagrams of the key neutrino-matter interaction rates during the post-bounce phase: i) charged-current emission and absorption by nucleons, ii) neutral-current scattering on nucleons, iii) neutrino-electron scattering, iv) nucleon-nucleon bremsstrahlung, v) 
    neutrino-electron and neutrino-neutrino pair processes. Legs that can represented electrons or neutrinos are represented by ``l''.
    Hollow lines are used for nucleons as ``dressed'' particles modified by in-medium effects. Some of the leptonic diagrams involve both $\mathrm{Z}^0$- and
    $\mathrm{W}^\pm$}-exchanges  and are drawn as four-particles interaction, with an implicit summation over all relevant diagrams in the composite vertex. 
    \label{fig:interactions}
\end{figure*}

\section{Nuclear Physics and Neutrino Emission}
\label{sec:neutrino_emission}
\subsection{Outline of Neutrino Emission}
The neutrino emission from the PNS is the key driver of the  post-bounce evolution in the supernova core. It is influenced by the high-density nuclear physics indirectly via the contraction of the PNS as highlighted in the previous section, but nuclear structure also affects neutrino emission more directly through the neutrino-matter interaction rates. A different perspective on the neutrino emission is found in Section~\ref{sec:neutrino_gw_signal}, where we shall discuss the potential of neutrinos as a multi-messenger probe into the supernova interior.

The dominant neutrino interaction processes in the supernova core during the post-bounce phase are sketched in Figure~\ref{fig:interactions}.
These include i) charged-current emission and absorption on nucleons, ii) neutral-scattering on nucleons iii) neutrino-electron scattering, iv) nucleon-nucleon bremsstrahlung, v) neutrino-electron and neutrino-neutrino pair processes. Among these, charged-current absorption and emission and neutral-current scattering on nucleons are the dominant processes for electron neutrinos ($\nu_\mathrm{e}$) and
antineutrinos ($\bar{\nu}_\mathrm{e}$). The pair and bremsstrahlung processes dominate the emission and absorption of heavy-flavor neutrinos
($\nu_\mu$, $\bar{\nu}_\mu$, $\nu_\tau$, $\bar{\nu}_\tau$, sometimes collectively referred to as $\nu_\mathrm{X}$), while scattering on nucleons is the dominant process that regulates the diffusion of these neutrinos.

Moreover, muonic processes also play a role at high temperatures \citep{bollig_17}. Interactions with other baryons or quark matter can become important once these appear at high densities and are of particular relevance on longer time scales for the (P)NS cooling phase \citep{pons_99,pons_01a,pons_01b}. Interactions with nuclei are of minor importance after the collapse phase (since there are at most light nuclei in the PNS and its environment), as is the neutrino plasmon process $\gamma^*\rightarrow \nu \bar{\nu}$.
Physics beyond the standard model, e.g., PNS cooling by axions and axion-like particles, could also indirectly affect neutrino emission (see, e.g.,
\citealp{keil_97,fischer_16a,lucente_20} and also \citealp{raffelt_26} for an overview). In this section, we focus on the more elementary topic of neutrino-nucleon interactions.

The neutrino interaction rates depend strongly on the matter density and neutrino energy. At high densities, neutrinos of all flavors are in equilibrium with the matter, and diffuse according to temperature and chemical potential gradients. 
Towards lower densities, electron flavor neutrinos ($\nu_\mathrm{e}$ and $\bar{\nu}_\mathrm{e}$) and heavy-flavor neutrinos ($\nu_\mathrm{\mu}$, $\bar{\nu}_\mathrm{\mu}$, $\nu_\mathrm{\tau}$, $\bar{\nu}_\mathrm{\tau}$) behave differently (Figure~\ref{fig:eos_neutrinos}). For electron-flavor neutrinos, the charged-current emission and absorption processes have similar cross sections as nucleon scattering. These neutrinos therefore remain close to thermal equilibrium throughout the mantle and the deeper layers of the atmosphere. At lower densities in the atmosphere, the scattering and absorption opacity $\kappa_\mathrm{s}$ and $\kappa_\mathrm{a}$ become low enough for electron-flavor neutrinos to escape. Most of the emitted $\nu_\mathrm{e}$'s and $\bar{\nu}_\mathrm{e}$'s originate from close to the surface of last scattering (neutrinosphere), where the total optical depth
$\tau=\int_r^\infty \kappa_\mathrm{s}+\kappa_\mathrm{a}\,\ud r'\approx 2/3$. As the absorption and scattering opacity 
increase with neutrino energy, high-energy neutrinos decouple further outside. 

A fraction of the neutrinos are reabsorbed in the gain layer. Typically the neutrino heating rate in the gain region amounts to about $5\texttt{-}10\%$ of the electron-flavor luminosity. 

Heavy-flavor neutrinos are only produced by various pair processes and by muonic processes. Emission and absorption of these neutrinos therefore freezes out at higher densities around the edge of the mantle. Scattering rates at still high below these densities. As the heavy-flavor neutrinos diffuse towards the surface of last scattering, they can still exchange energy with the matter through the recoil in neutrino-nucleon scattering and neutrino-electron scattering. Their contribution to heating in the gain region is negligible.
The different decoupling of the neutrino flavors affects their emergent fluxes and spectra (Section~\ref{sec:neutrino_signal}).

The formal framework for describing the absorption, emission and scattering of neutrinos is \emph{kinetic theory}. The classical Boltzmann transport equation for the neutrino distribution function $f$
as a function of time $t$, position $x^i$ and momentum $p^i$
can be written schematically as
\citep{janka_17b}
\begin{equation}
\label{eq:boltzmann}
    \frac{\pd f}{\pd t}+\dot{x}^i\frac{\pd f}{\pd x^i}+
    \dot{p}^i\frac{\pd f}{\pd p^i}=\mathfrak{C}[f],
\end{equation}
where the collision integral $\mathfrak{C}[f]$ describes interactions of neutrinos with matter and among themselves. In anticipation of relativistic effects like gravitational redshift and ray bending, Equation~(\ref{eq:boltzmann}) contains a term $\dot{p}^i$ for the rate of change of the neutrino momentum along its trajectory. The rigorous formulation of the relativistic transport equation is discussed by
\citet{lindquist_66,ehlers_71,cardall_13}. 

In recent years, the relevance of neutrino flavor conversion in dense astrophysical environments has increasingly been recognized. Taking into account flavor conversion requires a quantum kinetic approach, where the distribution function $f$ is replaced by a density matrix $\rho$ for the composition in flavor space. The transport equation is then supplemented by a quantum kinetic term \citep{zhang_13b},
\begin{equation}
    \frac{\pd \rho}{\pd t}+\dot{x}^i\frac{\pd \rho}{\pd x^i}+
    \dot{p}^i\frac{\pd \rho }{\pd p^i}=\mathfrak{C}[f]-i[H,\rho],
\end{equation}
where the Hamiltonian includes the vacuum mixing terms and forward scattering off electrons, positrons and other neutrinos. For details on neutrino transport and neutrino quantum kinetics we refer the reader, e.g., to Chapter~4, \citet{mezzacappa_20} and \citet{johns_25}.

\subsection{Nuclear  Physics and Neutrino Interaction Rates}
Similar to the collapse phase, nuclear interactions and the structure of the nucleon shape the neutrino emission from the PNS and thereby also influence the conditions for shock revival.

The nucleonic processes involve coupling of $\mathrm{W}^\pm$ and $\mathrm{Z}^{0}$ bosons to the difference of the vector current and axial current \citep{halzen_84,tubbs_75},
\begin{equation}
    J^\mu=\bar\psi \gamma^\mu(1-\gamma^5)\psi.
\end{equation}
Since nucleons are composite particle, effective coupling constants $g_\mathrm{v}$ and $g_\mathrm{a}$ for the vector and axial current appear in the interaction. For each two-particle interaction with a neutrino, these can be further related to effective couplings $C_\mathrm{V}$ and $C_\mathrm{A}$ for the entire interaction \citep{tubbs_75,bruenn_85,janka_17b}. For non-zero momentum exchange $q$, a Pauli form factor $F_2$, related to the anomalous magnetic moment of the nucleon, also appears \citep{horowitz_02}.

Correctly accounting for the nucleon structure in the effective couplings turns out to be quantitatively important in supernova simulations.
For example, a small contribution $g_\mathrm{a,s}$ of strange quarks to the axial coupling can reduce the neutrino scattering cross section \citep{horowitz_02}, leading to faster PNS contraction and higher luminosities, thereby aiding explosions \citep{melson_15b}.
Though this demonstration study assumed a 
$g_\mathrm{a,s}$ somewhat larger than the best experimental values \citep{airapetian_07,maas_17}, the reduction of the scattering opacity is important for precision modeling. The form factor $F_2$ for ``weak magnetism'' is also relevant for precision modeling, as it decreases antineutrino  cross sections relative to those of neutrinos, with larger effects for higher neutrino energies
\citep{horowitz_02}. Inclusion of weak magnetism in simulations tends to increase the electron fraction in the heating region and in outflows
\citep{buras_06a}, and is therefore relevant for precisely determining nucleosynthesis conditions.

At the high densities inside PNSs, interactions \emph{between} nucleons also influence the coupling of neutrinos to nuclear matter. This is most apparent from neutrino emission by nucleon-nucleon bremsstrahlung,
which is commonly treated in the free one-pion exchange approximation \citep{hannestad_98}, although deviation from this approximation again matters for precision modeling \citep{bartl_16,fischer_16b}.

In-medium effects are, however, also of major importance for emission, absorption and scattering of neutrinos by nucleons. Due to interactions with the medium, the nucleon appears as a ``dressed'' particle, which exhibits an effective mass different from vacuum, and experiences an interaction potential. The nucleon interactions also create correlations within the medium. Since the late 1990s, substantial efforts has gone into calculating neutrino rates that incorporate such in-medium interaction effects in nuclear matter
(e.g., \citealp{reddy_99,burrows_98, burrows_99,horowitz_17}; see also \citealp{mezzacappa_26} for a recent overview).

In-medium corrections to charged- and neutral-current neutrino-nucleon interactions turn out to have substantial effects in supernova simulations. Differences in the mean-field interaction potentials of neutrons and protons increase the difference in mean energy of the emitted electron neutrinos and antineutrinos, which once again affects the composition of outflows by typically lowering the electron fraction \citep{martinez_12,roberts_12c}. This implies that the composition of the outflows, in particular during the later phase of the neutrino-driven wind, is sensitive to the nuclear symmetry energy \citep{martinez_12}, which is directly related to the difference in the mean-field potentials.

Nucleon correlations \citep{reddy_99,burrows_98,burrows_99,horowitz_17} affect the overall transport opacity. They generally tend to reduce it compared to non-interacting matter \citep{horowitz_17}, though this need not be true under all conditions. The reduction of the opacities is more pronounced at higher densities; well above saturation density, the opacity can be lowered by a factor of several \citep{burrows_98}.

Since the effect is most pronounced at high densities, the most glaring consequence is a substantial shortening of the Kelvin-Helmholtz cooling phase of the PNS from $\gtrsim 20\,\mathrm{s}$ to $\lesssim 10\,\mathrm{s}$ \citep{huedepohl_10}. During the first hundreds of milliseconds, the effects is a small enhancement of the neutrino heating rate, similar to the effect of nucleon strangeness \citep{horowitz_17}.

\begin{figure*}
    \centering
    \includegraphics[width=\linewidth]{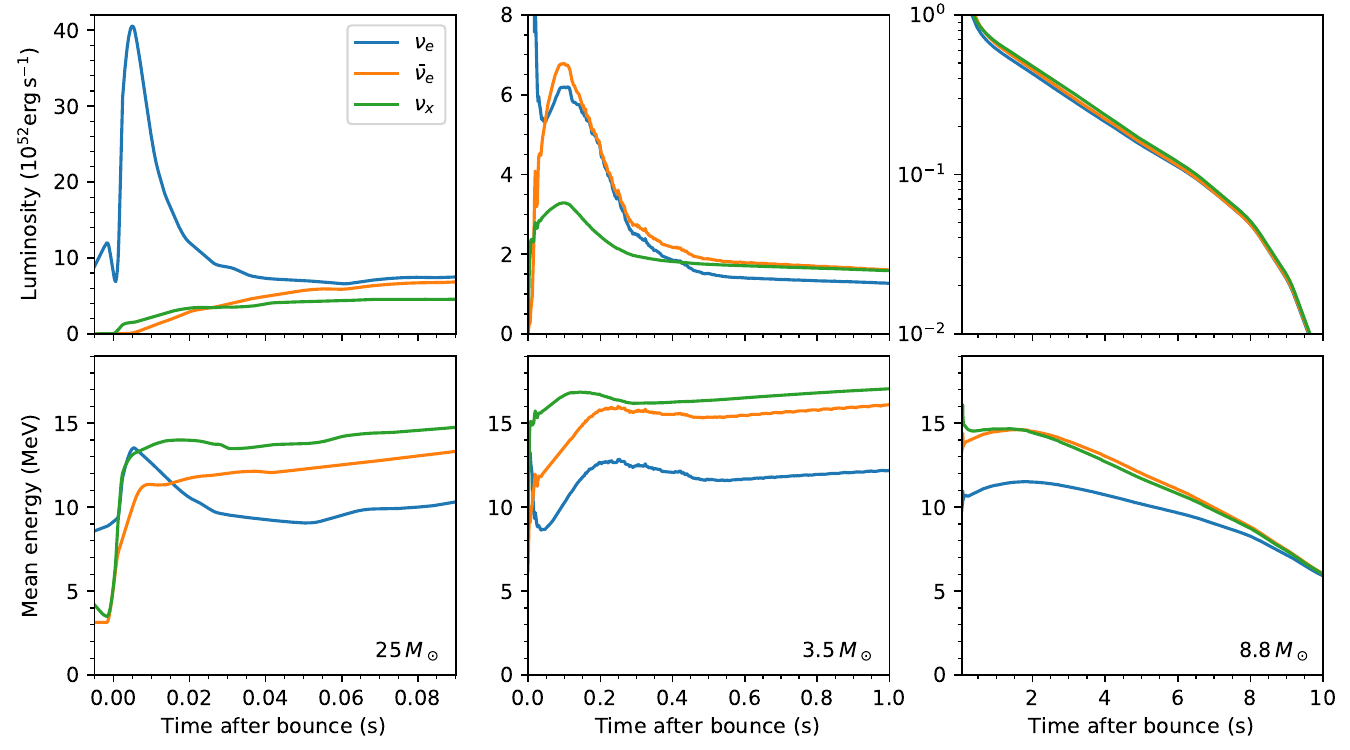}
    \caption{Phases of the CCSN neutrino signal, illustrated based on three different simulations. The left column shows the neutronization burst and the early post-bounce emission in a $25\msun$ model 
    \citep{mueller_14}. The middle column shows the accretion and early explosion phase in a CCSN simulation of a $3.5\msun$ stripped-envelope progenitor \citep{mueller_19a}. The right column shows the 
    cooling phase for an $8.8\msun$ model of an electron-capture supernovae
    \citep{mueller_10,mueller_19d}.
    The top row shows the luminosities of electron neutrinos ($\nu_\mathrm{e}$), electron antineutrinos ($\bar{\nu}_\mathrm{e}$), and heavy-flavor muon and tau neutrinos ($\nu_\mathrm{X}$), and the bottom row show the neutrino mean energies.}
    \label{fig:neutrino_lum}
\end{figure*}

\section{Neutrinos and Gravitational Waves as Probes into  the Supernova Core}
\label{sec:neutrino_gw_signal}

\subsection{Neutrino Signal}
\label{sec:neutrino_signal}
Let us now consider neutrinos as diagnostics of the PNS structure and evolution of supernova dynamics. In this chapter, we focus on how the macroscopic dynamics translates into the neutrino emission from the supernova core, ignoring the ramifications of flavor conversion in the core and on the way through the outer layers of the star (Chapter~16). The actual detection of supernova neutrinos on Earth will be dealt with in Chapter~5. 

The neutrino signal from  CCSNe can be divided into three distinct phases; these are evident in the sample data provided in Figure~\ref{fig:neutrino_lum}.
The first phase is the electron neutrino burst.
During the collapse, the neutrino luminosity and mean energy initially increase as the density and electron chemical potential (which mainly drives electron captures) increase. Due to the high degeneracy, mostly electron neutrinos are emitted during this phase. 
When densities become high enough, neutrinos become trapped, and there is a short dip in the neutrino emission. Once the core has bounced and the  shock wave formed moves into lower density material, neutrinos can escape freely from behind the shock, and there is rapid deleptonization by electron capture on free protons. This produces a massive breakout burst (also called neutronization burst) of electron neutrinos. The $\nu_\mathrm{e}$-burst reaches a peak luminosity of ${\sim}4\times 10^{53} \, \mathrm{erg \, s^{-1}}$ with the mean neutrino energy peaking at around $15 \, \mathrm{MeV}$. The burst decays rapidly from its peak, lasting only $10-20 \, \mathrm{ms}$. As the matter at the PNS surface deleptonizes it approaches $\beta$-equilibrium where the electron neutrino chemical potential is close to zero $\mu_{\nu_\mathrm{e}}=\mu_\mathrm{p}+\mu_{e}-\mu_\mathrm{n}\approx 0$. As a result, the electron antineutrino luminosity rises and becomes similar to that of electron neutrinos. The luminosity of heavy-flavor neutrinos rises earlier during the burst, as their production is mostly determined by temperature and density, and are less inhibited by strong electron degeneracy.

As a neutrino ``standard candle'' that is quite uniform across progenitors, the burst can be used as a diagnostic for determining the
supernova distance, neutrino physics parameters, and the progenitor
structure that affects flavor conversion on the way to the observer
\citep{kachelriess_05,duan_08_b,lunardini_08,serpico_12}.

Within tens of milliseconds the neutrino burst has subsided and is followed by an extended accretion phase, which continues over the entire pre-explosion phase and usually several hundreds of milliseconds up to a few seconds into the early explosion phase, depending on the mass of the progenitor. The luminosity of the neutrinos and antineutrinos of each flavor is a few $10^{52}\,\mathrm{erg}\,\mathrm{s}^{-1}$ during this phase, and their mean energies increase as the PNS contracts.
There are important differences in the emission of electron and heavy-flavor neutrinos, reflecting the different production
processes (Figure~\ref{fig:eos_neutrinos}). The emission of
$\nu_\mathrm{e}$ and $\bar{\nu}_\mathrm{e}$ is powered to a substantial degree by accretion of material onto the PNS. Electron neutrinos and electron antineutrinos are produced in almost equal measure by the charged-current interactions in the PNS surface region. In addition, diffusion of neutrinos from deeper layers in the PNS also contributes flux. This diffusion contribution is similar for both electron-flavor and heavy-flavor neutrinos. The total electron-flavor luminosity can therefore be related to PNS properties via \citep{mirizzi_16}, 
\begin{equation}
    L_{\nu_\mathrm{e}} + L_{\bar{\nu}_\mathrm{e}} = 2 \beta_{1} L_{\nu_{\mu/\tau}} + \beta_{2} \frac{GM\dot{M}}{R},
    \label{eqn:neutrino_lum_e}
\end{equation}
where $M$ and $R$ and the PNS mass and radius, $\dot{M}$ is the mass accretion rate onto the PNS, and $\beta_{1} \approx 1.25$ and $\beta_{2} \approx 0.5$ are dimensionless parameters which set the contributions of neutrino diffusion (first term on the RHS), and accretion (second term of RHS) to the total signal.

Heavy-flavor neutrinos are produced in fewer numbers from deeper layers, where densities and temperatures are sufficiently high for bremsstrahlung and various pair processes. The luminosity of heavy flavors is dictated by the temperature of the neutrinosphere, and is suitably modeled by a gray-body emission law,
\begin{equation}
    L_{\mu/\tau} = 4 \pi \phi \sigma_\mathrm{fermi} R^{2} T^{4},
    \label{eqn:neutrino_lum_heavy}
\end{equation}
with $\phi$ being a grayness factor of order $0.4-0.6$, and $\sigma_\mathrm{fermi}=4.50\times 10^{35} \, \mathrm{erg \, MeV^{-4} \, cm^{-2} \, s^{-1}}$ being the radiation constant for massless fermions \citep{huedepohl_10}. As before, $R$ is the PNS radius, while $T$ is its surface temperature. 

The mean energy of neutrinos is also evolving during this phase, tending to increase with the mass of the PNS and is thus driven by ongoing accretion onto the remnant, loosely obeying \citep{mueller_14},
\begin{equation}
\label{eq:eman}
    \langle E_\nu \rangle \propto M_\mathrm{PNS}.
\end{equation}
Because of the energy-dependence of the cross sections for neutrino detection, the detailed neutrino energy spectrum is relevant for interpreting observed neutrino fluxes. It is important to take into account that the neutrino spectra are generally not thermal because the neutrinos of different energy decouple from the matter at different depths and hence different temperatures. Especially for electron-flavor neutrinos, this leads to ``pinching'', as the neutrinos in the high-energy tail are from \emph{colder} layers further outside.
A convenient mathematical model for the neutrino energy distribution is given by \citet{keil_03}, in terms of the neutrino mean energy and a shape parameter $\alpha$,
\begin{equation}
    f_\nu \propto E_\nu e^{-(\alpha + 1) E_\nu / \langle E_\nu \rangle}.
\end{equation}
Subsequently, \citet{tamborra_12} calculated suitable values of $\alpha$, which differ between phases of collapse, neutrino flavors, and progenitors, but typically sits in the range $\alpha \approx 2-3$, with values towards the lower end for $\nu_\mu$ and $\nu_\tau$ and higher values for $\nu_\mathrm{e}$ and $\bar{\nu}_\mathrm{e}$.

\begin{figure*}[h]
    \centering
    \includegraphics[width=0.65\linewidth]{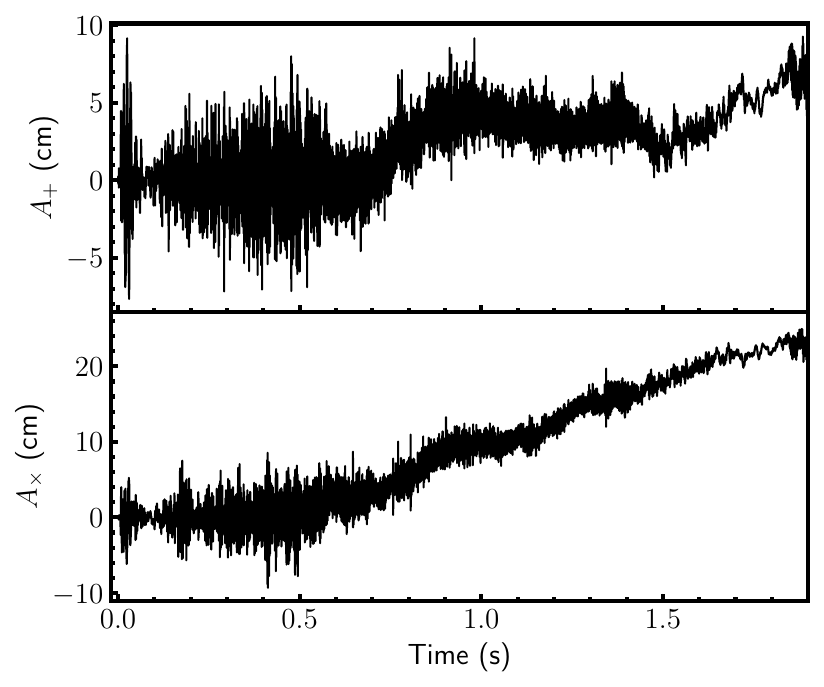}
    \caption{The GW signal from the simulation of a $15 \, \msun$ progenitor of supersolar metallicity. The top panel shows the `plus' polarization $A_+$ of the GW while  the bottom panel shows the `cross' polarization $A_\times$ . Both polarizations exhibit a signal from prompt convection in the first $\mathord{\sim}50\,\mathrm{ms}$ after bounce, followed by high-frequency emission from about $\mathord{\sim}100\,\mathrm{ms}$ after bounce. Later during the explosion phase, the high-frequency emission becomes weaker. The cross polarization shows a pronounced tail signal during this phase.
    }
    \label{fig:gw_pc}
\end{figure*}

\begin{figure*}[h]
    \centering
    \includegraphics[width=0.65\linewidth]{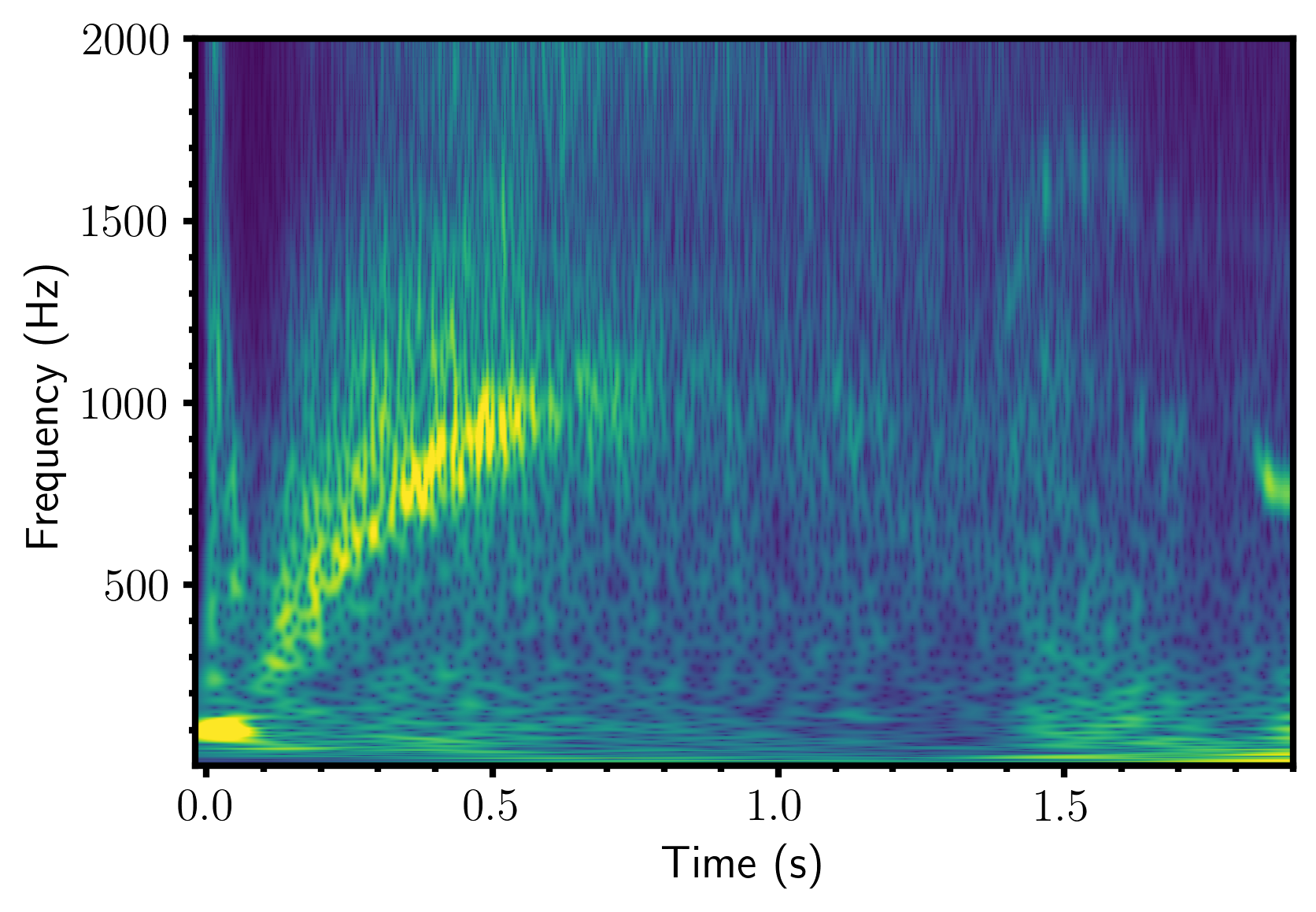}
    \caption{Spectrogram of the cross polarization in Figure~\ref{fig:gw_pc} showing a dominant rising mode during the accretion and early explosion phase. 
    }
    \label{fig:gw_spec}
\end{figure*}

Equations~(\ref{eqn:neutrino_lum_e}), (\ref{eqn:neutrino_lum_heavy}) 
and (\ref{eq:eman})
make it clear that CCSN neutrino emission could, in principle, constrain bulk PNS properties and mass accretion rates. Due to neutrino flavor conversion, it is not straightforward to reconstruct flux at the source from the observed flavor-dependent fluxes on Earth for this purpose. Nonetheless, meaningful constraints on the nature of the progenitor
(low vs.\ high compactness) are possible within different scenarios for neutrino flavor conversion, even from the small number of neutrinos observed from SN~1987A, and more so in case of a future Galactic supernova \citep{bruenn_87,burrows_88a,oconnor_13,horiuchi_18}.

Neutrinos can potentially even reveal multi-dimensional fluid flow in the supernova core. If the shock develops SASI oscillations, this will modulate the direction-dependent accretion flow onto the PNS and hence the neutrino emission. Analyses based on 2D and 3D simulations have shown that this modulation encodes the frequency of SASI shock oscillations, which is related to the shock and PNS radius. For a sufficiently high signal-to-noise ratio, one can reconstruct the time-dependent expansion and contraction of the shock \citep{mueller_14,mueller_19d}.

A first-order phase transition with a second collapse
to a more compact stable NS could also leave smoking guns in the neutrino emission. The formation of a second shock in the phase-transition
mechanism for CCSN explosions also produces a second breakout burst, this time for electron antineutrinos
\citep{sagert_09,dasgupta_10,fischer_18}. In other cases, where the second bounce is too weak to trigger an explosion, there could still be ringdown oscillations that produce an oscillatory neutrino signal \citep{zha_21,lin_24}. This signal would be clearly distinguishable from SASI-induced modulations by its much higher frequency.

In the explosion phase, accretion onto the PNS gradually subsides, and there is a smooth transition into the third phase of neutrino-emission, the Kelvin-Helmholtz cooling phase, which lasts for several seconds. During this phase, the neutrino emission of all flavors is fed by slow diffusion out of the PNS. The luminosities of all flavors become similar. The spectra also become more similar, though electron antineutrinos maintain lower mean energies and hence a higher number flux than the other flavors. As a result, the PNS cools down and approaches the more stable configuration of a cold NS in $\beta$-equilibrium.

A substantial fraction of the NS binding energy is radiated during the cooling phase. The total time-integrated luminosity in all flavors is a key observable from the accretion and cooling phase, as the binding energy constrains the NS mass and radius (and, via the radius, also the nuclear EoS).
Energy loss through hypothetical particles
such as axions, sterile neutrinos or Kaluza-Klein gravitons \citep{keil_97,hannestad_01,raffelt_11b,fischer_16b} would modify the total radiated neutrino energy, and thus the cooling phase is also a laboratory for non-standard particle physics. The time-dependence of the luminosities during the Kelvin-Helmholtz cooling phase is also sensitive to the properties of high-density nuclear matter such as the symmetry energy \citep{roberts_12b} and the appearance of additional hadrons or quark matter \citep{pons_99,pons_01a,pons_01b}. 

The Kelvin-Helmholtz cooling phase ends once the PNS becomes transparent to neutrinos and transitions from diffusive cooling to volume cooling. This much longer cooling phase is discussed in greater detail in Chapter~9.

Finally, the population-integrated neutrino emission -- the \emph{diffuse supernova neutrino background} (DSNB) -- is also an important probe of supernova physics and stellar physics. We refer the reader to specialist literature on this topic \citep{beacom_10,kresse_21,horiuchi_21,lunardini_26}.

\subsection{Gravitational Wave Signal}
While neutrinos have already been detected from one CCSN, namely SN~1987A, GWs are yet to be detected from a stellar explosion. In the event of a Galactic supernova, they would provide a valuable complementary diagnostic for the dynamics in the supernova core. GWs are sensitive to asymmetric mass motions that produce temporal variations in the mass quadrupole moment \citep{einstein}.

There are many processes that induce asymmetry in CCSNe: rotation, convection, the SASI, as well as the growth of seed asymmetries from the progenitor. The GW emission from CCSNe has been discussed extensively in recent overviews \citep{abdikamalov_22,mezzacappa_25,mueller_26}. We here provide only a brief outline.

An example of the GW signal from a CCSN is shown in Figure~\ref{fig:gw_pc}. Both plus and cross polarizations contain similar features, however there are some noteworthy differences too. Additionally, the $A_{+,\times}$ quantity plotted, with units of cm, is referred to as the GW amplitude and is independent of the observer distance. The \emph{strain} amplitude $h$, which is measured by GW detectors, scales
as $h\sim A_{+,\times}/r$.

The first component of the GW signal only occurs in relatively rapidly rotating stars.
Once the core reaches nuclear saturation densities, rotating stars produce a characteristic, high-amplitude bounce signal as the rotationally deformed core contracts, re-expands and then undergoes some ring-down oscillations \citep{dimmelmeier_07_a,dimmelmeier_08}. 
For the non-rotating star shown in Figure~\ref{fig:gw_pc}, the bounce signal is absent.

Following bounce, prompt convection in the post-shock region produces convective plumes and oscillations of the shock. While the convection quickly mixes the region, the ongoing shock oscillations it triggered radiate GWs with a typical frequency of order $100 \, \mathrm{Hz}$. An example of this prompt convection signal is present in the plus polarization in Figure~\ref{fig:gw_pc} for the first ${\sim} 50\, \mathrm{ms}$ after bounce.

The subsequent GW signal is often dominated by a high-frequency component which further increases in frequency with time, typically reaching somewhat above $1\,\mathrm{kHz}$ \citep[e.g.,][]{mueller_13,radice_19,mezzacappa_20b,mezzacappa_23}. Current modeling efforts suggest that this is the most robust signal from CCSNe. Its origin is oscillations of the PNS surface region, which are excited by aspherical driving forces from both the gain region and the convective region inside the PNS \citep{murphy_09,mueller_13}. More precisely, the dominant mode is a buoyancy-driven gravity mode, or g-mode for short; the dominant mode is also often called the fundamental, or f-mode \citep{torres_18,morozova_18}. Other g-modes exist, and are sometimes excited by fluid motions; these often have a different frequency evolution, such as decreasing in frequency with time. Pressure modes, or p-modes, are another type of oscillation which can produce GWs and are acoustic by nature. While g-modes are more strongly linked to the PNS, p-modes are more sensitive to the cavity formed by the gain region and shock; however, both types are broadly dependent on the structure of the entire post-shock region, down to the PNS core.

The dominant mode in the GW signal holds information about the PNS structure.
The frequency of this mode can be described
by a relatively simple but robust relation \citep{mueller_13},
\begin{equation}
    f_\mathrm{g} \approx \frac{1}{2 \pi} \frac{GM}{R^{2}} \sqrt{\frac{1.1 m_\mathrm{n} }{\langle E_{\bar{\nu}_\mathrm{e}} \rangle}}
    \left(1-\frac{G M}{R c^2}\right)^2,
\end{equation}
where $M$ and $R$ are the PNS mass and radius, $m_\mathrm{n}$ is the mass of a neutron, and $\langle E_{\bar{\nu}_\mathrm{e}} \rangle$ is the mean energy of electron antineutrinos as a proxy
for the PNS surface temperature. The last factor accounts for general relativistic corrections.

The SASI can also leave an imprint in the GW signal of CCSNe, typically at lower frequencies of $100-200 \, \mathrm{Hz}$. GWs from the SASI encode information about the radius of the shock and PNS, offered another view of the interior dynamics of CCSNe, assuming it can be detected.

Other low-frequency contributions to the GW signal are produced by anisotropic neutrino emission \citep{epstein_78}, and asymmetric shock expansion \citep{murphy_09,mueller_13}. These produce tails 
in the GW signal with amplitudes
that do not return to zero (memory effect). 
Such an effect is evident starting $200 \, \mathrm{ms}$ after bounce in the plus polarization in Figure~\ref{fig:gw_pc}, where the signal starts trending upwards.
Though the amplitude of the tail signal
can be very large, their detection is not straightforward. Good sensitivity at low frequencies in future detectors, possibly also in the deci-Hertz range \citep{choi_24} is required.

\section{Conclusions}
Core-collapse supernovae remain a unique laboratory for matter under extreme conditions and its interactions with neutrinos. Over the past decades, significant progress has already been made in incorporating sophisticated microphysics into CCSN simulations. This has enabled greater simulation fidelity and has contributed significantly to the advent of successful three-dimensional explosion models that are now able to explain important trends and tendencies among CCSN explosion and remnant properties. Nonetheless, the microphysical input in CCSN simulations is by no means fully settled. Important (albeit shrinking) uncertainties remain about the nuclear EoS, in-medium neutrino interaction rates and neutrino quantum kinetics. Macroscopic simulations have proved useful in identifying key uncertainties, and also in identifying diagnostics that can at least indirectly probe the microphysics, e.g., through the signatures of PNS contraction in a prospective GW signal from a Galactic CCSN. A key challenge for the future will be to formalize the quantification of uncertainties and sensitivities in simulations and predictions of multi-messenger observables to maximize the science that can be learned from a Galactic supernova.

\begin{ack}[Acknowledgments]
{}
BM acknowledges support by the Australian Research Council through grants DP240101786 and DP260104967,
by Australia Limited's ASTAC scheme, and by the National Computational Merit Allocation Scheme (NCMAS).
Some of the presented work was performed on the Gadi supercomputer with the assistance of resources and services from the National Computational Infrastructure (NCI), which is supported by the Australian Government.

\end{ack}

\seealso{
Astrophysics of supernovae and supernova observations: \citet{gal-yam_17,jerkstrand_17,jerkstrand_26};
Supernova mechanisms and simulations:
\citet{mueller_20,burrows_21,mueller_25b,janka_25};
Neutrino emission from supernovae: \citet{mirizzi_16,janka_17b,mueller_19d};
Neutrino transport and quantum kinetics:
\citet{duan_10,mirizzi_16,mezzacappa_20,johns_25,raffelt_26}
Gravitational waves: \citet{abdikamalov_22,mezzacappa_25,mueller_26}
}

\bibliographystyle{Harvard}
\onecolumn
\bibliography{paper}

\begin{thebibliography*}{218}
\providecommand{\bibtype}[1]{}
\providecommand{\natexlab}[1]{#1}
{\catcode`\|=0\catcode`\#=12\catcode`\@=11\catcode`\\=12
|immediate|write|@auxout{\expandafter\ifx\csname
  natexlab\endcsname\relax\gdef\natexlab#1{#1}\fi}}
\renewcommand{\url}[1]{{\tt #1}}
\providecommand{\urlprefix}{URL }
\expandafter\ifx\csname urlstyle\endcsname\relax
  \providecommand{\doi}[1]{doi:\discretionary{}{}{}#1}\else
  \providecommand{\doi}{doi:\discretionary{}{}{}\begingroup
  \urlstyle{rm}\Url}\fi
\providecommand{\bibinfo}[2]{#2}
\providecommand{\eprint}[2][]{\url{#2}}

\bibtype{Article}%
\bibitem[{Abbott} et al.(2018)]{abbott_18}
\bibinfo{author}{{Abbott} BP}, \bibinfo{author}{{Abbott} R},
  \bibinfo{author}{{Abbott} TD}, \bibinfo{author}{{Acernese} F},
  \bibinfo{author}{{Ackley} K}, \bibinfo{author}{{Adams} C},
  \bibinfo{author}{{Adams} T}, \bibinfo{author}{{Addesso} P},
  \bibinfo{author}{{Adhikari} RX}, \bibinfo{author}{{Adya} VB},
  \bibinfo{author}{{Affeldt} C}, \bibinfo{author}{{Agarwal} B},
  \bibinfo{author}{{Agathos} M}, \bibinfo{author}{{Agatsuma} K},
  \bibinfo{author}{{Aggarwal} N}, \bibinfo{author}{{Aguiar} OD},
  \bibinfo{author}{{Aiello} L}, \bibinfo{author}{{Ain} A},
  \bibinfo{author}{{Ajith} P}, \bibinfo{author}{{Allen} B},
  \bibinfo{author}{{Allen} G}, \bibinfo{author}{{Allocca} A},
  \bibinfo{author}{{Aloy} MA}, \bibinfo{author}{{Altin} PA},
  \bibinfo{author}{{Amato} A}, \bibinfo{author}{{Ananyeva} A},
  \bibinfo{author}{{Anderson} SB}, \bibinfo{author}{{Anderson} WG},
  \bibinfo{author}{{Angelova} SV}, \bibinfo{author}{{Antier} S},
  \bibinfo{author}{{Appert} S}, \bibinfo{author}{{Arai} K},
  \bibinfo{author}{{Araya} MC}, \bibinfo{author}{{Areeda} JS},
  \bibinfo{author}{{Ar{\`e}ne} M}, \bibinfo{author}{{Arnaud} N},
  \bibinfo{author}{{Arun} KG}, \bibinfo{author}{{Ascenzi} S},
  \bibinfo{author}{{Ashton} G}, \bibinfo{author}{{Ast} M},
  \bibinfo{author}{{Aston} SM}, \bibinfo{author}{{Astone} P},
  \bibinfo{author}{{Atallah} DV}, \bibinfo{author}{{Aubin} F},
  \bibinfo{author}{{Aufmuth} P}, \bibinfo{author}{{Aulbert} C},
  \bibinfo{author}{{AultONeal} K}, \bibinfo{author}{{Austin} C},
  \bibinfo{author}{{Avila-Alvarez} A}, \bibinfo{author}{{Babak} S},
  \bibinfo{author}{{Bacon} P}, \bibinfo{author}{{Badaracco} F},
  \bibinfo{author}{{Bader} MKM}, \bibinfo{author}{{Bae} S},
  \bibinfo{author}{{Baker} PT}, \bibinfo{author}{{Baldaccini} F},
  \bibinfo{author}{{Ballardin} G}, \bibinfo{author}{{Ballmer} SW},
  \bibinfo{author}{{Banagiri} S}, \bibinfo{author}{{Barayoga} JC},
  \bibinfo{author}{{Barclay} SE}, \bibinfo{author}{{Barish} BC},
  \bibinfo{author}{{Barker} D}, \bibinfo{author}{{Barkett} K},
  \bibinfo{author}{{Barnum} S}, \bibinfo{author}{{Barone} F},
  \bibinfo{author}{{Barr} B}, \bibinfo{author}{{Barsotti} L},
  \bibinfo{author}{{Barsuglia} M}, \bibinfo{author}{{Barta} D},
  \bibinfo{author}{{Bartlett} J}, \bibinfo{author}{{Bartos} I},
  \bibinfo{author}{{Bassiri} R}, \bibinfo{author}{{Basti} A},
  \bibinfo{author}{{Batch} JC}, \bibinfo{author}{{Bawaj} M},
  \bibinfo{author}{{Bayley} JC}, \bibinfo{author}{{Bazzan} M},
  \bibinfo{author}{{B{\'e}csy} B}, \bibinfo{author}{{Beer} C},
  \bibinfo{author}{{Bejger} M}, \bibinfo{author}{{Belahcene} I},
  \bibinfo{author}{{Bell} AS}, \bibinfo{author}{{Beniwal} D},
  \bibinfo{author}{{Bensch} M}, \bibinfo{author}{{Berger} BK},
  \bibinfo{author}{{Bergmann} G}, \bibinfo{author}{{Bernuzzi} S},
  \bibinfo{author}{{Bero} JJ}, \bibinfo{author}{{Berry} CPL},
  \bibinfo{author}{{Bersanetti} D}, \bibinfo{author}{{Bertolini} A},
  \bibinfo{author}{{Betzwieser} J}, \bibinfo{author}{{Bhandare} R},
  \bibinfo{author}{{Bilenko} IA}, \bibinfo{author}{{Bilgili} SA},
  \bibinfo{author}{{Billingsley} G}, \bibinfo{author}{{Billman} CR},
  \bibinfo{author}{{Birch} J}, \bibinfo{author}{{Birney} R},
  \bibinfo{author}{{Birnholtz} O}, \bibinfo{author}{{Biscans} S},
  \bibinfo{author}{{Biscoveanu} S}, \bibinfo{author}{{Bisht} A},
  \bibinfo{author}{{Bitossi} M}, \bibinfo{author}{{Bizouard} MA},
  \bibinfo{author}{{Blackburn} JK}, \bibinfo{author}{{Blackman} J},
  \bibinfo{author}{{Blair} CD}, \bibinfo{author}{{Blair} DG},
  \bibinfo{author}{{Blair} RM}, \bibinfo{author}{{Bloemen} S},
  \bibinfo{author}{{Bock} O}, \bibinfo{author}{{Bode} N},
  \bibinfo{author}{{Boer} M}, \bibinfo{author}{{Boetzel} Y},
  \bibinfo{author}{{Bogaert} G}, \bibinfo{author}{{Bohe} A},
  \bibinfo{author}{{Bondu} F}, \bibinfo{author}{{Bonilla} E},
  \bibinfo{author}{{Bonnand} R}, \bibinfo{author}{{Booker} P},
  \bibinfo{author}{{Boom} BA}, \bibinfo{author}{{Booth} CD},
  \bibinfo{author}{{Bork} R}, \bibinfo{author}{{Boschi} V},
  \bibinfo{author}{{Bose} S}, \bibinfo{author}{{Bossie} K},
  \bibinfo{author}{{Bossilkov} V}, \bibinfo{author}{{Bosveld} J},
  \bibinfo{author}{{Bouffanais} Y}, \bibinfo{author}{{Bozzi} A},
  \bibinfo{author}{{Bradaschia} C}, \bibinfo{author}{{Brady} PR},
  \bibinfo{author}{{Bramley} A}, \bibinfo{author}{{Branchesi} M},
  \bibinfo{author}{{Brau} JE}, \bibinfo{author}{{Briant} T},
  \bibinfo{author}{{Brighenti} F}, \bibinfo{author}{{Brillet} A},
  \bibinfo{author}{{Brinkmann} M}, \bibinfo{author}{{Brisson} V},
  \bibinfo{author}{{Brockill} P}, \bibinfo{author}{{Brooks} AF},
  \bibinfo{author}{{Brown} DD}, \bibinfo{author}{{Brunett} S},
  \bibinfo{author}{{Buchanan} CC}, \bibinfo{author}{{Buikema} A},
  \bibinfo{author}{{Bulik} T}, \bibinfo{author}{{Bulten} HJ},
  \bibinfo{author}{{Buonanno} A}, \bibinfo{author}{{Buskulic} D},
  \bibinfo{author}{{Buy} C}, \bibinfo{author}{{Byer} RL},
  \bibinfo{author}{{Cabero} M}, \bibinfo{author}{{Cadonati} L},
  \bibinfo{author}{{Cagnoli} G}, \bibinfo{author}{{Cahillane} C},
  \bibinfo{author}{{Calder{\'o}n Bustillo} J}, \bibinfo{author}{{Callister}
  TA}, \bibinfo{author}{{Calloni} E}, \bibinfo{author}{{Camp} JB},
  \bibinfo{author}{{Canepa} M}, \bibinfo{author}{{Canizares} P},
  \bibinfo{author}{{Cannon} KC}, \bibinfo{author}{{Cao} H},
  \bibinfo{author}{{Cao} J}, \bibinfo{author}{{Capano} CD},
  \bibinfo{author}{{Capocasa} E}, \bibinfo{author}{{Carbognani} F},
  \bibinfo{author}{{Caride} S}, \bibinfo{author}{{Carney} MF},
  \bibinfo{author}{{Carullo} G}, \bibinfo{author}{{Casanueva Diaz} J},
  \bibinfo{author}{{Casentini} C}, \bibinfo{author}{{Caudill} S},
  \bibinfo{author}{{Cavagli{\`a}} M}, \bibinfo{author}{{Cavalier} F},
  \bibinfo{author}{{Cavalieri} R}, \bibinfo{author}{{Cella} G},
  \bibinfo{author}{{Cepeda} CB}, \bibinfo{author}{{Cerd{\'a}-Dur{\'a}n} P},
  \bibinfo{author}{{Cerretani} G}, \bibinfo{author}{{Cesarini} E},
  \bibinfo{author}{{Chaibi} O}, \bibinfo{author}{{Chamberlin} SJ},
  \bibinfo{author}{{Chan} M}, \bibinfo{author}{{Chao} S},
  \bibinfo{author}{{Charlton} P}, \bibinfo{author}{{Chase} E},
  \bibinfo{author}{{Chassande-Mottin} E}, \bibinfo{author}{{Chatterjee} D},
  \bibinfo{author}{{Chatziioannou} K}, \bibinfo{author}{{Cheeseboro} BD},
  \bibinfo{author}{{Chen} HY}, \bibinfo{author}{{Chen} X},
  \bibinfo{author}{{Chen} Y}, \bibinfo{author}{{Cheng} HP},
  \bibinfo{author}{{Chia} HY} and  \bibinfo{author}{{Chincarini} A}
  (\bibinfo{year}{2018}), \bibinfo{month}{Oct.}
\bibinfo{title}{{GW170817: Measurements of Neutron Star Radii and Equation of
  State}}.
\bibinfo{journal}{{\em \prl}} \bibinfo{volume}{121} (\bibinfo{number}{16}),
  \bibinfo{eid}{161101}. \bibinfo{doi}{\doi{10.1103/PhysRevLett.121.161101}}.
\eprint{1805.11581}.

\bibtype{incollection}%
\bibitem[{Abdikamalov} et al.(2022)]{abdikamalov_22}
\bibinfo{author}{{Abdikamalov} E}, \bibinfo{author}{{Pagliaroli} G} and
  \bibinfo{author}{{Radice} D} (\bibinfo{year}{2022}),
  \bibinfo{title}{{Gravitational Waves from Core-Collapse Supernovae}},
  \bibinfo{editor}{C.~Bambi S.~Katsanevas KDK}, (Ed.),
  \bibinfo{booktitle}{Handbook of Gravitational Wave Astronomy},
  \bibinfo{publisher}{Springer Nature}, \bibinfo{address}{Singapore},
  pp.~\bibinfo{pages}{21}.

\bibtype{Article}%
\bibitem[{Adams} et al.(2017)]{adams_17}
\bibinfo{author}{{Adams} SM}, \bibinfo{author}{{Kochanek} CS},
  \bibinfo{author}{{Gerke} JR}, \bibinfo{author}{{Stanek} KZ} and
  \bibinfo{author}{{Dai} X} (\bibinfo{year}{2017}), \bibinfo{month}{Jul.}
\bibinfo{title}{{The search for failed supernovae with the Large Binocular
  Telescope: confirmation of a disappearing star}}.
\bibinfo{journal}{{\em \mnras}} \bibinfo{volume}{468} (\bibinfo{number}{4}):
  \bibinfo{pages}{4968--4981}. \bibinfo{doi}{\doi{10.1093/mnras/stx816}}.
\eprint{1609.01283}.

\bibtype{Article}%
\bibitem[{Airapetian} et al.(2007)]{airapetian_07}
\bibinfo{author}{{Airapetian} A}, \bibinfo{author}{{Akopov} N},
  \bibinfo{author}{{Akopov} Z}, \bibinfo{author}{{Andrus} A},
  \bibinfo{author}{{Aschenauer} EC}, \bibinfo{author}{{Augustyniak} W} and
  \bibinfo{author}{{Avakian} R} (\bibinfo{year}{2007}), \bibinfo{month}{Jan.}
\bibinfo{title}{{Precise determination of the spin structure function g$_{1}$
  of the proton, deuteron, and neutron}}.
\bibinfo{journal}{{\em \prd}} \bibinfo{volume}{75} (\bibinfo{number}{1}):
  \bibinfo{pages}{012007}. \bibinfo{doi}{\doi{10.1103/PhysRevD.75.012007}}.

\bibtype{Article}%
\bibitem[{Alekseev} et al.(1987)]{alekseev_87}
\bibinfo{author}{{Alekseev} EN}, \bibinfo{author}{{Alekseeva} LN},
  \bibinfo{author}{{Volchenko} VI} and  \bibinfo{author}{{Krivosheina} IV}
  (\bibinfo{year}{1987}), \bibinfo{month}{May}.
\bibinfo{title}{{Possible detection of a neutrino signal on 23 February 1987 at
  the Baksan underground scintillation telescope of the Institute of Nuclear
  Research}}.
\bibinfo{journal}{{\em Soviet Journal of Experimental and Theoretical Physics
  Letters}} \bibinfo{volume}{45}: \bibinfo{pages}{589}.

\bibtype{Article}%
\bibitem[{Arnett} et al.(1989)]{arnett_89}
\bibinfo{author}{{Arnett} WD}, \bibinfo{author}{{Bahcall} JN},
  \bibinfo{author}{{Kirshner} RP} and  \bibinfo{author}{{Woosley} SE}
  (\bibinfo{year}{1989}).
\bibinfo{title}{{Supernova 1987A}}.
\bibinfo{journal}{{\em \araa}} \bibinfo{volume}{27}: \bibinfo{pages}{629--700}.
  \bibinfo{doi}{\doi{10.1146/annurev.aa.27.090189.003213}}.

\bibtype{Article}%
\bibitem[{Baade} and {Zwicky}(1934{\natexlab{a}})]{baade_34b}
\bibinfo{author}{{Baade} W} and  \bibinfo{author}{{Zwicky} F}
  (\bibinfo{year}{1934}{\natexlab{a}}).
\bibinfo{title}{{Cosmic Rays from Super-novae}}.
\bibinfo{journal}{{\em Proceedings of the National Academy of Science}}
  \bibinfo{volume}{20}: \bibinfo{pages}{259--263}.
  \bibinfo{doi}{\doi{10.1073/pnas.20.5.259}}.

\bibtype{Article}%
\bibitem[{Baade} and {Zwicky}(1934{\natexlab{b}})]{baade_34a}
\bibinfo{author}{{Baade} W} and  \bibinfo{author}{{Zwicky} F}
  (\bibinfo{year}{1934}{\natexlab{b}}).
\bibinfo{title}{{On Super-novae}}.
\bibinfo{journal}{{\em Proceedings of the National Academy of Science}}
  \bibinfo{volume}{20}: \bibinfo{pages}{254--259}.
  \bibinfo{doi}{\doi{10.1073/pnas.20.5.254}}.

\bibtype{Article}%
\bibitem[{Baade} and {Zwicky}(1934{\natexlab{c}})]{baade_34c}
\bibinfo{author}{{Baade} W} and  \bibinfo{author}{{Zwicky} F}
  (\bibinfo{year}{1934}{\natexlab{c}}), \bibinfo{month}{Jul.}
\bibinfo{title}{{Remarks on Super-Novae and Cosmic Rays}}.
\bibinfo{journal}{{\em Physical Review}} \bibinfo{volume}{46}
  (\bibinfo{number}{1}): \bibinfo{pages}{76--77}.
  \bibinfo{doi}{\doi{10.1103/PhysRev.46.76.2}}.

\bibtype{Article}%
\bibitem[{Barkat} et al.(1967)]{barkat_67}
\bibinfo{author}{{Barkat} Z}, \bibinfo{author}{{Rakavy} G} and
  \bibinfo{author}{{Sack} N} (\bibinfo{year}{1967}), \bibinfo{month}{Mar.}
\bibinfo{title}{{Dynamics of Supernova Explosion Resulting from Pair
  Formation}}.
\bibinfo{journal}{{\em \prl}} \bibinfo{volume}{18} (\bibinfo{number}{10}):
  \bibinfo{pages}{379--381}. \bibinfo{doi}{\doi{10.1103/PhysRevLett.18.379}}.

\bibtype{Article}%
\bibitem[{Bartl} et al.(2016)]{bartl_16}
\bibinfo{author}{{Bartl} A}, \bibinfo{author}{{Bollig} R},
  \bibinfo{author}{{Janka} HT} and  \bibinfo{author}{{Schwenk} A}
  (\bibinfo{year}{2016}), \bibinfo{month}{Oct.}
\bibinfo{title}{{Impact of nucleon-nucleon bremsstrahlung rates beyond one-pion
  exchange}}.
\bibinfo{journal}{{\em \prd}} \bibinfo{volume}{94} (\bibinfo{number}{8}),
  \bibinfo{eid}{083009}. \bibinfo{doi}{\doi{10.1103/PhysRevD.94.083009}}.
\eprint{1608.05037}.

\bibtype{Article}%
\bibitem[{Beacom}(2010)]{beacom_10}
\bibinfo{author}{{Beacom} JF} (\bibinfo{year}{2010}), \bibinfo{month}{Nov.}
\bibinfo{title}{{The Diffuse Supernova Neutrino Background}}.
\bibinfo{journal}{{\em Annual Review of Nuclear and Particle Science}}
  \bibinfo{volume}{60}: \bibinfo{pages}{439--462}.
  \bibinfo{doi}{\doi{10.1146/annurev.nucl.010909.083331}}.

\bibtype{Article}%
\bibitem[{Beasor} et al.(2025)]{beasor_25}
\bibinfo{author}{{Beasor} ER}, \bibinfo{author}{{Smith} N} and
  \bibinfo{author}{{Jencson} JE} (\bibinfo{year}{2025}), \bibinfo{month}{Feb.}
\bibinfo{title}{{The Red Supergiant Progenitor Luminosity Problem}}.
\bibinfo{journal}{{\em \apj}} \bibinfo{volume}{979} (\bibinfo{number}{2}),
  \bibinfo{eid}{117}. \bibinfo{doi}{\doi{10.3847/1538-4357/ad8f3f}}.
\eprint{2410.14027}.

\bibtype{Article}%
\bibitem[{Belczynski} et al.(2016)]{belczynski_16}
\bibinfo{author}{{Belczynski} K}, \bibinfo{author}{{Heger} A},
  \bibinfo{author}{{Gladysz} W}, \bibinfo{author}{{Ruiter} AJ},
  \bibinfo{author}{{Woosley} S}, \bibinfo{author}{{Wiktorowicz} G},
  \bibinfo{author}{{Chen} HY}, \bibinfo{author}{{Bulik} T},
  \bibinfo{author}{{O'Shaughnessy} R}, \bibinfo{author}{{Holz} DE},
  \bibinfo{author}{{Fryer} CL} and  \bibinfo{author}{{Berti} E}
  (\bibinfo{year}{2016}), \bibinfo{month}{Oct.}
\bibinfo{title}{{The effect of pair-instability mass loss on black-hole
  mergers}}.
\bibinfo{journal}{{\em \aap}} \bibinfo{volume}{594}, \bibinfo{eid}{A97}.
  \bibinfo{doi}{\doi{10.1051/0004-6361/201628980}}.
\eprint{1607.03116}.

\bibtype{Article}%
\bibitem[{Bethe} and {Wilson}(1985)]{bethe_85}
\bibinfo{author}{{Bethe} HA} and  \bibinfo{author}{{Wilson} JR}
  (\bibinfo{year}{1985}), \bibinfo{month}{Aug.}
\bibinfo{title}{{Revival of a stalled supernova shock by neutrino heating}}.
\bibinfo{journal}{{\em \apj}} \bibinfo{volume}{295}: \bibinfo{pages}{14--23}.
  \bibinfo{doi}{\doi{10.1086/163343}}.

\bibtype{Article}%
\bibitem[{Bionta} et al.(1987)]{bionta_87}
\bibinfo{author}{{Bionta} RM}, \bibinfo{author}{{Blewitt} G},
  \bibinfo{author}{{Bratton} CB}, \bibinfo{author}{{Casper} D} and
  \bibinfo{author}{{Ciocio} A} (\bibinfo{year}{1987}), \bibinfo{month}{Apr.}
\bibinfo{title}{{Observation of a neutrino burst in coincidence with supernova
  1987A in the Large Magellanic Cloud}}.
\bibinfo{journal}{{\em Physical Review Letters}} \bibinfo{volume}{58}:
  \bibinfo{pages}{1494--1496}.
  \bibinfo{doi}{\doi{10.1103/PhysRevLett.58.1494}}.

\bibtype{Article}%
\bibitem[{Blondin} et al.(2003)]{blondin_03}
\bibinfo{author}{{Blondin} JM}, \bibinfo{author}{{Mezzacappa} A} and
  \bibinfo{author}{{DeMarino} C} (\bibinfo{year}{2003}), \bibinfo{month}{Feb.}
\bibinfo{title}{{Stability of Standing Accretion Shocks, with an Eye toward
  Core-Collapse Supernovae}}.
\bibinfo{journal}{{\em \apj}} \bibinfo{volume}{584}: \bibinfo{pages}{971--980}.
  \bibinfo{doi}{\doi{10.1086/345812}}.

\bibtype{Article}%
\bibitem[{Bodansky} et al.(1968)]{bodansky_68}
\bibinfo{author}{{Bodansky} D}, \bibinfo{author}{{Clayton} DD} and
  \bibinfo{author}{{Fowler} WA} (\bibinfo{year}{1968}), \bibinfo{month}{Nov}.
\bibinfo{title}{{Nuclear Quasi-Equilibrium during Silicon Burning}}.
\bibinfo{journal}{{\em \apjs}} \bibinfo{volume}{16}: \bibinfo{pages}{299}.
  \bibinfo{doi}{\doi{10.1086/190176}}.

\bibtype{Article}%
\bibitem[{Bollig} et al.(2017)]{bollig_17}
\bibinfo{author}{{Bollig} R}, \bibinfo{author}{{Janka} HT},
  \bibinfo{author}{{Lohs} A}, \bibinfo{author}{{Mart{\'{\i}}nez-Pinedo} G},
  \bibinfo{author}{{Horowitz} CJ} and  \bibinfo{author}{{Melson} T}
  (\bibinfo{year}{2017}), \bibinfo{month}{Dec.}
\bibinfo{title}{{Muon Creation in Supernova Matter Facilitates Neutrino-Driven
  Explosions}}.
\bibinfo{journal}{{\em Physical Review Letters}} \bibinfo{volume}{119}
  (\bibinfo{number}{24}): \bibinfo{pages}{242702}.
  \bibinfo{doi}{\doi{10.1103/PhysRevLett.119.242702}}.

\bibtype{Article}%
\bibitem[{Bollig} et al.(2021)]{bollig_21}
\bibinfo{author}{{Bollig} R}, \bibinfo{author}{{Yadav} N},
  \bibinfo{author}{{Kresse} D}, \bibinfo{author}{{Janka} HT},
  \bibinfo{author}{{M{\"u}ller} B} and  \bibinfo{author}{{Heger} A}
  (\bibinfo{year}{2021}), \bibinfo{month}{Jul.}
\bibinfo{title}{{Self-consistent 3D Supernova Models From -7 Minutes to +7 s: A
  1-bethe Explosion of a 19 M$_{{\ensuremath{\odot}}}$ Progenitor}}.
\bibinfo{journal}{{\em \apj}} \bibinfo{volume}{915} (\bibinfo{number}{1}),
  \bibinfo{eid}{28}. \bibinfo{doi}{\doi{10.3847/1538-4357/abf82e}}.

\bibtype{Article}%
\bibitem[{Bruenn}(1985)]{bruenn_85}
\bibinfo{author}{{Bruenn} SW} (\bibinfo{year}{1985}), \bibinfo{month}{Aug.}
\bibinfo{title}{{Stellar core collapse - Numerical model and infall epoch}}.
\bibinfo{journal}{{\em \apjs}} \bibinfo{volume}{58}: \bibinfo{pages}{771--841}.
  \bibinfo{doi}{\doi{10.1086/191056}}.

\bibtype{Article}%
\bibitem[{Bruenn}(1987)]{bruenn_87}
\bibinfo{author}{{Bruenn} SW} (\bibinfo{year}{1987}), \bibinfo{month}{Aug.}
\bibinfo{title}{{Neutrinos from SN1987A and current models of stellar-core
  collapse}}.
\bibinfo{journal}{{\em Physical Review Letters}} \bibinfo{volume}{59}:
  \bibinfo{pages}{938--941}. \bibinfo{doi}{\doi{10.1103/PhysRevLett.59.938}}.

\bibtype{Article}%
\bibitem[{Bruenn} et al.(1995)]{bruenn_95}
\bibinfo{author}{{Bruenn} SW}, \bibinfo{author}{{Mezzacappa} A} and
  \bibinfo{author}{{Dineva} T} (\bibinfo{year}{1995}), \bibinfo{month}{May}.
\bibinfo{title}{{Dynamic and diffusive instabilities in core collapse
  supernovae.}}
\bibinfo{journal}{{\em \physrep}} \bibinfo{volume}{256}:
  \bibinfo{pages}{69--94}. \bibinfo{doi}{\doi{10.1016/0370-1573(94)00102-9}}.

\bibtype{Article}%
\bibitem[{Buras} et al.(2006{\natexlab{a}})]{buras_06b}
\bibinfo{author}{{Buras} R}, \bibinfo{author}{{Janka} HT},
  \bibinfo{author}{{Rampp} M} and  \bibinfo{author}{{Kifonidis} K}
  (\bibinfo{year}{2006}{\natexlab{a}}), \bibinfo{month}{Oct.}
\bibinfo{title}{{Two-dimensional hydrodynamic core-collapse supernova
  simulations with spectral neutrino transport. II. Models for different
  progenitor stars}}.
\bibinfo{journal}{{\em \aap}} \bibinfo{volume}{457}: \bibinfo{pages}{281--308}.
  \bibinfo{doi}{\doi{10.1051/0004-6361:20054654}}.

\bibtype{Article}%
\bibitem[{Buras} et al.(2006{\natexlab{b}})]{buras_06a}
\bibinfo{author}{{Buras} R}, \bibinfo{author}{{Rampp} M},
  \bibinfo{author}{{Janka} HT} and  \bibinfo{author}{{Kifonidis} K}
  (\bibinfo{year}{2006}{\natexlab{b}}), \bibinfo{month}{Mar.}
\bibinfo{title}{{Two-dimensional hydrodynamic core-collapse supernova
  simulations with spectral neutrino transport. I. Numerical method and results
  for a $15 M_\odot$ star}}.
\bibinfo{journal}{{\em \aap}} \bibinfo{volume}{447}:
  \bibinfo{pages}{1049--1092}. \bibinfo{doi}{\doi{10.1051/0004-6361:20053783}}.

\bibtype{Article}%
\bibitem[{Burrows}(1988)]{burrows_88a}
\bibinfo{author}{{Burrows} A} (\bibinfo{year}{1988}), \bibinfo{month}{Nov.}
\bibinfo{title}{{Supernova neutrinos}}.
\bibinfo{journal}{{\em \apj}} \bibinfo{volume}{334}: \bibinfo{pages}{891--908}.
  \bibinfo{doi}{\doi{10.1086/166885}}.

\bibtype{Article}%
\bibitem[{Burrows} and {Goshy}(1993)]{burrows_93}
\bibinfo{author}{{Burrows} A} and  \bibinfo{author}{{Goshy} J}
  (\bibinfo{year}{1993}), \bibinfo{month}{Oct.}
\bibinfo{title}{{A Theory of Supernova Explosions}}.
\bibinfo{journal}{{\em \apjl}} \bibinfo{volume}{416}: \bibinfo{pages}{L75+}.
  \bibinfo{doi}{\doi{10.1086/187074}}.

\bibtype{Article}%
\bibitem[{Burrows} and {Sawyer}(1998)]{burrows_98}
\bibinfo{author}{{Burrows} A} and  \bibinfo{author}{{Sawyer} RF}
  (\bibinfo{year}{1998}), \bibinfo{month}{Jul.}
\bibinfo{title}{{Effects of correlations on neutrino opacities in nuclear
  matter}}.
\bibinfo{journal}{{\em \prc}} \bibinfo{volume}{58}: \bibinfo{pages}{554--571}.
  \bibinfo{doi}{\doi{10.1103/PhysRevC.58.554}}.

\bibtype{Article}%
\bibitem[{Burrows} and {Sawyer}(1999)]{burrows_99}
\bibinfo{author}{{Burrows} A} and  \bibinfo{author}{{Sawyer} RF}
  (\bibinfo{year}{1999}), \bibinfo{month}{Jan.}
\bibinfo{title}{{Many-body corrections to charged-current neutrino absorption
  rates in nuclear matter}}.
\bibinfo{journal}{{\em \prc}} \bibinfo{volume}{59}: \bibinfo{pages}{510--514}.
  \bibinfo{doi}{\doi{10.1103/PhysRevC.59.510}}.

\bibtype{Article}%
\bibitem[{Burrows} and {Vartanyan}(2021)]{burrows_21}
\bibinfo{author}{{Burrows} A} and  \bibinfo{author}{{Vartanyan} D}
  (\bibinfo{year}{2021}), \bibinfo{month}{Jan.}
\bibinfo{title}{{Core-collapse supernova explosion theory}}.
\bibinfo{journal}{{\em \nat}} \bibinfo{volume}{589} (\bibinfo{number}{7840}):
  \bibinfo{pages}{29--39}. \bibinfo{doi}{\doi{10.1038/s41586-020-03059-w}}.

\bibtype{Article}%
\bibitem[{Burrows} et al.(1995)]{burrows_95}
\bibinfo{author}{{Burrows} A}, \bibinfo{author}{{Hayes} J} and
  \bibinfo{author}{{Fryxell} BA} (\bibinfo{year}{1995}), \bibinfo{month}{Sep.}
\bibinfo{title}{{On the Nature of Core-Collapse Supernova Explosions}}.
\bibinfo{journal}{{\em \apj}} \bibinfo{volume}{450}: \bibinfo{pages}{830--850}.
  \bibinfo{doi}{\doi{10.1086/176188}}.

\bibtype{Article}%
\bibitem[{Burrows} et al.(2019)]{burrowS_19}
\bibinfo{author}{{Burrows} A}, \bibinfo{author}{{Radice} D} and
  \bibinfo{author}{{Vartanyan} D} (\bibinfo{year}{2019}), \bibinfo{month}{May}.
\bibinfo{title}{{Three-dimensional supernova explosion simulations of 9-, 10-,
  11-, 12-, and 13-$\mathrm{M}_\odot$ stars}}.
\bibinfo{journal}{{\em \mnras}} \bibinfo{volume}{485} (\bibinfo{number}{3}):
  \bibinfo{pages}{3153--3168}. \bibinfo{doi}{\doi{10.1093/mnras/stz543}}.

\bibtype{Article}%
\bibitem[{Burrows} et al.(2020)]{burrows_20}
\bibinfo{author}{{Burrows} A}, \bibinfo{author}{{Radice} D},
  \bibinfo{author}{{Vartanyan} D}, \bibinfo{author}{{Nagakura} H},
  \bibinfo{author}{{Skinner} MA} and  \bibinfo{author}{{Dolence} JC}
  (\bibinfo{year}{2020}), \bibinfo{month}{Jan}.
\bibinfo{title}{{The overarching framework of core-collapse supernova
  explosions as revealed by 3D FORNAX simulations}}.
\bibinfo{journal}{{\em \mnras}} \bibinfo{volume}{491} (\bibinfo{number}{2}):
  \bibinfo{pages}{2715--2735}. \bibinfo{doi}{\doi{10.1093/mnras/stz3223}}.

\bibtype{Article}%
\bibitem[{Burrows} et al.(2024)]{burrows_24a}
\bibinfo{author}{{Burrows} A}, \bibinfo{author}{{Wang} T} and
  \bibinfo{author}{{Vartanyan} D} (\bibinfo{year}{2024}), \bibinfo{month}{Mar.}
\bibinfo{title}{{Physical Correlations and Predictions Emerging from Modern
  Core-collapse Supernova Theory}}.
\bibinfo{journal}{{\em \apjl}} \bibinfo{volume}{964} (\bibinfo{number}{1}),
  \bibinfo{eid}{L16}. \bibinfo{doi}{\doi{10.3847/2041-8213/ad319e}}.

\bibtype{Article}%
\bibitem[{Burrows} et al.(2025)]{burrows_25}
\bibinfo{author}{{Burrows} A}, \bibinfo{author}{{Wang} T} and
  \bibinfo{author}{{Vartanyan} D} (\bibinfo{year}{2025}), \bibinfo{month}{Jul.}
\bibinfo{title}{{Channels of Stellar-mass Black Hole Formation}}.
\bibinfo{journal}{{\em \apj}} \bibinfo{volume}{987} (\bibinfo{number}{2}),
  \bibinfo{eid}{164}. \bibinfo{doi}{\doi{10.3847/1538-4357/addd04}}.
\eprint{2412.07831}.

\bibtype{Article}%
\bibitem[{Cabez{\'o}n} et al.(2018)]{cabezon_18}
\bibinfo{author}{{Cabez{\'o}n} RM}, \bibinfo{author}{{Pan} KC},
  \bibinfo{author}{{Liebend{\"o}rfer} M}, \bibinfo{author}{{Kuroda} T},
  \bibinfo{author}{{Ebinger} K}, \bibinfo{author}{{Heinimann} O},
  \bibinfo{author}{{Perego} A} and  \bibinfo{author}{{Thielemann} FK}
  (\bibinfo{year}{2018}), \bibinfo{month}{Nov.}
\bibinfo{title}{{Core-collapse supernovae in the hall of mirrors. A
  three-dimensional code-comparison project}}.
\bibinfo{journal}{{\em \aap}} \bibinfo{volume}{619}, \bibinfo{eid}{A118}.
  \bibinfo{doi}{\doi{10.1051/0004-6361/201833705}}.

\bibtype{Article}%
\bibitem[{Cardall} et al.(2013)]{cardall_13}
\bibinfo{author}{{Cardall} CY}, \bibinfo{author}{{Endeve} E} and
  \bibinfo{author}{{Mezzacappa} A} (\bibinfo{year}{2013}),
  \bibinfo{month}{Jul.}
\bibinfo{title}{{Conservative 3+1 general relativistic Boltzmann equation}}.
\bibinfo{journal}{{\em \prd}} \bibinfo{volume}{88} (\bibinfo{number}{2}),
  \bibinfo{eid}{023011}. \bibinfo{doi}{\doi{10.1103/PhysRevD.88.023011}}.
\eprint{1305.0037}.

\bibtype{Article}%
\bibitem[{Chan} et al.(2018)]{chan_18}
\bibinfo{author}{{Chan} C}, \bibinfo{author}{{M{\"u}ller} B},
  \bibinfo{author}{{Heger} A}, \bibinfo{author}{{Pakmor} R} and
  \bibinfo{author}{{Springel} V} (\bibinfo{year}{2018}), \bibinfo{month}{Jan.}
\bibinfo{title}{{Black Hole Formation and Fallback during the Supernova
  Explosion of a $40\,\mathrm{M}_\odot$ Star}}.
\bibinfo{journal}{{\em \apjl}} \bibinfo{volume}{852}: \bibinfo{pages}{L19}.
  \bibinfo{doi}{\doi{10.3847/2041-8213/aaa28c}}.

\bibtype{Article}%
\bibitem[{Chan} et al.(2020)]{chan_20b}
\bibinfo{author}{{Chan} C}, \bibinfo{author}{{M{\"u}ller} B} and
  \bibinfo{author}{{Heger} A} (\bibinfo{year}{2020}), \bibinfo{month}{Jul.}
\bibinfo{title}{{The impact of fallback on the compact remnants and chemical
  yields of core-collapse supernovae}}.
\bibinfo{journal}{{\em \mnras}} \bibinfo{volume}{495} (\bibinfo{number}{4}):
  \bibinfo{pages}{3751--3762}. \bibinfo{doi}{\doi{10.1093/mnras/staa1431}}.
\eprint{2003.04320}.

\bibtype{Article}%
\bibitem[{Choi} et al.(2024)]{choi_24}
\bibinfo{author}{{Choi} L}, \bibinfo{author}{{Burrows} A} and
  \bibinfo{author}{{Vartanyan} D} (\bibinfo{year}{2024}), \bibinfo{month}{Nov.}
\bibinfo{title}{{Gravitational-wave and Gravitational-wave Memory Signatures of
  Core-collapse Supernovae}}.
\bibinfo{journal}{{\em \apj}} \bibinfo{volume}{975} (\bibinfo{number}{1}),
  \bibinfo{eid}{12}. \bibinfo{doi}{\doi{10.3847/1538-4357/ad74f8}}.

\bibtype{Article}%
\bibitem[{Cinquegrana} et al.(2023)]{cinquegrana_23}
\bibinfo{author}{{Cinquegrana} GC}, \bibinfo{author}{{Joyce} M} and
  \bibinfo{author}{{Karakas} AI} (\bibinfo{year}{2023}), \bibinfo{month}{Nov.}
\bibinfo{title}{{Bridging the gap between intermediate and massive stars II:
  M$_{mas}$ for the most metal-rich stars and implications for Fe CCSNe
  rates}}.
\bibinfo{journal}{{\em \mnras}} \bibinfo{volume}{525} (\bibinfo{number}{3}):
  \bibinfo{pages}{3216--3235}. \bibinfo{doi}{\doi{10.1093/mnras/stad2461}}.
\eprint{2308.06002}.

\bibtype{Article}%
\bibitem[{Colgate} and {White}(1966)]{colgate_66}
\bibinfo{author}{{Colgate} SA} and  \bibinfo{author}{{White} RH}
  (\bibinfo{year}{1966}), \bibinfo{month}{Mar.}
\bibinfo{title}{{The Hydrodynamic Behavior of Supernovae Explosions}}.
\bibinfo{journal}{{\em \apj}} \bibinfo{volume}{143}: \bibinfo{pages}{626--681}.
  \bibinfo{doi}{\doi{10.1086/148549}}.

\bibtype{Article}%
\bibitem[{Cowan} et al.(2021)]{cowan_21}
\bibinfo{author}{{Cowan} JJ}, \bibinfo{author}{{Sneden} C},
  \bibinfo{author}{{Lawler} JE}, \bibinfo{author}{{Aprahamian} A},
  \bibinfo{author}{{Wiescher} M}, \bibinfo{author}{{Langanke} K},
  \bibinfo{author}{{Mart{\'\i}nez-Pinedo} G} and  \bibinfo{author}{{Thielemann}
  FK} (\bibinfo{year}{2021}), \bibinfo{month}{Jan.}
\bibinfo{title}{{Origin of the heaviest elements: The rapid neutron-capture
  process}}.
\bibinfo{journal}{{\em Reviews of Modern Physics}} \bibinfo{volume}{93}
  (\bibinfo{number}{1}), \bibinfo{eid}{015002}.
  \bibinfo{doi}{\doi{10.1103/RevModPhys.93.015002}}.
\eprint{1901.01410}.

\bibtype{Article}%
\bibitem[{Dasgupta} et al.(2010)]{dasgupta_10}
\bibinfo{author}{{Dasgupta} B}, \bibinfo{author}{{Fischer} T},
  \bibinfo{author}{{Horiuchi} S}, \bibinfo{author}{{Liebend{\"o}rfer} M},
  \bibinfo{author}{{Mirizzi} A}, \bibinfo{author}{{Sagert} I} and
  \bibinfo{author}{{Schaffner-Bielich} J} (\bibinfo{year}{2010}),
  \bibinfo{month}{May}.
\bibinfo{title}{{Detecting the QCD phase transition in the next Galactic
  supernova neutrino burst}}.
\bibinfo{journal}{{\em \prd}} \bibinfo{volume}{81} (\bibinfo{number}{10}),
  \bibinfo{eid}{103005}. \bibinfo{doi}{\doi{10.1103/PhysRevD.81.103005}}.
\eprint{0912.2568}.

\bibtype{Article}%
\bibitem[{De} et al.(2026)]{de_26}
\bibinfo{author}{{De} K}, \bibinfo{author}{{MacLeod} M},
  \bibinfo{author}{{Jencson} JE}, \bibinfo{author}{{Lovegrove} E},
  \bibinfo{author}{{Antoni} A}, \bibinfo{author}{{Kara} E},
  \bibinfo{author}{{Kasliwal} MM}, \bibinfo{author}{{Lau} RM},
  \bibinfo{author}{{Loeb} A}, \bibinfo{author}{{Masterson} M},
  \bibinfo{author}{{Meisner} AM}, \bibinfo{author}{{Panagiotou} C},
  \bibinfo{author}{{Quataert} E} and  \bibinfo{author}{{Simcoe} R}
  (\bibinfo{year}{2026}), \bibinfo{month}{Feb.}
\bibinfo{title}{{Disappearance of a massive star in the Andromeda Galaxy due to
  formation of a black hole}}.
\bibinfo{journal}{{\em Science}} \bibinfo{volume}{391}
  (\bibinfo{number}{6786}): \bibinfo{pages}{689--693}.
  \bibinfo{doi}{\doi{10.1126/science.adt4853}}.
\eprint{2410.14778}.

\bibtype{Article}%
\bibitem[{Diehl} et al.(2006)]{diehl_06}
\bibinfo{author}{{Diehl} R}, \bibinfo{author}{{Halloin} H},
  \bibinfo{author}{{Kretschmer} K}, \bibinfo{author}{{Lichti} GG},
  \bibinfo{author}{{Sch{\"o}nfelder} V}, \bibinfo{author}{{Strong} AW},
  \bibinfo{author}{{von Kienlin} A}, \bibinfo{author}{{Wang} W},
  \bibinfo{author}{{Jean} P}, \bibinfo{author}{{Kn{\"o}dlseder} J},
  \bibinfo{author}{{Roques} JP}, \bibinfo{author}{{Weidenspointner} G},
  \bibinfo{author}{{Schanne} S}, \bibinfo{author}{{Hartmann} DH},
  \bibinfo{author}{{Winkler} C} and  \bibinfo{author}{{Wunderer} C}
  (\bibinfo{year}{2006}), \bibinfo{month}{Jan.}
\bibinfo{title}{{Radioactive $^{26}$Al from massive stars in the Galaxy}}.
\bibinfo{journal}{{\em \nat}} \bibinfo{volume}{439} (\bibinfo{number}{7072}):
  \bibinfo{pages}{45--47}. \bibinfo{doi}{\doi{10.1038/nature04364}}.
\eprint{astro-ph/0601015}.

\bibtype{Article}%
\bibitem[{Dimmelmeier} et al.(2007)]{dimmelmeier_07_a}
\bibinfo{author}{{Dimmelmeier} H}, \bibinfo{author}{{Ott} CD},
  \bibinfo{author}{{Janka} HT}, \bibinfo{author}{{Marek} A} and
  \bibinfo{author}{{M{\"u}ller} E} (\bibinfo{year}{2007}),
  \bibinfo{month}{Jun.}
\bibinfo{title}{{Generic Gravitational-Wave Signals from the Collapse of
  Rotating Stellar Cores}}.
\bibinfo{journal}{{\em \prl}} \bibinfo{volume}{98} (\bibinfo{number}{25}):
  \bibinfo{pages}{251101:1--4}.
  \bibinfo{doi}{\doi{10.1103/PhysRevLett.98.251101}}.

\bibtype{Article}%
\bibitem[{Dimmelmeier} et al.(2008)]{dimmelmeier_08}
\bibinfo{author}{{Dimmelmeier} H}, \bibinfo{author}{{Ott} CD},
  \bibinfo{author}{{Marek} A} and  \bibinfo{author}{{Janka} HT}
  (\bibinfo{year}{2008}), \bibinfo{month}{Sep.}
\bibinfo{title}{{Gravitational wave burst signal from core collapse of rotating
  stars}}.
\bibinfo{journal}{{\em \prd}} \bibinfo{volume}{78} (\bibinfo{number}{6}):
  \bibinfo{pages}{064056:1--28}.
  \bibinfo{doi}{\doi{10.1103/PhysRevD.78.064056}}.

\bibtype{Article}%
\bibitem[{Duan} et al.(2008)]{duan_08_b}
\bibinfo{author}{{Duan} H}, \bibinfo{author}{{Fuller} GM},
  \bibinfo{author}{{Carlson} J} and  \bibinfo{author}{{Qian} YZ}
  (\bibinfo{year}{2008}), \bibinfo{month}{Jan.}
\bibinfo{title}{{Flavor Evolution of the Neutronization Neutrino Burst From an
  O-Ne-Mg Core-Collapse Supernova}}.
\bibinfo{journal}{{\em \prl}} \bibinfo{volume}{100} (\bibinfo{number}{2}):
  \bibinfo{pages}{021101:1--4}.
  \bibinfo{doi}{\doi{10.1103/PhysRevLett.100.021101}}.

\bibtype{Article}%
\bibitem[{Duan} et al.(2010)]{duan_10}
\bibinfo{author}{{Duan} H}, \bibinfo{author}{{Fuller} GM} and
  \bibinfo{author}{{Qian} YZ} (\bibinfo{year}{2010}), \bibinfo{month}{Nov.}
\bibinfo{title}{{Collective Neutrino Oscillations}}.
\bibinfo{journal}{{\em Annual Review of Nuclear and Particle Science}}
  \bibinfo{volume}{60}: \bibinfo{pages}{569--594}.
  \bibinfo{doi}{\doi{10.1146/annurev.nucl.012809.104524}}.

\bibtype{Article}%
\bibitem[{Duncan} et al.(1986)]{duncan_86}
\bibinfo{author}{{Duncan} RC}, \bibinfo{author}{{Shapiro} SL} and
  \bibinfo{author}{{Wasserman} I} (\bibinfo{year}{1986}), \bibinfo{month}{Oct.}
\bibinfo{title}{{Neutrino-driven winds from young, hot neutron stars}}.
\bibinfo{journal}{{\em \apj}} \bibinfo{volume}{309}: \bibinfo{pages}{141--160}.
  \bibinfo{doi}{\doi{10.1086/164587}}.

\bibtype{Article}%
\bibitem[{Ebinger} et al.(2019)]{ebinger_19}
\bibinfo{author}{{Ebinger} K}, \bibinfo{author}{{Curtis} S},
  \bibinfo{author}{{Fr{\"o}hlich} C}, \bibinfo{author}{{Hempel} M},
  \bibinfo{author}{{Perego} A}, \bibinfo{author}{{Liebend{\"o}rfer} M} and
  \bibinfo{author}{{Thielemann} FK} (\bibinfo{year}{2019}),
  \bibinfo{month}{Jan.}
\bibinfo{title}{{PUSHing Core-collapse Supernovae to Explosions in Spherical
  Symmetry. II. Explodability and Remnant Properties}}.
\bibinfo{journal}{{\em \apj}} \bibinfo{volume}{870}: \bibinfo{pages}{1}.
  \bibinfo{doi}{\doi{10.3847/1538-4357/aae7c9}}.

\bibtype{Article}%
\bibitem[Ehlers(1971)]{ehlers_71}
\bibinfo{author}{Ehlers J} (\bibinfo{year}{1971}), \bibinfo{month}{01}.
\bibinfo{title}{General relativity and kinetic theory.}
\bibinfo{journal}{{\em Rend. Scu. Int. Fis. Enrico Fermi}}
  \bibinfo{volume}{47}: \bibinfo{pages}{1--70}.
\bibinfo{url}{\url{https://www.osti.gov/biblio/4707856}}.

\bibtype{Article}%
\bibitem[{Einstein}(1918)]{einstein}
\bibinfo{author}{{Einstein} A} (\bibinfo{year}{1918}).
\bibinfo{title}{{{\"U}ber Gravitationswellen}}.
\bibinfo{journal}{{\em Sitzungsberichte der K{\"o}niglich Preu{\ss}ischen
  Akademie der Wissenschaften (Berlin), Seite 154-167}} :
  \bibinfo{pages}{154--167}.

\bibtype{Article}%
\bibitem[{Eldridge} and {Tout}(2004)]{eldridge_04}
\bibinfo{author}{{Eldridge} JJ} and  \bibinfo{author}{{Tout} CA}
  (\bibinfo{year}{2004}), \bibinfo{month}{Sep.}
\bibinfo{title}{{The progenitors of core-collapse supernovae}}.
\bibinfo{journal}{{\em \mnras}} \bibinfo{volume}{353} (\bibinfo{number}{1}):
  \bibinfo{pages}{87--97}.
  \bibinfo{doi}{\doi{10.1111/j.1365-2966.2004.08041.x}}.
\eprint{astro-ph/0405408}.

\bibtype{Article}%
\bibitem[{Eldridge} et al.(2013)]{eldridge_13}
\bibinfo{author}{{Eldridge} JJ}, \bibinfo{author}{{Fraser} M},
  \bibinfo{author}{{Smartt} SJ}, \bibinfo{author}{{Maund} JR} and
  \bibinfo{author}{{Crockett} RM} (\bibinfo{year}{2013}), \bibinfo{month}{Nov.}
\bibinfo{title}{{The death of massive stars - II. Observational constraints on
  the progenitors of Type Ibc supernovae}}.
\bibinfo{journal}{{\em \mnras}} \bibinfo{volume}{436}:
  \bibinfo{pages}{774--795}. \bibinfo{doi}{\doi{10.1093/mnras/stt1612}}.

\bibtype{Article}%
\bibitem[{Epstein}(1978)]{epstein_78}
\bibinfo{author}{{Epstein} R} (\bibinfo{year}{1978}), \bibinfo{month}{Aug.}
\bibinfo{title}{{The generation of gravitational radiation by escaping
  supernova neutrinos}}.
\bibinfo{journal}{{\em \apj}} \bibinfo{volume}{223}:
  \bibinfo{pages}{1037--1045}. \bibinfo{doi}{\doi{10.1086/156337}}.

\bibtype{Article}%
\bibitem[{Fern{\'a}ndez}(2012)]{fernandez_12}
\bibinfo{author}{{Fern{\'a}ndez} R} (\bibinfo{year}{2012}),
  \bibinfo{month}{Apr.}
\bibinfo{title}{{Hydrodynamics of Core-collapse Supernovae at the Transition to
  Explosion. I. Spherical Symmetry}}.
\bibinfo{journal}{{\em \apj}} \bibinfo{volume}{749}: \bibinfo{pages}{142}.
  \bibinfo{doi}{\doi{10.1088/0004-637X/749/2/142}}.

\bibtype{Article}%
\bibitem[{Filippenko}(1997)]{filippenko_97}
\bibinfo{author}{{Filippenko} AV} (\bibinfo{year}{1997}), \bibinfo{month}{Jan.}
\bibinfo{title}{{Optical Spectra of Supernovae}}.
\bibinfo{journal}{{\em \araa}} \bibinfo{volume}{35}: \bibinfo{pages}{309--355}.
  \bibinfo{doi}{\doi{10.1146/annurev.astro.35.1.309}}.

\bibtype{Article}%
\bibitem[{Fischer}(2016)]{fischer_16b}
\bibinfo{author}{{Fischer} T} (\bibinfo{year}{2016}), \bibinfo{month}{Sep.}
\bibinfo{title}{{The role of medium modifications for neutrino-pair processes
  from nucleon-nucleon bremsstrahlung. Impact on the protoneutron star
  deleptonization}}.
\bibinfo{journal}{{\em \aap}} \bibinfo{volume}{593}, \bibinfo{eid}{A103}.
  \bibinfo{doi}{\doi{10.1051/0004-6361/201628991}}.
\eprint{1608.05004}.

\bibtype{Article}%
\bibitem[{Fischer} et al.(2010)]{fischer_10}
\bibinfo{author}{{Fischer} T}, \bibinfo{author}{{Whitehouse} SC},
  \bibinfo{author}{{Mezzacappa} A}, \bibinfo{author}{{Thielemann} F} and
  \bibinfo{author}{{Liebend{\"o}rfer} M} (\bibinfo{year}{2010}),
  \bibinfo{month}{Jul.}
\bibinfo{title}{{Protoneutron star evolution and the neutrino-driven wind in
  general relativistic neutrino radiation hydrodynamics simulations}}.
\bibinfo{journal}{{\em \aap}} \bibinfo{volume}{517}: \bibinfo{pages}{A80+}.
  \bibinfo{doi}{\doi{10.1051/0004-6361/200913106}}.

\bibtype{Article}%
\bibitem[{Fischer} et al.(2011)]{fischer_11}
\bibinfo{author}{{Fischer} T}, \bibinfo{author}{{Sagert} I},
  \bibinfo{author}{{Pagliara} G}, \bibinfo{author}{{Hempel} M},
  \bibinfo{author}{{Schaffner-Bielich} J}, \bibinfo{author}{{Rauscher} T},
  \bibinfo{author}{{Thielemann} FK}, \bibinfo{author}{{K{\"a}ppeli} R},
  \bibinfo{author}{{Mart{\'\i}nez-Pinedo} G} and
  \bibinfo{author}{{Liebend{\"o}rfer} M} (\bibinfo{year}{2011}),
  \bibinfo{month}{Jun.}
\bibinfo{title}{{Core-collapse Supernova Explosions Triggered by a Quark-Hadron
  Phase Transition During the Early Post-bounce Phase}}.
\bibinfo{journal}{{\em \apjs}} \bibinfo{volume}{194} (\bibinfo{number}{2}),
  \bibinfo{eid}{39}. \bibinfo{doi}{\doi{10.1088/0067-0049/194/2/39}}.

\bibtype{Article}%
\bibitem[{Fischer} et al.(2016)]{fischer_16a}
\bibinfo{author}{{Fischer} T}, \bibinfo{author}{{Chakraborty} S},
  \bibinfo{author}{{Giannotti} M}, \bibinfo{author}{{Mirizzi} A},
  \bibinfo{author}{{Payez} A} and  \bibinfo{author}{{Ringwald} A}
  (\bibinfo{year}{2016}), \bibinfo{month}{Oct.}
\bibinfo{title}{{Probing axions with the neutrino signal from the next Galactic
  supernova}}.
\bibinfo{journal}{{\em \prd}} \bibinfo{volume}{94} (\bibinfo{number}{8}),
  \bibinfo{eid}{085012}. \bibinfo{doi}{\doi{10.1103/PhysRevD.94.085012}}.
\eprint{1605.08780}.

\bibtype{Article}%
\bibitem[{Fischer} et al.(2017)]{fischer_17}
\bibinfo{author}{{Fischer} T}, \bibinfo{author}{{Bastian} NU},
  \bibinfo{author}{{Blaschke} D}, \bibinfo{author}{{Cierniak} M},
  \bibinfo{author}{{Hempel} M}, \bibinfo{author}{{Kl{\"a}hn} T},
  \bibinfo{author}{{Mart{\'\i}nez-Pinedo} G}, \bibinfo{author}{{Newton} WG},
  \bibinfo{author}{{R{\"o}pke} G} and  \bibinfo{author}{{Typel} S}
  (\bibinfo{year}{2017}), \bibinfo{month}{Dec}.
\bibinfo{title}{{The State of Matter in Simulations of Core-Collapse
  supernovae{\textemdash}Reflections and Recent Developments}}.
\bibinfo{journal}{{\em \pasa}} \bibinfo{volume}{34}: \bibinfo{pages}{e067}.
  \bibinfo{doi}{\doi{10.1017/pasa.2017.63}}.

\bibtype{Article}%
\bibitem[{Fischer} et al.(2018)]{fischer_18}
\bibinfo{author}{{Fischer} T}, \bibinfo{author}{{Bastian} NUF},
  \bibinfo{author}{{Wu} MR}, \bibinfo{author}{{Baklanov} P},
  \bibinfo{author}{{Sorokina} E}, \bibinfo{author}{{Blinnikov} S},
  \bibinfo{author}{{Typel} S}, \bibinfo{author}{{Kl{\"a}hn} T} and
  \bibinfo{author}{{Blaschke} DB} (\bibinfo{year}{2018}), \bibinfo{month}{Oct.}
\bibinfo{title}{{Quark deconfinement as a supernova explosion engine for
  massive blue supergiant stars}}.
\bibinfo{journal}{{\em Nature Astronomy}} \bibinfo{volume}{2}:
  \bibinfo{pages}{980--986}. \bibinfo{doi}{\doi{10.1038/s41550-018-0583-0}}.

\bibtype{Article}%
\bibitem[{Fischer} et al.(2020)]{fischer_20}
\bibinfo{author}{{Fischer} T}, \bibinfo{author}{{Typel} S},
  \bibinfo{author}{{R{\"o}pke} G}, \bibinfo{author}{{Bastian} NUF} and
  \bibinfo{author}{{Mart{\'\i}nez-Pinedo} G} (\bibinfo{year}{2020}),
  \bibinfo{month}{Nov.}
\bibinfo{title}{{Medium modifications for light and heavy nuclear clusters in
  simulations of core collapse supernovae: Impact on equation of state and weak
  interactions}}.
\bibinfo{journal}{{\em \prc}} \bibinfo{volume}{102} (\bibinfo{number}{5}),
  \bibinfo{eid}{055807}. \bibinfo{doi}{\doi{10.1103/PhysRevC.102.055807}}.
\eprint{2008.13608}.

\bibtype{Article}%
\bibitem[{Foglizzo} et al.(2007)]{foglizzo_07}
\bibinfo{author}{{Foglizzo} T}, \bibinfo{author}{{Galletti} P},
  \bibinfo{author}{{Scheck} L} and  \bibinfo{author}{{Janka} HT}
  (\bibinfo{year}{2007}), \bibinfo{month}{Jan.}
\bibinfo{title}{{Instability of a Stalled Accretion Shock: Evidence for the
  Advective-Acoustic Cycle}}.
\bibinfo{journal}{{\em \apj}} \bibinfo{volume}{654}:
  \bibinfo{pages}{1006--1021}. \bibinfo{doi}{\doi{10.1086/509612}}.

\bibtype{Article}%
\bibitem[{Fonseca} et al.(2021)]{fonseca_21}
\bibinfo{author}{{Fonseca} E}, \bibinfo{author}{{Cromartie} HT},
  \bibinfo{author}{{Pennucci} TT}, \bibinfo{author}{{Ray} PS},
  \bibinfo{author}{{Kirichenko} AY}, \bibinfo{author}{{Ransom} SM},
  \bibinfo{author}{{Demorest} PB}, \bibinfo{author}{{Stairs} IH},
  \bibinfo{author}{{Arzoumanian} Z}, \bibinfo{author}{{Guillemot} L},
  \bibinfo{author}{{Parthasarathy} A}, \bibinfo{author}{{Kerr} M},
  \bibinfo{author}{{Cognard} I}, \bibinfo{author}{{Baker} PT},
  \bibinfo{author}{{Blumer} H}, \bibinfo{author}{{Brook} PR},
  \bibinfo{author}{{DeCesar} M}, \bibinfo{author}{{Dolch} T},
  \bibinfo{author}{{Dong} FA}, \bibinfo{author}{{Ferrara} EC},
  \bibinfo{author}{{Fiore} W}, \bibinfo{author}{{Garver-Daniels} N},
  \bibinfo{author}{{Good} DC}, \bibinfo{author}{{Jennings} R},
  \bibinfo{author}{{Jones} ML}, \bibinfo{author}{{Kaspi} VM},
  \bibinfo{author}{{Lam} MT}, \bibinfo{author}{{Lorimer} DR},
  \bibinfo{author}{{Luo} J}, \bibinfo{author}{{McEwen} A},
  \bibinfo{author}{{McKee} JW}, \bibinfo{author}{{McLaughlin} MA},
  \bibinfo{author}{{McMann} N}, \bibinfo{author}{{Meyers} BW},
  \bibinfo{author}{{Naidu} A}, \bibinfo{author}{{Ng} C},
  \bibinfo{author}{{Nice} DJ}, \bibinfo{author}{{Pol} N},
  \bibinfo{author}{{Radovan} HA}, \bibinfo{author}{{Shapiro-Albert} B},
  \bibinfo{author}{{Tan} CM}, \bibinfo{author}{{Tendulkar} SP},
  \bibinfo{author}{{Swiggum} JK}, \bibinfo{author}{{Wahl} HM} and
  \bibinfo{author}{{Zhu} WW} (\bibinfo{year}{2021}), \bibinfo{month}{Jul.}
\bibinfo{title}{{Refined Mass and Geometric Measurements of the High-mass PSR
  J0740+6620}}.
\bibinfo{journal}{{\em \apjl}} \bibinfo{volume}{915} (\bibinfo{number}{1}),
  \bibinfo{eid}{L12}. \bibinfo{doi}{\doi{10.3847/2041-8213/ac03b8}}.
\eprint{2104.00880}.

\bibtype{Article}%
\bibitem[{Fransson} et al.(2024)]{fransson_24}
\bibinfo{author}{{Fransson} C}, \bibinfo{author}{{Barlow} MJ},
  \bibinfo{author}{{Kavanagh} PJ}, \bibinfo{author}{{Larsson} J},
  \bibinfo{author}{{Jones} OC}, \bibinfo{author}{{Sargent} B},
  \bibinfo{author}{{Meixner} M}, \bibinfo{author}{{Bouchet} P},
  \bibinfo{author}{{Temim} T}, \bibinfo{author}{{Wright} GS},
  \bibinfo{author}{{Blommaert} JADL}, \bibinfo{author}{{Habel} N},
  \bibinfo{author}{{Hirschauer} AS}, \bibinfo{author}{{Hjorth} J},
  \bibinfo{author}{{Lenki{\'c}} L}, \bibinfo{author}{{Tikkanen} T},
  \bibinfo{author}{{Wesson} R}, \bibinfo{author}{{Coulais} A},
  \bibinfo{author}{{Fox} OD}, \bibinfo{author}{{Gastaud} R},
  \bibinfo{author}{{Glasse} A}, \bibinfo{author}{{Jaspers} J},
  \bibinfo{author}{{Krause} O}, \bibinfo{author}{{Lau} RM},
  \bibinfo{author}{{Nayak} O}, \bibinfo{author}{{Rest} A},
  \bibinfo{author}{{Colina} L}, \bibinfo{author}{{van Dishoeck} EF},
  \bibinfo{author}{{G{\"u}del} M}, \bibinfo{author}{{Henning} T},
  \bibinfo{author}{{Lagage} PO}, \bibinfo{author}{{{\"O}stlin} G},
  \bibinfo{author}{{Ray} TP} and  \bibinfo{author}{{Vandenbussche} B}
  (\bibinfo{year}{2024}), \bibinfo{month}{Feb.}
\bibinfo{title}{{Emission lines due to ionizing radiation from a compact object
  in the remnant of Supernova 1987A}}.
\bibinfo{journal}{{\em Science}} \bibinfo{volume}{383}
  (\bibinfo{number}{6685}): \bibinfo{pages}{898--903}.
  \bibinfo{doi}{\doi{10.1126/science.adj5796}}.
\eprint{2403.04386}.

\bibtype{Article}%
\bibitem[{Fuller} et al.(1982)]{fuller_82}
\bibinfo{author}{{Fuller} GM}, \bibinfo{author}{{Fowler} WA} and
  \bibinfo{author}{{Newman} MJ} (\bibinfo{year}{1982}), \bibinfo{month}{Mar.}
\bibinfo{title}{{Stellar weak interaction rates for intermediate mass nuclei.
  III - Rate tables for the free nucleons and nuclei with A = 21 to A = 60}}.
\bibinfo{journal}{{\em \apjs}} \bibinfo{volume}{48}: \bibinfo{pages}{279--319}.
  \bibinfo{doi}{\doi{10.1086/190779}}.

\bibtype{incollection}%
\bibitem[{Gal-Yam}(2017)]{gal-yam_17}
\bibinfo{author}{{Gal-Yam} A} (\bibinfo{year}{2017}),
  \bibinfo{title}{{Observational and Physical Classification of Supernovae}},
  \bibinfo{editor}{{Alsabti} AW} and  \bibinfo{editor}{{Murdin} P}, (Eds.),
  \bibinfo{booktitle}{Handbook of Supernovae}, \bibinfo{publisher}{Springer
  International Publishing}, \bibinfo{address}{Cham}, pp. \bibinfo{pages}{195}.

\bibtype{Article}%
\bibitem[{Ghosh} et al.(2022)]{ghosh_22}
\bibinfo{author}{{Ghosh} S}, \bibinfo{author}{{Wolfe} N} and
  \bibinfo{author}{{Fr{\"o}hlich} C} (\bibinfo{year}{2022}),
  \bibinfo{month}{Apr.}
\bibinfo{title}{{PUSHing Core-collapse Supernovae to Explosions in Spherical
  Symmetry. V. Equation of State Dependency of Explosion Properties,
  Nucleosynthesis Yields, and Compact Remnants}}.
\bibinfo{journal}{{\em \apj}} \bibinfo{volume}{929} (\bibinfo{number}{1}),
  \bibinfo{eid}{43}. \bibinfo{doi}{\doi{10.3847/1538-4357/ac4d20}}.

\bibtype{Article}%
\bibitem[{Glas} et al.(2019)]{glas_19}
\bibinfo{author}{{Glas} R}, \bibinfo{author}{{Janka} HT},
  \bibinfo{author}{{Melson} T}, \bibinfo{author}{{Stockinger} G} and
  \bibinfo{author}{{Just} O} (\bibinfo{year}{2019}), \bibinfo{month}{Aug}.
\bibinfo{title}{{Effects of LESA in Three-dimensional Supernova Simulations
  with Multidimensional and Ray-by-ray-plus Neutrino Transport}}.
\bibinfo{journal}{{\em \apj}} \bibinfo{volume}{881} (\bibinfo{number}{1}):
  \bibinfo{pages}{36}. \bibinfo{doi}{\doi{10.3847/1538-4357/ab275c}}.

\bibtype{Book}%
\bibitem[{Halzen} and {Martin}(1984)]{halzen_84}
\bibinfo{author}{{Halzen} F} and  \bibinfo{author}{{Martin} AD}
  (\bibinfo{year}{1984}).
\bibinfo{title}{{Quarks and leptons: an introductory course in modern particle
  physics}}, \bibinfo{publisher}{John Wiley \& Sons}, \bibinfo{address}{New
  York}.

\bibtype{Article}%
\bibitem[{Hannestad} and {Raffelt}(1998)]{hannestad_98}
\bibinfo{author}{{Hannestad} S} and  \bibinfo{author}{{Raffelt} G}
  (\bibinfo{year}{1998}), \bibinfo{month}{Nov.}
\bibinfo{title}{{Supernova Neutrino Opacity from Nucleon-Nucleon Bremsstrahlung
  and Related Processes}}.
\bibinfo{journal}{{\em \apj}} \bibinfo{volume}{507}: \bibinfo{pages}{339--352}.
  \bibinfo{doi}{\doi{10.1086/306303}}.

\bibtype{Article}%
\bibitem[{Hannestad} and {Raffelt}(2001)]{hannestad_01}
\bibinfo{author}{{Hannestad} S} and  \bibinfo{author}{{Raffelt} GG}
  (\bibinfo{year}{2001}), \bibinfo{month}{Jul.}
\bibinfo{title}{{New Supernova Limit on Large Extra Dimensions: Bounds on
  Kaluza-Klein Graviton Production}}.
\bibinfo{journal}{{\em \prl}} \bibinfo{volume}{87} (\bibinfo{number}{5}),
  \bibinfo{eid}{051301}. \bibinfo{doi}{\doi{10.1103/PhysRevLett.87.051301}}.
\eprint{hep-ph/0103201}.

\bibtype{Article}%
\bibitem[{Heger} and {Woosley}(2002)]{heger_02}
\bibinfo{author}{{Heger} A} and  \bibinfo{author}{{Woosley} SE}
  (\bibinfo{year}{2002}), \bibinfo{month}{Mar.}
\bibinfo{title}{{The Nucleosynthetic Signature of Population III}}.
\bibinfo{journal}{{\em \apj}} \bibinfo{volume}{567}: \bibinfo{pages}{532--543}.
  \bibinfo{doi}{\doi{10.1086/338487}}.

\bibtype{Article}%
\bibitem[{Heger} et al.(2003)]{heger_03}
\bibinfo{author}{{Heger} A}, \bibinfo{author}{{Fryer} CL},
  \bibinfo{author}{{Woosley} SE}, \bibinfo{author}{{Langer} N} and
  \bibinfo{author}{{Hartmann} DH} (\bibinfo{year}{2003}), \bibinfo{month}{Jul.}
\bibinfo{title}{{How Massive Single Stars End Their Life}}.
\bibinfo{journal}{{\em \apj}} \bibinfo{volume}{591}: \bibinfo{pages}{288--300}.
  \bibinfo{doi}{\doi{10.1086/375341}}.

\bibtype{incollection}%
\bibitem[{Heger} et al.(2023)]{heger_23}
\bibinfo{author}{{Heger} A}, \bibinfo{author}{{M{\"u}ller} B} and
  \bibinfo{author}{{Mandel} I} (\bibinfo{year}{2023}), \bibinfo{title}{Black
  holes as the end state of stellar evolution: Theory and simulations},
  \bibinfo{editor}{Haiman Z}, (Ed.), \bibinfo{booktitle}{The Encyclopedia of
  Cosmology}, \bibinfo{publisher}{World Scientific Series in Astrophysics},
  \bibinfo{pages}{61--111}.

\bibtype{Article}%
\bibitem[{Herant} et al.(1994)]{herant_94}
\bibinfo{author}{{Herant} M}, \bibinfo{author}{{Benz} W},
  \bibinfo{author}{{Hix} WR}, \bibinfo{author}{{Fryer} CL} and
  \bibinfo{author}{{Colgate} SA} (\bibinfo{year}{1994}), \bibinfo{month}{Nov.}
\bibinfo{title}{{Inside the supernova: A powerful convective engine}}.
\bibinfo{journal}{{\em \apj}} \bibinfo{volume}{435}: \bibinfo{pages}{339--361}.
  \bibinfo{doi}{\doi{10.1086/174817}}.

\bibtype{Article}%
\bibitem[{Hirata} et al.(1987)]{hirata_87}
\bibinfo{author}{{Hirata} K}, \bibinfo{author}{{Kajita} T},
  \bibinfo{author}{{Koshiba} M}, \bibinfo{author}{{Nakahata} M} and
  \bibinfo{author}{{Oyama} Y} (\bibinfo{year}{1987}), \bibinfo{month}{Apr.}
\bibinfo{title}{{Observation of a neutrino burst from the supernova SN1987A}}.
\bibinfo{journal}{{\em Physical Review Letters}} \bibinfo{volume}{58}:
  \bibinfo{pages}{1490--1493}.
  \bibinfo{doi}{\doi{10.1103/PhysRevLett.58.1490}}.

\bibtype{Article}%
\bibitem[{Hix} and {Thielemann}(1999)]{hix_99}
\bibinfo{author}{{Hix} WR} and  \bibinfo{author}{{Thielemann} FK}
  (\bibinfo{year}{1999}), \bibinfo{month}{Feb.}
\bibinfo{title}{{Silicon Burning. II. Quasi-Equilibrium and Explosive
  Burning}}.
\bibinfo{journal}{{\em \apj}} \bibinfo{volume}{511} (\bibinfo{number}{2}):
  \bibinfo{pages}{862--875}. \bibinfo{doi}{\doi{10.1086/306692}}.
\eprint{astro-ph/9808203}.

\bibtype{Article}%
\bibitem[{Hix} et al.(2003)]{hix_03}
\bibinfo{author}{{Hix} WR}, \bibinfo{author}{{Messer} OE},
  \bibinfo{author}{{Mezzacappa} A}, \bibinfo{author}{{Liebend{\"o}rfer} M},
  \bibinfo{author}{{Sampaio} J}, \bibinfo{author}{{Langanke} K},
  \bibinfo{author}{{Dean} DJ} and  \bibinfo{author}{{Mart{\'{\i}}nez-Pinedo} G}
  (\bibinfo{year}{2003}), \bibinfo{month}{Nov.}
\bibinfo{title}{{Consequences of Nuclear Electron Capture in Core Collapse
  Supernovae}}.
\bibinfo{journal}{{\em Physical Review Letters}} \bibinfo{volume}{91}
  (\bibinfo{number}{20}): \bibinfo{pages}{201102}.
  \bibinfo{doi}{\doi{10.1103/PhysRevLett.91.201102}}.

\bibtype{Article}%
\bibitem[{Horiuchi} and {Kneller}(2018)]{horiuchi_18}
\bibinfo{author}{{Horiuchi} S} and  \bibinfo{author}{{Kneller} JP}
  (\bibinfo{year}{2018}), \bibinfo{month}{Apr.}
\bibinfo{title}{{What can be learned from a future supernova neutrino
  detection?}}
\bibinfo{journal}{{\em Journal of Physics G Nuclear Physics}}
  \bibinfo{volume}{45} (\bibinfo{number}{4}): \bibinfo{pages}{043002}.
  \bibinfo{doi}{\doi{10.1088/1361-6471/aaa90a}}.
\eprint{1709.01515}.

\bibtype{Article}%
\bibitem[{Horiuchi} et al.(2021)]{horiuchi_21}
\bibinfo{author}{{Horiuchi} S}, \bibinfo{author}{{Kinugawa} T},
  \bibinfo{author}{{Takiwaki} T}, \bibinfo{author}{{Takahashi} K} and
  \bibinfo{author}{{Kotake} K} (\bibinfo{year}{2021}), \bibinfo{month}{Feb.}
\bibinfo{title}{{Impact of binary interactions on the diffuse supernova
  neutrino background}}.
\bibinfo{journal}{{\em \prd}} \bibinfo{volume}{103} (\bibinfo{number}{4}),
  \bibinfo{eid}{043003}. \bibinfo{doi}{\doi{10.1103/PhysRevD.103.043003}}.
\eprint{2012.08524}.

\bibtype{Article}%
\bibitem[{Horowitz}(2002)]{horowitz_02}
\bibinfo{author}{{Horowitz} CJ} (\bibinfo{year}{2002}), \bibinfo{month}{Feb.}
\bibinfo{title}{{Weak magnetism for antineutrinos in supernovae}}.
\bibinfo{journal}{{\em \prd}} \bibinfo{volume}{65} (\bibinfo{number}{4}):
  \bibinfo{pages}{043001--+}. \bibinfo{doi}{\doi{10.1103/PhysRevD.65.043001}}.

\bibtype{Article}%
\bibitem[{Horowitz} et al.(2017)]{horowitz_17}
\bibinfo{author}{{Horowitz} CJ}, \bibinfo{author}{{Caballero} OL},
  \bibinfo{author}{{Lin} Z}, \bibinfo{author}{{O'Connor} E} and
  \bibinfo{author}{{Schwenk} A} (\bibinfo{year}{2017}), \bibinfo{month}{Feb.}
\bibinfo{title}{{Neutrino-nucleon scattering in supernova matter from the
  virial expansion}}.
\bibinfo{journal}{{\em \prc}} \bibinfo{volume}{95} (\bibinfo{number}{2}):
  \bibinfo{pages}{025801}. \bibinfo{doi}{\doi{10.1103/PhysRevC.95.025801}}.

\bibtype{Article}%
\bibitem[{Hubble}(1928)]{hubble_28}
\bibinfo{author}{{Hubble} EP} (\bibinfo{year}{1928}).
\bibinfo{title}{{Novae or Temporary Stars}}.
\bibinfo{journal}{{\em Leaflet of the Astronomical Society of the Pacific}}
  \bibinfo{volume}{1}: \bibinfo{pages}{55--58}.

\bibtype{Article}%
\bibitem[{H\"udepohl} et al.(2010)]{huedepohl_10}
\bibinfo{author}{{H\"udepohl} L}, \bibinfo{author}{{M\"uller} B},
  \bibinfo{author}{{Janka} H}, \bibinfo{author}{{Marek} A} and
  \bibinfo{author}{{Raffelt} GG} (\bibinfo{year}{2010}), \bibinfo{month}{Jun.}
\bibinfo{title}{{Neutrino Signal of Electron-Capture Supernovae from Core
  Collapse to Cooling}}.
\bibinfo{journal}{{\em \prl}} \bibinfo{volume}{104} (\bibinfo{number}{25}):
  \bibinfo{pages}{251101}. \bibinfo{doi}{\doi{10.1103/PhysRevLett.104.251101}}.

\bibtype{Article}%
\bibitem[{Ibeling} and {Heger}(2013)]{ibeling_13}
\bibinfo{author}{{Ibeling} D} and  \bibinfo{author}{{Heger} A}
  (\bibinfo{year}{2013}), \bibinfo{month}{Mar.}
\bibinfo{title}{{The Metallicity Dependence of the Minimum Mass for
  Core-collapse Supernovae}}.
\bibinfo{journal}{{\em \apjl}} \bibinfo{volume}{765} (\bibinfo{number}{2}),
  \bibinfo{eid}{L43}. \bibinfo{doi}{\doi{10.1088/2041-8205/765/2/L43}}.
\eprint{1301.5783}.

\bibtype{Article}%
\bibitem[{Iwamoto} et al.(1998)]{iwamoto_98}
\bibinfo{author}{{Iwamoto} K}, \bibinfo{author}{{Mazzali} PA},
  \bibinfo{author}{{Nomoto} K}, \bibinfo{author}{{Umeda} H},
  \bibinfo{author}{{Nakamura} T}, \bibinfo{author}{{Patat} F},
  \bibinfo{author}{{Danziger} IJ}, \bibinfo{author}{{Young} TR},
  \bibinfo{author}{{Suzuki} T}, \bibinfo{author}{{Shigeyama} T},
  \bibinfo{author}{{Augusteijn} T}, \bibinfo{author}{{Doublier} V},
  \bibinfo{author}{{Gonzalez} JF}, \bibinfo{author}{{Boehnhardt} H},
  \bibinfo{author}{{Brewer} J}, \bibinfo{author}{{Hainaut} OR},
  \bibinfo{author}{{Lidman} C}, \bibinfo{author}{{Leibundgut} B},
  \bibinfo{author}{{Cappellaro} E}, \bibinfo{author}{{Turatto} M},
  \bibinfo{author}{{Galama} TJ}, \bibinfo{author}{{Vreeswijk} PM},
  \bibinfo{author}{{Kouveliotou} C}, \bibinfo{author}{{van Paradijs} J},
  \bibinfo{author}{{Pian} E}, \bibinfo{author}{{Palazzi} E} and
  \bibinfo{author}{{Frontera} F} (\bibinfo{year}{1998}), \bibinfo{month}{Oct.}
\bibinfo{title}{{A hypernova model for the supernova associated with the
  {\ensuremath{\gamma}}-ray burst of 25 April 1998}}.
\bibinfo{journal}{{\em \nat}} \bibinfo{volume}{395} (\bibinfo{number}{6703}):
  \bibinfo{pages}{672--674}. \bibinfo{doi}{\doi{10.1038/27155}}.
\eprint{astro-ph/9806382}.

\bibtype{Article}%
\bibitem[{Jakobus} et al.(2022)]{jakobus_22}
\bibinfo{author}{{Jakobus} P}, \bibinfo{author}{{M{\"u}ller} B},
  \bibinfo{author}{{Heger} A}, \bibinfo{author}{{Motornenko} A},
  \bibinfo{author}{{Steinheimer} J} and  \bibinfo{author}{{Stoecker} H}
  (\bibinfo{year}{2022}), \bibinfo{month}{Oct.}
\bibinfo{title}{{The role of the hadron-quark phase transition in core-collapse
  supernovae}}.
\bibinfo{journal}{{\em \mnras}} \bibinfo{volume}{516} (\bibinfo{number}{2}):
  \bibinfo{pages}{2554--2574}. \bibinfo{doi}{\doi{10.1093/mnras/stac2352}}.

\bibtype{Article}%
\bibitem[{Jakobus} et al.(2025)]{jakobus_25}
\bibinfo{author}{{Jakobus} P}, \bibinfo{author}{{M{\"u}ller} B} and
  \bibinfo{author}{{Heger} A} (\bibinfo{year}{2025}), \bibinfo{month}{Jul.}
\bibinfo{title}{{Convection and the core g mode in proto-compact stars -- a
  detailed analysis}}.
\bibinfo{journal}{{\em \mnras}} \bibinfo{volume}{540} (\bibinfo{number}{4}):
  \bibinfo{pages}{3008--3031}. \bibinfo{doi}{\doi{10.1093/mnras/staf868}}.

\bibtype{Article}%
\bibitem[{Janka}(2001)]{janka_01}
\bibinfo{author}{{Janka} HT} (\bibinfo{year}{2001}), \bibinfo{month}{Mar.}
\bibinfo{title}{{Conditions for shock revival by neutrino heating in
  core-collapse supernovae}}.
\bibinfo{journal}{{\em \aap}} \bibinfo{volume}{368}: \bibinfo{pages}{527--560}.
  \bibinfo{doi}{\doi{10.1051/0004-6361:20010012}}.

\bibtype{Article}%
\bibitem[{Janka}(2012)]{janka_12}
\bibinfo{author}{{Janka} HT} (\bibinfo{year}{2012}), \bibinfo{month}{Nov.}
\bibinfo{title}{{Explosion Mechanisms of Core-Collapse Supernovae}}.
\bibinfo{journal}{{\em Annual Review of Nuclear and Particle Science}}
  \bibinfo{volume}{62}: \bibinfo{pages}{407--451}.
  \bibinfo{doi}{\doi{10.1146/annurev-nucl-102711-094901}}.

\bibtype{incollection}%
\bibitem[{Janka}(2017)]{janka_17b}
\bibinfo{author}{{Janka} HT} (\bibinfo{year}{2017}), \bibinfo{title}{{Neutrino
  Emission from Supernovae}}, \bibinfo{editor}{{Alsabti} AW} and
  \bibinfo{editor}{{Murdin} P}, (Eds.), \bibinfo{booktitle}{Handbook of
  Supernovae}, pp. \bibinfo{pages}{1575}.

\bibtype{Article}%
\bibitem[{Janka}(2025)]{janka_25}
\bibinfo{author}{{Janka} HT} (\bibinfo{year}{2025}), \bibinfo{month}{Sep.}
\bibinfo{title}{{Long-Term Multidimensional Models of Core-Collapse Supernovae:
  Progress and Challenges}}.
\bibinfo{journal}{{\em Annual Review of Nuclear and Particle Science}}
  \bibinfo{volume}{75} (\bibinfo{number}{1}): \bibinfo{pages}{425--461}.
  \bibinfo{doi}{\doi{10.1146/annurev-nucl-121423-100945}}.

\bibtype{Article}%
\bibitem[{Janka} and {Kresse}(2024)]{janka_24}
\bibinfo{author}{{Janka} HT} and  \bibinfo{author}{{Kresse} D}
  (\bibinfo{year}{2024}), \bibinfo{month}{Aug.}
\bibinfo{title}{{Interplay between neutrino kicks and hydrodynamic kicks of
  neutron stars and black holes}}.
\bibinfo{journal}{{\em \apss}} \bibinfo{volume}{369} (\bibinfo{number}{8}),
  \bibinfo{eid}{80}. \bibinfo{doi}{\doi{10.1007/s10509-024-04343-1}}.

\bibtype{Article}%
\bibitem[{Janka} and {M\"uller}(1996)]{janka_96}
\bibinfo{author}{{Janka} HT} and  \bibinfo{author}{{M\"uller} E}
  (\bibinfo{year}{1996}), \bibinfo{month}{Feb.}
\bibinfo{title}{{Neutrino heating, convection, and the mechanism of Type-II
  supernova explosions.}}
\bibinfo{journal}{{\em \aap}} \bibinfo{volume}{306}: \bibinfo{pages}{167--198}.

\bibtype{Article}%
\bibitem[{Janka} et al.(2007)]{janka_07}
\bibinfo{author}{{Janka} HT}, \bibinfo{author}{{Langanke} K},
  \bibinfo{author}{{Marek} A}, \bibinfo{author}{{Mart{\'{\i}}nez-Pinedo} G} and
   \bibinfo{author}{{M{\"u}ller} B} (\bibinfo{year}{2007}),
  \bibinfo{month}{Apr.}
\bibinfo{title}{{Theory of core-collapse supernovae}}.
\bibinfo{journal}{{\em \physrep}} \bibinfo{volume}{442}:
  \bibinfo{pages}{38--74}. \bibinfo{doi}{\doi{10.1016/j.physrep.2007.02.002}}.

\bibtype{Article}%
\bibitem[{Janka} et al.(2012)]{janka_12b}
\bibinfo{author}{{Janka} HT}, \bibinfo{author}{{Hanke} F},
  \bibinfo{author}{{H{\"u}depohl} L}, \bibinfo{author}{{Marek} A},
  \bibinfo{author}{{M{\"u}ller} B} and  \bibinfo{author}{{Obergaulinger} M}
  (\bibinfo{year}{2012}), \bibinfo{month}{Dec.}
\bibinfo{title}{{Core-collapse supernovae: Reflections and directions}}.
\bibinfo{journal}{{\em Progress of Theoretical and Experimental Physics}}
  \bibinfo{volume}{2012} (\bibinfo{number}{1}): \bibinfo{pages}{010000}.
  \bibinfo{doi}{\doi{10.1093/ptep/pts067}}.

\bibtype{incollection}%
\bibitem[{Jerkstrand}(2017)]{jerkstrand_17}
\bibinfo{author}{{Jerkstrand} A} (\bibinfo{year}{2017}),
  \bibinfo{title}{{Spectra of Supernovae in the Nebular Phase}},
  \bibinfo{editor}{{Alsabti} AW} and  \bibinfo{editor}{{Murdin} P}, (Eds.),
  \bibinfo{booktitle}{Handbook of Supernovae}, pp. \bibinfo{pages}{795}.

\bibtype{Article}%
\bibitem[{Jerkstrand} et al.(2020)]{jerkstrand_20}
\bibinfo{author}{{Jerkstrand} A}, \bibinfo{author}{{Wongwathanarat} A},
  \bibinfo{author}{{Janka} HT}, \bibinfo{author}{{Gabler} M},
  \bibinfo{author}{{Alp} D}, \bibinfo{author}{{Diehl} R},
  \bibinfo{author}{{Maeda} K}, \bibinfo{author}{{Larsson} J},
  \bibinfo{author}{{Fransson} C}, \bibinfo{author}{{Menon} A} and
  \bibinfo{author}{{Heger} A} (\bibinfo{year}{2020}), \bibinfo{month}{May}.
\bibinfo{title}{{Properties of gamma-ray decay lines in 3D core-collapse
  supernova models, with application to SN 1987A and Cas A}}.
\bibinfo{journal}{{\em \mnras}} \bibinfo{volume}{494} (\bibinfo{number}{2}):
  \bibinfo{pages}{2471--2497}. \bibinfo{doi}{\doi{10.1093/mnras/staa736}}.
\eprint{2003.05156}.

\bibtype{incollection}%
\bibitem[Jerkstrand et al.(2026)]{jerkstrand_26}
\bibinfo{author}{Jerkstrand A}, \bibinfo{author}{Milisavljevic D} and
  \bibinfo{author}{Müller B} (\bibinfo{year}{2026}),
  \bibinfo{title}{Core-collapse supernovae}, \bibinfo{editor}{Mandel I}, (Ed.),
  \bibinfo{booktitle}{Encyclopedia of Astrophysics (First Edition)},
  \bibinfo{publisher}{Elsevier}, \bibinfo{address}{Oxford},
  \bibinfo{pages}{639--668},
  \bibinfo{url}{\url{https://www.sciencedirect.com/science/article/pii/B9780443214394000900}}.

\bibtype{Article}%
\bibitem[{Johns} et al.(2025)]{johns_25}
\bibinfo{author}{{Johns} L}, \bibinfo{author}{{Richers} S} and
  \bibinfo{author}{{Wu} MR} (\bibinfo{year}{2025}), \bibinfo{month}{Sep.}
\bibinfo{title}{{Neutrino Oscillations in Core-Collapse Supernovae and Neutron
  Star Mergers}}.
\bibinfo{journal}{{\em Annual Review of Nuclear and Particle Science}}
  \bibinfo{volume}{75} (\bibinfo{number}{1}): \bibinfo{pages}{399--423}.
  \bibinfo{doi}{\doi{10.1146/annurev-nucl-121423-100853}}.
\eprint{2503.05959}.

\bibtype{Article}%
\bibitem[{Jones} et al.(2014)]{jones_14}
\bibinfo{author}{{Jones} S}, \bibinfo{author}{{Hirschi} R} and
  \bibinfo{author}{{Nomoto} K} (\bibinfo{year}{2014}), \bibinfo{month}{Dec.}
\bibinfo{title}{{The Final Fate of Stars that Ignite Neon and Oxygen
  Off-center: Electron Capture or Iron Core-collapse Supernova?}}
\bibinfo{journal}{{\em \apj}} \bibinfo{volume}{797}: \bibinfo{pages}{83}.
  \bibinfo{doi}{\doi{10.1088/0004-637X/797/2/83}}.

\bibtype{Article}%
\bibitem[{Jones} et al.(2016)]{jones_16}
\bibinfo{author}{{Jones} S}, \bibinfo{author}{{R{\"o}pke} FK},
  \bibinfo{author}{{Pakmor} R}, \bibinfo{author}{{Seitenzahl} IR},
  \bibinfo{author}{{Ohlmann} ST} and  \bibinfo{author}{{Edelmann} PVF}
  (\bibinfo{year}{2016}), \bibinfo{month}{Sep}.
\bibinfo{title}{{Do electron-capture supernovae make neutron stars?. First
  multidimensional hydrodynamic simulations of the oxygen deflagration}}.
\bibinfo{journal}{{\em \aap}} \bibinfo{volume}{593}: \bibinfo{pages}{A72}.
  \bibinfo{doi}{\doi{10.1051/0004-6361/201628321}}.

\bibtype{Article}%
\bibitem[{Just} et al.(2018)]{just_18}
\bibinfo{author}{{Just} O}, \bibinfo{author}{{Bollig} R},
  \bibinfo{author}{{Janka} HT}, \bibinfo{author}{{Obergaulinger} M},
  \bibinfo{author}{{Glas} R} and  \bibinfo{author}{{Nagataki} S}
  (\bibinfo{year}{2018}), \bibinfo{month}{Dec}.
\bibinfo{title}{{Core-collapse supernova simulations in one and two dimensions:
  comparison of codes and approximations}}.
\bibinfo{journal}{{\em \mnras}} \bibinfo{volume}{481} (\bibinfo{number}{4}):
  \bibinfo{pages}{4786--4814}. \bibinfo{doi}{\doi{10.1093/mnras/sty2578}}.

\bibtype{Article}%
\bibitem[{Kachelrie{\ss}} et al.(2005)]{kachelriess_05}
\bibinfo{author}{{Kachelrie{\ss}} M}, \bibinfo{author}{{Tom{\`a}s} R},
  \bibinfo{author}{{Buras} R}, \bibinfo{author}{{Janka} HT},
  \bibinfo{author}{{Marek} A} and  \bibinfo{author}{{Rampp} M}
  (\bibinfo{year}{2005}), \bibinfo{month}{Mar.}
\bibinfo{title}{{Exploiting the neutronization burst of a galactic supernova}}.
\bibinfo{journal}{{\em \prd}} \bibinfo{volume}{71} (\bibinfo{number}{6}):
  \bibinfo{pages}{063003}. \bibinfo{doi}{\doi{10.1103/PhysRevD.71.063003}}.

\bibtype{Article}%
\bibitem[{Kasen} and {Woosley}(2009)]{kasen_09}
\bibinfo{author}{{Kasen} D} and  \bibinfo{author}{{Woosley} SE}
  (\bibinfo{year}{2009}), \bibinfo{month}{Oct.}
\bibinfo{title}{{Type II Supernovae: Model Light Curves and Standard Candle
  Relationships}}.
\bibinfo{journal}{{\em \apj}} \bibinfo{volume}{703}:
  \bibinfo{pages}{2205--2216}.
  \bibinfo{doi}{\doi{10.1088/0004-637X/703/2/2205}}.

\bibtype{Article}%
\bibitem[{Keil} et al.(1997)]{keil_97}
\bibinfo{author}{{Keil} W}, \bibinfo{author}{{Janka} HT},
  \bibinfo{author}{{Schramm} DN}, \bibinfo{author}{{Sigl} G},
  \bibinfo{author}{{Turner} MS} and  \bibinfo{author}{{Ellis} J}
  (\bibinfo{year}{1997}), \bibinfo{month}{Aug.}
\bibinfo{title}{{Fresh look at axions and SN 1987A}}.
\bibinfo{journal}{{\em \prd}} \bibinfo{volume}{56}:
  \bibinfo{pages}{2419--2432}. \bibinfo{doi}{\doi{10.1103/PhysRevD.56.2419}}.

\bibtype{Article}%
\bibitem[{Keil} et al.(2003)]{keil_03}
\bibinfo{author}{{Keil} MT}, \bibinfo{author}{{Raffelt} GG} and
  \bibinfo{author}{{Janka} HT} (\bibinfo{year}{2003}), \bibinfo{month}{Jun.}
\bibinfo{title}{{Monte Carlo Study of Supernova Neutrino Spectra Formation}}.
\bibinfo{journal}{{\em \apj}} \bibinfo{volume}{590}: \bibinfo{pages}{971--991}.
  \bibinfo{doi}{\doi{10.1086/375130}}.

\bibtype{Article}%
\bibitem[{Kirsebom} et al.(2019)]{kirsebom_19}
\bibinfo{author}{{Kirsebom} OS}, \bibinfo{author}{{Jones} S},
  \bibinfo{author}{{Str{\"o}mberg} DF}, \bibinfo{author}{{Mart{\'\i}nez-Pinedo}
  G}, \bibinfo{author}{{Langanke} K}, \bibinfo{author}{{R{\"o}pke} FK},
  \bibinfo{author}{{Brown} BA}, \bibinfo{author}{{Eronen} T},
  \bibinfo{author}{{Fynbo} HOU}, \bibinfo{author}{{Hukkanen} M},
  \bibinfo{author}{{Idini} A}, \bibinfo{author}{{Jokinen} A},
  \bibinfo{author}{{Kankainen} A}, \bibinfo{author}{{Kostensalo} J},
  \bibinfo{author}{{Moore} I}, \bibinfo{author}{{M{\"o}ller} H},
  \bibinfo{author}{{Ohlmann} ST}, \bibinfo{author}{{Penttil{\"a}} H},
  \bibinfo{author}{{Riisager} K}, \bibinfo{author}{{Rinta-Antila} S},
  \bibinfo{author}{{Srivastava} PC}, \bibinfo{author}{{Suhonen} J},
  \bibinfo{author}{{Trzaska} WH} and  \bibinfo{author}{{{\'n}yst{\"o}} J}
  (\bibinfo{year}{2019}), \bibinfo{month}{Dec.}
\bibinfo{title}{{Discovery of an Exceptionally Strong {\ensuremath{\beta}}
  -Decay Transition of $^{20}$F and Implications for the Fate of
  Intermediate-Mass Stars}}.
\bibinfo{journal}{{\em \prl}} \bibinfo{volume}{123} (\bibinfo{number}{26}),
  \bibinfo{eid}{262701}. \bibinfo{doi}{\doi{10.1103/PhysRevLett.123.262701}}.

\bibtype{Article}%
\bibitem[{Kitaura} et al.(2006)]{kitaura_06}
\bibinfo{author}{{Kitaura} FS}, \bibinfo{author}{{Janka} HT} and
  \bibinfo{author}{{Hillebrandt} W} (\bibinfo{year}{2006}),
  \bibinfo{month}{Apr.}
\bibinfo{title}{{Explosions of O-Ne-Mg cores, the Crab supernova, and
  subluminous type II-P supernovae}}.
\bibinfo{journal}{{\em \aap}} \bibinfo{volume}{450}: \bibinfo{pages}{345--350}.
  \bibinfo{doi}{\doi{10.1051/0004-6361:20054703}}.

\bibtype{Article}%
\bibitem[{Kovalenko} et al.(2026)]{kovalenko_26}
\bibinfo{author}{{Kovalenko} L}, \bibinfo{author}{{O'Connor} E},
  \bibinfo{author}{{Andresen} H} and  \bibinfo{author}{{Couch} SM}
  (\bibinfo{year}{2026}), \bibinfo{month}{May}.
\bibinfo{title}{{A Three-Dimensional Exploration of Magnetic Fields, Rotation,
  and Shock Revival in a $39 M_\odot$ Core-Collapse Supernova Progenitor}}.
\bibinfo{journal}{{\em arXiv e-prints}} ,
  \bibinfo{eid}{arXiv:2605.18347}\bibinfo{doi}{\doi{10.48550/arXiv.2605.18347}}.
\eprint{2605.18347}.

\bibtype{Article}%
\bibitem[{Kresse} et al.(2021)]{kresse_21}
\bibinfo{author}{{Kresse} D}, \bibinfo{author}{{Ertl} T} and
  \bibinfo{author}{{Janka} HT} (\bibinfo{year}{2021}), \bibinfo{month}{Mar.}
\bibinfo{title}{{Stellar Collapse Diversity and the Diffuse Supernova Neutrino
  Background}}.
\bibinfo{journal}{{\em \apj}} \bibinfo{volume}{909} (\bibinfo{number}{2}),
  \bibinfo{eid}{169}. \bibinfo{doi}{\doi{10.3847/1538-4357/abd54e}}.
\eprint{2010.04728}.

\bibtype{Article}%
\bibitem[{Kuroda} et al.(2018)]{kuroda_18}
\bibinfo{author}{{Kuroda} T}, \bibinfo{author}{{Kotake} K},
  \bibinfo{author}{{Takiwaki} T} and  \bibinfo{author}{{Thielemann} FK}
  (\bibinfo{year}{2018}), \bibinfo{month}{Jun.}
\bibinfo{title}{{A full general relativistic neutrino radiation-hydrodynamics
  simulation of a collapsing very massive star and the formation of a black
  hole}}.
\bibinfo{journal}{{\em \mnras}} \bibinfo{volume}{477}:
  \bibinfo{pages}{L80--L84}. \bibinfo{doi}{\doi{10.1093/mnrasl/sly059}}.

\bibtype{Article}%
\bibitem[{Langanke} et al.(2003)]{langanke_03}
\bibinfo{author}{{Langanke} K}, \bibinfo{author}{{Mart{\'{\i}}nez--Pinedo} G},
  \bibinfo{author}{{Sampaio} JM}, \bibinfo{author}{{Dean} DJ},
  \bibinfo{author}{{Hix} WR}, \bibinfo{author}{{Messer} OE},
  \bibinfo{author}{{Mezzacappa} A}, \bibinfo{author}{{Liebend{\"o}rfer} M},
  \bibinfo{author}{{Janka} HT} and  \bibinfo{author}{{Rampp} M}
  (\bibinfo{year}{2003}), \bibinfo{month}{Jun.}
\bibinfo{title}{Electron capture rates on nuclei and implications for stellar
  core collapse}.
\bibinfo{journal}{{\em Physical Review Letters}} \bibinfo{volume}{90}
  (\bibinfo{number}{24}): \bibinfo{pages}{241102--+}.
  \bibinfo{doi}{\doi{10.1103/PhysRevLett.90.241102}}.

\bibtype{Article}%
\bibitem[{Lattimer} and {Swesty}(1991)]{lattimer_91}
\bibinfo{author}{{Lattimer} JM} and  \bibinfo{author}{{Swesty} FD}
  (\bibinfo{year}{1991}), \bibinfo{month}{Dec.}
\bibinfo{title}{{A generalized equation of state for hot, dense matter}}.
\bibinfo{journal}{{\em Nuclear Physics A}} \bibinfo{volume}{535}:
  \bibinfo{pages}{331--376}. \bibinfo{doi}{\doi{10.1016/0375-9474(91)90452-C}}.

\bibtype{Article}%
\bibitem[{Lentz} et al.(2015)]{lentz_15}
\bibinfo{author}{{Lentz} EJ}, \bibinfo{author}{{Bruenn} SW},
  \bibinfo{author}{{Hix} WR}, \bibinfo{author}{{Mezzacappa} A},
  \bibinfo{author}{{Messer} OEB}, \bibinfo{author}{{Endeve} E},
  \bibinfo{author}{{Blondin} JM}, \bibinfo{author}{{Harris} JA},
  \bibinfo{author}{{Marronetti} P} and  \bibinfo{author}{{Yakunin} KN}
  (\bibinfo{year}{2015}), \bibinfo{month}{Jul.}
\bibinfo{title}{{Three-dimensional Core-collapse Supernova Simulated Using a
  $15\,\mathrm{M}_\odot$ Progenitor}}.
\bibinfo{journal}{{\em \apjl}} \bibinfo{volume}{807}: \bibinfo{pages}{L31}.
  \bibinfo{doi}{\doi{10.1088/2041-8205/807/2/L31}}.

\bibtype{Article}%
\bibitem[{Li} et al.(2011)]{li_11}
\bibinfo{author}{{Li} W}, \bibinfo{author}{{Chornock} R},
  \bibinfo{author}{{Leaman} J}, \bibinfo{author}{{Filippenko} AV},
  \bibinfo{author}{{Poznanski} D}, \bibinfo{author}{{Wang} X},
  \bibinfo{author}{{Ganeshalingam} M} and  \bibinfo{author}{{Mannucci} F}
  (\bibinfo{year}{2011}), \bibinfo{month}{Apr.}
\bibinfo{title}{{Nearby supernova rates from the Lick Observatory Supernova
  Search - III. The rate-size relation, and the rates as a function of galaxy
  Hubble type and colour}}.
\bibinfo{journal}{{\em \mnras}} \bibinfo{volume}{412} (\bibinfo{number}{3}):
  \bibinfo{pages}{1473--1507}.
  \bibinfo{doi}{\doi{10.1111/j.1365-2966.2011.18162.x}}.
\eprint{1006.4613}.

\bibtype{Article}%
\bibitem[{Liebend{\"o}rfer} et al.(2001)]{liebendoerfer_00}
\bibinfo{author}{{Liebend{\"o}rfer} M}, \bibinfo{author}{{Mezzacappa} A},
  \bibinfo{author}{{Thielemann} FK}, \bibinfo{author}{{Messer} OE},
  \bibinfo{author}{{Hix} WR} and  \bibinfo{author}{{Bruenn} SW}
  (\bibinfo{year}{2001}), \bibinfo{month}{May}.
\bibinfo{title}{{Probing the gravitational well: No supernova explosion in
  spherical symmetry with general relativistic Boltzmann neutrino transport}}.
\bibinfo{journal}{{\em \prd}} \bibinfo{volume}{63} (\bibinfo{number}{10}):
  \bibinfo{pages}{103004:1--13}.
  \bibinfo{doi}{\doi{10.1103/PhysRevD.63.103004}}.

\bibtype{Article}%
\bibitem[{Liebend{\"o}rfer} et al.(2005)]{liebendoerfer_05}
\bibinfo{author}{{Liebend{\"o}rfer} M}, \bibinfo{author}{{Rampp} M},
  \bibinfo{author}{{Janka} HT} and  \bibinfo{author}{{Mezzacappa} A}
  (\bibinfo{year}{2005}).
\bibinfo{title}{{Supernova simulations with boltzmann neutrino transport: A
  comparison of methods}}.
\bibinfo{journal}{{\em \apj}} \bibinfo{volume}{620}: \bibinfo{pages}{840--860}.
  \bibinfo{doi}{\doi{10.1086/427203}}.

\bibtype{Article}%
\bibitem[{Lin} et al.(2024)]{lin_24}
\bibinfo{author}{{Lin} Z}, \bibinfo{author}{{Zha} S},
  \bibinfo{author}{{O'Connor} EP} and  \bibinfo{author}{{Steiner} AW}
  (\bibinfo{year}{2024}), \bibinfo{month}{Jan.}
\bibinfo{title}{{Detectability of neutrino-signal fluctuations induced by the
  hadron-quark phase transition in failing core-collapse supernovae}}.
\bibinfo{journal}{{\em \prd}} \bibinfo{volume}{109} (\bibinfo{number}{2}),
  \bibinfo{eid}{023005}. \bibinfo{doi}{\doi{10.1103/PhysRevD.109.023005}}.
\eprint{2203.05141}.

\bibtype{Article}%
\bibitem[{Lindquist}(1966)]{lindquist_66}
\bibinfo{author}{{Lindquist} RW} (\bibinfo{year}{1966}), \bibinfo{month}{May}.
\bibinfo{title}{{Relativistic transport theory}}.
\bibinfo{journal}{{\em {\it Ann.~Phys.}}} \bibinfo{volume}{37}:
  \bibinfo{pages}{487--518}.

\bibtype{Article}%
\bibitem[{Lovegrove} and {Woosley}(2013)]{Lovegrove_Woosley:2013}
\bibinfo{author}{{Lovegrove} E} and  \bibinfo{author}{{Woosley} SE}
  (\bibinfo{year}{2013}), \bibinfo{month}{Jun.}
\bibinfo{title}{{Very Low Energy Supernovae from Neutrino Mass Loss}}.
\bibinfo{journal}{{\em \apj}} \bibinfo{volume}{769} (\bibinfo{number}{2}),
  \bibinfo{eid}{109}. \bibinfo{doi}{\doi{10.1088/0004-637X/769/2/109}}.
\eprint{1303.5055}.

\bibtype{Article}%
\bibitem[{Lucente} et al.(2020)]{lucente_20}
\bibinfo{author}{{Lucente} G}, \bibinfo{author}{{Carenza} P},
  \bibinfo{author}{{Fischer} T}, \bibinfo{author}{{Giannotti} M} and
  \bibinfo{author}{{Mirizzi} A} (\bibinfo{year}{2020}), \bibinfo{month}{Dec.}
\bibinfo{title}{{Heavy axion-like particles and core-collapse supernovae:
  constraints and impact on the explosion mechanism}}.
\bibinfo{journal}{{\em \jcap}} \bibinfo{volume}{2020} (\bibinfo{number}{12}),
  \bibinfo{eid}{008}. \bibinfo{doi}{\doi{10.1088/1475-7516/2020/12/008}}.
\eprint{2008.04918}.

\bibtype{Article}%
\bibitem[{Lunardini} et al.(2008)]{lunardini_08}
\bibinfo{author}{{Lunardini} C}, \bibinfo{author}{{M{\"u}ller} B} and
  \bibinfo{author}{{Janka} HT} (\bibinfo{year}{2008}), \bibinfo{month}{Jul.}
\bibinfo{title}{{Neutrino oscillation signatures of oxygen-neon-magnesium
  supernovae}}.
\bibinfo{journal}{{\em \prd}} \bibinfo{volume}{78} (\bibinfo{number}{2}):
  \bibinfo{pages}{023016:1--13}.
  \bibinfo{doi}{\doi{10.1103/PhysRevD.78.023016}}.

\bibtype{Article}%
\bibitem[{Lunardini} et al.(2026)]{lunardini_26}
\bibinfo{author}{{Lunardini} C}, \bibinfo{author}{{Takiwaki} T},
  \bibinfo{author}{{Kinugawa} T}, \bibinfo{author}{{Horiuchi} S} and
  \bibinfo{author}{{Kotake} K} (\bibinfo{year}{2026}), \bibinfo{month}{Mar.}
\bibinfo{title}{{Diffuse supernova neutrino background: An update with modern
  population synthesis and core-collapse simulations}}.
\bibinfo{journal}{{\em \prd}} \bibinfo{volume}{113} (\bibinfo{number}{6}),
  \bibinfo{eid}{063044}. \bibinfo{doi}{\doi{10.1103/yt5x-cbnp}}.
\eprint{2506.22699}.

\bibtype{Article}%
\bibitem[{Maas} and {Paschke}(2017)]{maas_17}
\bibinfo{author}{{Maas} FE} and  \bibinfo{author}{{Paschke} KD}
  (\bibinfo{year}{2017}), \bibinfo{month}{Jul.}
\bibinfo{title}{{Strange nucleon form-factors}}.
\bibinfo{journal}{{\em Progress in Particle and Nuclear Physics}}
  \bibinfo{volume}{95}: \bibinfo{pages}{209--244}.
  \bibinfo{doi}{\doi{10.1016/j.ppnp.2016.11.001}}.

\bibtype{Article}%
\bibitem[{Mart{\'{\i}}nez-Pinedo} et al.(2012)]{martinez_12}
\bibinfo{author}{{Mart{\'{\i}}nez-Pinedo} G}, \bibinfo{author}{{Fischer} T},
  \bibinfo{author}{{Lohs} A} and  \bibinfo{author}{{Huther} L}
  (\bibinfo{year}{2012}), \bibinfo{month}{Dec.}
\bibinfo{title}{{Charged-Current Weak Interaction Processes in Hot and Dense
  Matter and its Impact on the Spectra of Neutrinos Emitted from Protoneutron
  Star Cooling}}.
\bibinfo{journal}{{\em Physical Review Letters}} \bibinfo{volume}{109}
  (\bibinfo{number}{25}): \bibinfo{pages}{251104}.
  \bibinfo{doi}{\doi{10.1103/PhysRevLett.109.251104}}.

\bibtype{Article}%
\bibitem[{Matsumoto} et al.(2024)]{matsumoto_24}
\bibinfo{author}{{Matsumoto} J}, \bibinfo{author}{{Takiwaki} T} and
  \bibinfo{author}{{Kotake} K} (\bibinfo{year}{2024}), \bibinfo{month}{Feb.}
\bibinfo{title}{{Neutrino-driven massive stellar explosions in 3D fostered by
  magnetic fields via turbulent {\ensuremath{\alpha}}-effect}}.
\bibinfo{journal}{{\em \mnras}} \bibinfo{volume}{528} (\bibinfo{number}{1}):
  \bibinfo{pages}{L96--L101}. \bibinfo{doi}{\doi{10.1093/mnrasl/slad173}}.

\bibtype{Article}%
\bibitem[{McCray}(1993)]{mccray_93}
\bibinfo{author}{{McCray} R} (\bibinfo{year}{1993}), \bibinfo{month}{Jan}.
\bibinfo{title}{{Supernova 1987A revisited.}}
\bibinfo{journal}{{\em \araa}} \bibinfo{volume}{31}: \bibinfo{pages}{175--216}.
  \bibinfo{doi}{\doi{10.1146/annurev.aa.31.090193.001135}}.

\bibtype{Article}%
\bibitem[{Melson} et al.(2015)]{melson_15b}
\bibinfo{author}{{Melson} T}, \bibinfo{author}{{Janka} HT},
  \bibinfo{author}{{Bollig} R}, \bibinfo{author}{{Hanke} F},
  \bibinfo{author}{{Marek} A} and  \bibinfo{author}{{M{\"u}ller} B}
  (\bibinfo{year}{2015}), \bibinfo{month}{Aug.}
\bibinfo{title}{{Neutrino-driven Explosion of a 20 Solar-mass Star in Three
  Dimensions Enabled by Strange-quark Contributions to Neutrino-Nucleon
  Scattering}}.
\bibinfo{journal}{{\em \apjl}} \bibinfo{volume}{808}: \bibinfo{pages}{L42}.
  \bibinfo{doi}{\doi{10.1088/2041-8205/808/2/L42}}.

\bibtype{Article}%
\bibitem[{Mezzacappa}(2026)]{mezzacappa_26}
\bibinfo{author}{{Mezzacappa} A} (\bibinfo{year}{2026}), \bibinfo{month}{Apr.}
\bibinfo{title}{{Core Collapse Supernova Modeling: The Next Ten Years}}.
\bibinfo{journal}{{\em arXiv e-prints}} ,
  \bibinfo{eid}{arXiv:2604.24970}\bibinfo{doi}{\doi{10.48550/arXiv.2604.24970}}.
\eprint{2604.24970}.

\bibtype{Article}%
\bibitem[{Mezzacappa} and {Zanolin}(2025)]{mezzacappa_25}
\bibinfo{author}{{Mezzacappa} A} and  \bibinfo{author}{{Zanolin} M}
  (\bibinfo{year}{2025}), \bibinfo{month}{Oct.}
\bibinfo{title}{{Colloquium: Gravitational waves from neutrino-driven core
  collapse supernovae}}.
\bibinfo{journal}{{\em Reviews of Modern Physics}} \bibinfo{volume}{97}
  (\bibinfo{number}{4}), \bibinfo{eid}{041002}.
  \bibinfo{doi}{\doi{10.1103/pv6p-dtr2}}.

\bibtype{Article}%
\bibitem[{Mezzacappa} et al.(2020{\natexlab{a}})]{mezzacappa_20}
\bibinfo{author}{{Mezzacappa} A}, \bibinfo{author}{{Endeve} E},
  \bibinfo{author}{{Messer} OEB} and  \bibinfo{author}{{Bruenn} SW}
  (\bibinfo{year}{2020}{\natexlab{a}}), \bibinfo{month}{Dec.}
\bibinfo{title}{{Physical, numerical, and computational challenges of modeling
  neutrino transport in core-collapse supernovae}}.
\bibinfo{journal}{{\em Living Reviews in Computational Astrophysics}}
  \bibinfo{volume}{6} (\bibinfo{number}{1}), \bibinfo{eid}{4}.
  \bibinfo{doi}{\doi{10.1007/s41115-020-00010-8}}.

\bibtype{Article}%
\bibitem[{Mezzacappa} et al.(2020{\natexlab{b}})]{mezzacappa_20b}
\bibinfo{author}{{Mezzacappa} A}, \bibinfo{author}{{Marronetti} P},
  \bibinfo{author}{{Landfield} RE}, \bibinfo{author}{{Lentz} EJ},
  \bibinfo{author}{{Yakunin} KN}, \bibinfo{author}{{Bruenn} SW},
  \bibinfo{author}{{Hix} WR}, \bibinfo{author}{{Messer} OEB},
  \bibinfo{author}{{Endeve} E}, \bibinfo{author}{{Blondin} JM} and
  \bibinfo{author}{{Harris} JA} (\bibinfo{year}{2020}{\natexlab{b}}),
  \bibinfo{month}{Jul.}
\bibinfo{title}{{Gravitational-wave signal of a core-collapse supernova
  explosion of a 15 M$_{{\ensuremath{\odot}}}$ star}}.
\bibinfo{journal}{{\em \prd}} \bibinfo{volume}{102} (\bibinfo{number}{2}),
  \bibinfo{eid}{023027}. \bibinfo{doi}{\doi{10.1103/PhysRevD.102.023027}}.

\bibtype{Article}%
\bibitem[{Mezzacappa} et al.(2023)]{mezzacappa_23}
\bibinfo{author}{{Mezzacappa} A}, \bibinfo{author}{{Marronetti} P},
  \bibinfo{author}{{Landfield} RE}, \bibinfo{author}{{Lentz} EJ},
  \bibinfo{author}{{Murphy} RD}, \bibinfo{author}{{Raphael Hix} W},
  \bibinfo{author}{{Harris} JA}, \bibinfo{author}{{Bruenn} SW},
  \bibinfo{author}{{Blondin} JM}, \bibinfo{author}{{Bronson Messer} OE},
  \bibinfo{author}{{Casanova} J} and  \bibinfo{author}{{Kronzer} LL}
  (\bibinfo{year}{2023}), \bibinfo{month}{Feb.}
\bibinfo{title}{{Core collapse supernova gravitational wave emission for
  progenitors of 9.6, 15, and 25M{\ensuremath{\odot}}}}.
\bibinfo{journal}{{\em \prd}} \bibinfo{volume}{107} (\bibinfo{number}{4}),
  \bibinfo{eid}{043008}. \bibinfo{doi}{\doi{10.1103/PhysRevD.107.043008}}.

\bibtype{Article}%
\bibitem[{Miller} et al.(2021)]{miller_21}
\bibinfo{author}{{Miller} MC}, \bibinfo{author}{{Lamb} FK},
  \bibinfo{author}{{Dittmann} AJ}, \bibinfo{author}{{Bogdanov} S},
  \bibinfo{author}{{Arzoumanian} Z}, \bibinfo{author}{{Gendreau} KC},
  \bibinfo{author}{{Guillot} S}, \bibinfo{author}{{Ho} WCG},
  \bibinfo{author}{{Lattimer} JM}, \bibinfo{author}{{Loewenstein} M},
  \bibinfo{author}{{Morsink} SM}, \bibinfo{author}{{Ray} PS},
  \bibinfo{author}{{Wolff} MT}, \bibinfo{author}{{Baker} CL},
  \bibinfo{author}{{Cazeau} T}, \bibinfo{author}{{Manthripragada} S},
  \bibinfo{author}{{Markwardt} CB}, \bibinfo{author}{{Okajima} T},
  \bibinfo{author}{{Pollard} S}, \bibinfo{author}{{Cognard} I},
  \bibinfo{author}{{Cromartie} HT}, \bibinfo{author}{{Fonseca} E},
  \bibinfo{author}{{Guillemot} L}, \bibinfo{author}{{Kerr} M},
  \bibinfo{author}{{Parthasarathy} A}, \bibinfo{author}{{Pennucci} TT},
  \bibinfo{author}{{Ransom} S} and  \bibinfo{author}{{Stairs} I}
  (\bibinfo{year}{2021}), \bibinfo{month}{Sep.}
\bibinfo{title}{{The Radius of PSR J0740+6620 from NICER and XMM-Newton Data}}.
\bibinfo{journal}{{\em \apjl}} \bibinfo{volume}{918} (\bibinfo{number}{2}),
  \bibinfo{eid}{L28}. \bibinfo{doi}{\doi{10.3847/2041-8213/ac089b}}.
\eprint{2105.06979}.

\bibtype{Article}%
\bibitem[{Mirizzi} et al.(2016)]{mirizzi_16}
\bibinfo{author}{{Mirizzi} A}, \bibinfo{author}{{Tamborra} I},
  \bibinfo{author}{{Janka} HT}, \bibinfo{author}{{Saviano} N},
  \bibinfo{author}{{Scholberg} K}, \bibinfo{author}{{Bollig} R},
  \bibinfo{author}{{H{\"u}depohl} L} and  \bibinfo{author}{{Chakraborty} S}
  (\bibinfo{year}{2016}).
\bibinfo{title}{{Supernova neutrinos: production, oscillations and detection}}.
\bibinfo{journal}{{\em Nuovo Cimento Rivista Serie}} \bibinfo{volume}{39}:
  \bibinfo{pages}{1--112}. \bibinfo{doi}{\doi{10.1393/ncr/i2016-10120-8}}.

\bibtype{Article}%
\bibitem[{Miyaji} and {Nomoto}(1987)]{miyaji_87}
\bibinfo{author}{{Miyaji} S} and  \bibinfo{author}{{Nomoto} K}
  (\bibinfo{year}{1987}), \bibinfo{month}{Jul.}
\bibinfo{title}{{On the Collapse of 8--10 M$_{sun}$ Stars Due to Electron
  Capture}}.
\bibinfo{journal}{{\em \apj}} \bibinfo{volume}{318}: \bibinfo{pages}{307}.
  \bibinfo{doi}{\doi{10.1086/165368}}.

\bibtype{Article}%
\bibitem[{Morozova} et al.(2018)]{morozova_18}
\bibinfo{author}{{Morozova} V}, \bibinfo{author}{{Radice} D},
  \bibinfo{author}{{Burrows} A} and  \bibinfo{author}{{Vartanyan} D}
  (\bibinfo{year}{2018}), \bibinfo{month}{Jul}.
\bibinfo{title}{{The Gravitational Wave Signal from Core-collapse Supernovae}}.
\bibinfo{journal}{{\em \apj}} \bibinfo{volume}{861} (\bibinfo{number}{1}):
  \bibinfo{pages}{10}. \bibinfo{doi}{\doi{10.3847/1538-4357/aac5f1}}.

\bibtype{Article}%
\bibitem[{Motornenko} et al.(2020)]{motornenko_20}
\bibinfo{author}{{Motornenko} A}, \bibinfo{author}{{Steinheimer} J},
  \bibinfo{author}{{Vovchenko} V}, \bibinfo{author}{{Schramm} S} and
  \bibinfo{author}{{Stoecker} H} (\bibinfo{year}{2020}), \bibinfo{month}{Mar.}
\bibinfo{title}{{Equation of state for hot QCD and compact stars from a
  mean-field approach}}.
\bibinfo{journal}{{\em \prc}} \bibinfo{volume}{101} (\bibinfo{number}{3}),
  \bibinfo{eid}{034904}. \bibinfo{doi}{\doi{10.1103/PhysRevC.101.034904}}.

\bibtype{Article}%
\bibitem[{M{\"u}ller}(2019)]{mueller_19d}
\bibinfo{author}{{M{\"u}ller} B} (\bibinfo{year}{2019}), \bibinfo{month}{Oct.}
\bibinfo{title}{{Neutrino Emission as Diagnostics of Core-Collapse
  Supernovae}}.
\bibinfo{journal}{{\em Annual Review of Nuclear and Particle Science}}
  \bibinfo{volume}{69}: \bibinfo{pages}{253--278}.
  \bibinfo{doi}{\doi{10.1146/annurev-nucl-101918-023434}}.

\bibtype{Article}%
\bibitem[{M{\"u}ller}(2020)]{mueller_20}
\bibinfo{author}{{M{\"u}ller} B} (\bibinfo{year}{2020}), \bibinfo{month}{Jun.}
\bibinfo{title}{{Hydrodynamics of core-collapse supernovae and their
  progenitors}}.
\bibinfo{journal}{{\em Living Reviews in Computational Astrophysics}}
  \bibinfo{volume}{6} (\bibinfo{number}{1}), \bibinfo{eid}{3}.
  \bibinfo{doi}{\doi{10.1007/s41115-020-0008-5}}.

\bibtype{incollection}%
\bibitem[M{\"u}ller(2025)]{mueller_25b}
\bibinfo{author}{M{\"u}ller B} (\bibinfo{year}{2025}),
  \bibinfo{title}{Supernova simulations}, \bibinfo{editor}{Bambi C},
  \bibinfo{editor}{Mizuno Y}, \bibinfo{editor}{Shashank S} and
  \bibinfo{editor}{Yuan F}, (Eds.), \bibinfo{booktitle}{New Frontiers in GRMHD
  Simulations}, \bibinfo{publisher}{Springer Nature Singapore},
  \bibinfo{address}{Singapore},  \bibinfo{pages}{663--698}.

\bibtype{Article}%
\bibitem[{M{\"u}ller}(2026)]{mueller_26}
\bibinfo{author}{{M{\"u}ller} B} (\bibinfo{year}{2026}), \bibinfo{month}{Mar.}
\bibinfo{title}{{Core-Collapse Supernovae and their Gravitational Wave Signals:
  The Status of Theory and Modeling}}.
\bibinfo{journal}{{\em arXiv e-prints}} ,
  \bibinfo{eid}{arXiv:2603.24243}\bibinfo{doi}{\doi{10.48550/arXiv.2603.24243}}.
\eprint{2603.24243}.

\bibtype{Article}%
\bibitem[{M{\"u}ller} and {Janka}(2014)]{mueller_14}
\bibinfo{author}{{M{\"u}ller} B} and  \bibinfo{author}{{Janka} HT}
  (\bibinfo{year}{2014}), \bibinfo{month}{Jun.}
\bibinfo{title}{{A New Multi-dimensional General Relativistic Neutrino
  Hydrodynamics Code for Core-collapse Supernovae. IV. The Neutrino Signal}}.
\bibinfo{journal}{{\em \apj}} \bibinfo{volume}{788}: \bibinfo{pages}{82}.
  \bibinfo{doi}{\doi{10.1088/0004-637X/788/1/82}}.

\bibtype{Article}%
\bibitem[{M{\"u}ller} and {Janka}(2015)]{mueller_15a}
\bibinfo{author}{{M{\"u}ller} B} and  \bibinfo{author}{{Janka} HT}
  (\bibinfo{year}{2015}), \bibinfo{month}{Apr.}
\bibinfo{title}{{Non-radial instabilities and progenitor asphericities in
  core-collapse supernovae}}.
\bibinfo{journal}{{\em \mnras}} \bibinfo{volume}{448}:
  \bibinfo{pages}{2141--2174}. \bibinfo{doi}{\doi{10.1093/mnras/stv101}}.

\bibtype{Article}%
\bibitem[{M{\"u}ller} et al.(2010)]{mueller_10}
\bibinfo{author}{{M{\"u}ller} B}, \bibinfo{author}{{Janka} H} and
  \bibinfo{author}{{Dimmelmeier} H} (\bibinfo{year}{2010}),
  \bibinfo{month}{Jul.}
\bibinfo{title}{{A New Multi-dimensional General Relativistic Neutrino
  Hydrodynamic Code for Core-collapse Supernovae. I. Method and Code Tests in
  Spherical Symmetry}}.
\bibinfo{journal}{{\em \apjs}} \bibinfo{volume}{189}:
  \bibinfo{pages}{104--133}. \bibinfo{doi}{\doi{10.1088/0067-0049/189/1/104}}.

\bibtype{Article}%
\bibitem[{M{\"u}ller} et al.(2013)]{mueller_13}
\bibinfo{author}{{M{\"u}ller} B}, \bibinfo{author}{{Janka} HT} and
  \bibinfo{author}{{Marek} A} (\bibinfo{year}{2013}), \bibinfo{month}{Mar.}
\bibinfo{title}{{A New Multi-dimensional General Relativistic Neutrino
  Hydrodynamics Code of Core-collapse Supernovae. III. Gravitational Wave
  Signals from Supernova Explosion Models}}.
\bibinfo{journal}{{\em \apj}} \bibinfo{volume}{766}: \bibinfo{pages}{43}.
  \bibinfo{doi}{\doi{10.1088/0004-637X/766/1/43}}.

\bibtype{Article}%
\bibitem[{M{\"u}ller} et al.(2016)]{mueller_16a}
\bibinfo{author}{{M{\"u}ller} B}, \bibinfo{author}{{Heger} A},
  \bibinfo{author}{{Liptai} D} and  \bibinfo{author}{{Cameron} JB}
  (\bibinfo{year}{2016}), \bibinfo{month}{Jul.}
\bibinfo{title}{{A simple approach to the supernova progenitor-explosion
  connection}}.
\bibinfo{journal}{{\em \mnras}} \bibinfo{volume}{460}:
  \bibinfo{pages}{742--764}. \bibinfo{doi}{\doi{10.1093/mnras/stw1083}}.

\bibtype{Article}%
\bibitem[{M{\"u}ller} et al.(2017{\natexlab{a}})]{mueller_17}
\bibinfo{author}{{M{\"u}ller} B}, \bibinfo{author}{{Melson} T},
  \bibinfo{author}{{Heger} A} and  \bibinfo{author}{{Janka} HT}
  (\bibinfo{year}{2017}{\natexlab{a}}), \bibinfo{month}{Nov.}
\bibinfo{title}{{Supernova simulations from a 3D progenitor model - Impact of
  perturbations and evolution of explosion properties}}.
\bibinfo{journal}{{\em \mnras}} \bibinfo{volume}{472}:
  \bibinfo{pages}{491--513}. \bibinfo{doi}{\doi{10.1093/mnras/stx1962}}.

\bibtype{Article}%
\bibitem[{M{\"u}ller} et al.(2017{\natexlab{b}})]{mueller_t_17}
\bibinfo{author}{{M{\"u}ller} T}, \bibinfo{author}{{Prieto} JL},
  \bibinfo{author}{{Pejcha} O} and  \bibinfo{author}{{Clocchiatti} A}
  (\bibinfo{year}{2017}{\natexlab{b}}), \bibinfo{month}{Jun.}
\bibinfo{title}{{The Nickel Mass Distribution of Normal Type II Supernovae}}.
\bibinfo{journal}{{\em \apj}} \bibinfo{volume}{841}: \bibinfo{pages}{127}.
  \bibinfo{doi}{\doi{10.3847/1538-4357/aa72f1}}.

\bibtype{Article}%
\bibitem[{M{\"u}ller} et al.(2019)]{mueller_19a}
\bibinfo{author}{{M{\"u}ller} B}, \bibinfo{author}{{Tauris} TM},
  \bibinfo{author}{{Heger} A}, \bibinfo{author}{{Banerjee} P},
  \bibinfo{author}{{Qian} YZ}, \bibinfo{author}{{Powell} J},
  \bibinfo{author}{{Chan} C}, \bibinfo{author}{{Gay} DW} and
  \bibinfo{author}{{Langer} N} (\bibinfo{year}{2019}), \bibinfo{month}{Apr.}
\bibinfo{title}{{Three-dimensional simulations of neutrino-driven core-collapse
  supernovae from low-mass single and binary star progenitors}}.
\bibinfo{journal}{{\em \mnras}} \bibinfo{volume}{484}:
  \bibinfo{pages}{3307--3324}. \bibinfo{doi}{\doi{10.1093/mnras/stz216}}.

\bibtype{Article}%
\bibitem[{Murphy} and {Meakin}(2011)]{murphy_11}
\bibinfo{author}{{Murphy} JW} and  \bibinfo{author}{{Meakin} C}
  (\bibinfo{year}{2011}), \bibinfo{month}{Dec.}
\bibinfo{title}{{A Global Turbulence Model for Neutrino-driven Convection in
  Core-collapse Supernovae}}.
\bibinfo{journal}{{\em \apj}} \bibinfo{volume}{742}: \bibinfo{pages}{74}.
  \bibinfo{doi}{\doi{10.1088/0004-637X/742/2/74}}.

\bibtype{Article}%
\bibitem[{Murphy} et al.(2009)]{murphy_09}
\bibinfo{author}{{Murphy} JW}, \bibinfo{author}{{Ott} CD} and
  \bibinfo{author}{{Burrows} A} (\bibinfo{year}{2009}), \bibinfo{month}{Dec.}
\bibinfo{title}{{A Model for Gravitational Wave Emission from Neutrino-Driven
  Core-Collapse Supernovae}}.
\bibinfo{journal}{{\em \apj}} \bibinfo{volume}{707}:
  \bibinfo{pages}{1173--1190}.
  \bibinfo{doi}{\doi{10.1088/0004-637X/707/2/1173}}.

\bibtype{Article}%
\bibitem[{Nadezhin}(1980)]{Nadezhin:1980}
\bibinfo{author}{{Nadezhin} DK} (\bibinfo{year}{1980}), \bibinfo{month}{May}.
\bibinfo{title}{{Some Secondary Indications of Gravitational Collapse}}.
\bibinfo{journal}{{\em \apss}} \bibinfo{volume}{69} (\bibinfo{number}{1}):
  \bibinfo{pages}{115--125}. \bibinfo{doi}{\doi{10.1007/BF00638971}}.

\bibtype{Article}%
\bibitem[{Nakamura} et al.(2022)]{nakamura_22}
\bibinfo{author}{{Nakamura} K}, \bibinfo{author}{{Takiwaki} T} and
  \bibinfo{author}{{Kotake} K} (\bibinfo{year}{2022}), \bibinfo{month}{Aug.}
\bibinfo{title}{{Three-dimensional simulation of a core-collapse supernova for
  a binary star progenitor of SN 1987A}}.
\bibinfo{journal}{{\em \mnras}} \bibinfo{volume}{514} (\bibinfo{number}{3}):
  \bibinfo{pages}{3941--3952}. \bibinfo{doi}{\doi{10.1093/mnras/stac1586}}.
\eprint{2202.06295}.

\bibtype{Article}%
\bibitem[{Nakamura} et al.(2025)]{nakamura_25}
\bibinfo{author}{{Nakamura} K}, \bibinfo{author}{{Takiwaki} T},
  \bibinfo{author}{{Matsumoto} J} and  \bibinfo{author}{{Kotake} K}
  (\bibinfo{year}{2025}), \bibinfo{month}{Jan.}
\bibinfo{title}{{Three-dimensional magnetohydrodynamic simulations of
  core-collapse supernovae - I. Hydrodynamic evolution and protoneutron star
  properties}}.
\bibinfo{journal}{{\em \mnras}} \bibinfo{volume}{536} (\bibinfo{number}{1}):
  \bibinfo{pages}{280--294}. \bibinfo{doi}{\doi{10.1093/mnras/stae2611}}.
\eprint{2405.08367}.

\bibtype{Article}%
\bibitem[{Nomoto}(1984)]{nomoto_84}
\bibinfo{author}{{Nomoto} K} (\bibinfo{year}{1984}), \bibinfo{month}{Feb.}
\bibinfo{title}{{Evolution of 8-10 solar mass stars toward electron capture
  supernovae. I - Formation of electron-degenerate O + NE + MG cores}}.
\bibinfo{journal}{{\em \apj}} \bibinfo{volume}{277}: \bibinfo{pages}{791--805}.
  \bibinfo{doi}{\doi{10.1086/161749}}.

\bibtype{Article}%
\bibitem[{Nomoto}(1987)]{nomoto_87}
\bibinfo{author}{{Nomoto} K} (\bibinfo{year}{1987}), \bibinfo{month}{Nov.}
\bibinfo{title}{{Evolution of 8-10 solar mass stars toward electron capture
  supernovae. II - Collapse of an O + NE + MG core}}.
\bibinfo{journal}{{\em \apj}} \bibinfo{volume}{322}: \bibinfo{pages}{206--214}.
  \bibinfo{doi}{\doi{10.1086/165716}}.

\bibtype{Article}%
\bibitem[{O'Connor} and {Couch}(2018)]{oconnor_18b}
\bibinfo{author}{{O'Connor} EP} and  \bibinfo{author}{{Couch} SM}
  (\bibinfo{year}{2018}), \bibinfo{month}{Oct.}
\bibinfo{title}{{Exploring Fundamentally Three-dimensional Phenomena in
  High-fidelity Simulations of Core-collapse Supernovae}}.
\bibinfo{journal}{{\em \apj}} \bibinfo{volume}{865}: \bibinfo{pages}{81}.
  \bibinfo{doi}{\doi{10.3847/1538-4357/aadcf7}}.

\bibtype{Article}%
\bibitem[{O'Connor} and {Ott}(2011)]{oconnor_11}
\bibinfo{author}{{O'Connor} E} and  \bibinfo{author}{{Ott} CD}
  (\bibinfo{year}{2011}), \bibinfo{month}{Apr.}
\bibinfo{title}{{Black Hole Formation in Failing Core-Collapse Supernovae}}.
\bibinfo{journal}{{\em \apj}} \bibinfo{volume}{730}: \bibinfo{pages}{70}.
  \bibinfo{doi}{\doi{10.1088/0004-637X/730/2/70}}.

\bibtype{Article}%
\bibitem[{O'Connor} and {Ott}(2013)]{oconnor_13}
\bibinfo{author}{{O'Connor} E} and  \bibinfo{author}{{Ott} CD}
  (\bibinfo{year}{2013}), \bibinfo{month}{Jan.}
\bibinfo{title}{{The Progenitor Dependence of the Pre-explosion Neutrino
  Emission in Core-collapse Supernovae}}.
\bibinfo{journal}{{\em \apj}} \bibinfo{volume}{762}: \bibinfo{pages}{126}.
  \bibinfo{doi}{\doi{10.1088/0004-637X/762/2/126}}.

\bibtype{Article}%
\bibitem[{O'Connor} et al.(2018)]{oconnor_18c}
\bibinfo{author}{{O'Connor} E}, \bibinfo{author}{{Bollig} R},
  \bibinfo{author}{{Burrows} A}, \bibinfo{author}{{Couch} S},
  \bibinfo{author}{{Fischer} T}, \bibinfo{author}{{Janka} HT},
  \bibinfo{author}{{Kotake} K}, \bibinfo{author}{{Lentz} EJ},
  \bibinfo{author}{{Liebend{\"o}rfer} M}, \bibinfo{author}{{Messer} OEB},
  \bibinfo{author}{{Mezzacappa} A}, \bibinfo{author}{{Takiwaki} T} and
  \bibinfo{author}{{Vartanyan} D} (\bibinfo{year}{2018}), \bibinfo{month}{Oct.}
\bibinfo{title}{{Global comparison of core-collapse supernova simulations in
  spherical symmetry}}.
\bibinfo{journal}{{\em Journal of Physics G Nuclear Physics}}
  \bibinfo{volume}{45} (\bibinfo{number}{10}): \bibinfo{pages}{104001}.
  \bibinfo{doi}{\doi{10.1088/1361-6471/aadeae}}.

\bibtype{Article}%
\bibitem[{Pejcha} and {Prieto}(2015)]{pejcha_15c}
\bibinfo{author}{{Pejcha} O} and  \bibinfo{author}{{Prieto} JL}
  (\bibinfo{year}{2015}), \bibinfo{month}{Jun.}
\bibinfo{title}{{On the Intrinsic Diversity of Type II-Plateau Supernovae}}.
\bibinfo{journal}{{\em \apj}} \bibinfo{volume}{806}: \bibinfo{pages}{225}.
  \bibinfo{doi}{\doi{10.1088/0004-637X/806/2/225}}.

\bibtype{Article}%
\bibitem[{Pons} et al.(1999)]{pons_99}
\bibinfo{author}{{Pons} JA}, \bibinfo{author}{{Reddy} S},
  \bibinfo{author}{{Prakash} M}, \bibinfo{author}{{Lattimer} JM} and
  \bibinfo{author}{{Miralles} JA} (\bibinfo{year}{1999}), \bibinfo{month}{Mar.}
\bibinfo{title}{{Evolution of Proto-Neutron Stars}}.
\bibinfo{journal}{{\em \apj}} \bibinfo{volume}{513}: \bibinfo{pages}{780--804}.
  \bibinfo{doi}{\doi{10.1086/306889}}.

\bibtype{Article}%
\bibitem[{Pons} et al.(2001{\natexlab{a}})]{pons_01a}
\bibinfo{author}{{Pons} JA}, \bibinfo{author}{{Miralles} JA},
  \bibinfo{author}{{Prakash} M} and  \bibinfo{author}{{Lattimer} JM}
  (\bibinfo{year}{2001}{\natexlab{a}}), \bibinfo{month}{May}.
\bibinfo{title}{{Evolution of Proto-Neutron Stars with Kaon Condensates}}.
\bibinfo{journal}{{\em \apj}} \bibinfo{volume}{553}: \bibinfo{pages}{382--393}.
  \bibinfo{doi}{\doi{10.1086/320642}}.
\eprint{astro-ph/0008389}.

\bibtype{Article}%
\bibitem[{Pons} et al.(2001{\natexlab{b}})]{pons_01b}
\bibinfo{author}{{Pons} JA}, \bibinfo{author}{{Steiner} AW},
  \bibinfo{author}{{Prakash} M} and  \bibinfo{author}{{Lattimer} JM}
  (\bibinfo{year}{2001}{\natexlab{b}}), \bibinfo{month}{Jun.}
\bibinfo{title}{{Evolution of Proto-Neutron Stars with Quarks}}.
\bibinfo{journal}{{\em \prl}} \bibinfo{volume}{86}:
  \bibinfo{pages}{5223--5226}.
  \bibinfo{doi}{\doi{10.1103/PhysRevLett.86.5223}}.
\eprint{astro-ph/0102015}.

\bibtype{Article}%
\bibitem[{Popov}(1993)]{popov_93}
\bibinfo{author}{{Popov} DV} (\bibinfo{year}{1993}), \bibinfo{month}{Sep.}
\bibinfo{title}{{An analytical model for the plateau stage of Type II
  supernovae}}.
\bibinfo{journal}{{\em \apj}} \bibinfo{volume}{414}: \bibinfo{pages}{712--716}.
  \bibinfo{doi}{\doi{10.1086/173117}}.

\bibtype{Article}%
\bibitem[{Popov} et al.(2025)]{popov_25}
\bibinfo{author}{{Popov} S}, \bibinfo{author}{{M{\"u}ller} B} and
  \bibinfo{author}{{Mandel} I} (\bibinfo{year}{2025}), \bibinfo{month}{Dec.}
\bibinfo{title}{{Natal kicks of compact objects}}.
\bibinfo{journal}{{\em \nar}} \bibinfo{volume}{101}, \bibinfo{eid}{101734}.
  \bibinfo{doi}{\doi{10.1016/j.newar.2025.101734}}.
\eprint{2509.01430}.

\bibtype{Article}%
\bibitem[{Powell} and {M{\"u}ller}(2019)]{powell_19}
\bibinfo{author}{{Powell} J} and  \bibinfo{author}{{M{\"u}ller} B}
  (\bibinfo{year}{2019}), \bibinfo{month}{Jul}.
\bibinfo{title}{{Gravitational wave emission from 3D explosion models of
  core-collapse supernovae with low and normal explosion energies}}.
\bibinfo{journal}{{\em \mnras}} \bibinfo{volume}{487} (\bibinfo{number}{1}):
  \bibinfo{pages}{1178--1190}. \bibinfo{doi}{\doi{10.1093/mnras/stz1304}}.

\bibtype{Article}%
\bibitem[{Powell} and {M{\"u}ller}(2025)]{powell_26}
\bibinfo{author}{{Powell} J} and  \bibinfo{author}{{M{\"u}ller} B}
  (\bibinfo{year}{2025}), \bibinfo{month}{Oct.}
\bibinfo{title}{{Impact of the nuclear equation of state on the explodability
  of massive stars}}.
\bibinfo{journal}{{\em arXiv e-prints}} ,
  \bibinfo{eid}{arXiv:2510.20076}\bibinfo{doi}{\doi{10.48550/arXiv.2510.20076}}.

\bibtype{Article}%
\bibitem[{Powell} et al.(2021)]{powell_21}
\bibinfo{author}{{Powell} J}, \bibinfo{author}{{M{\"u}ller} B} and
  \bibinfo{author}{{Heger} A} (\bibinfo{year}{2021}), \bibinfo{month}{May}.
\bibinfo{title}{{The final core collapse of pulsational pair instability
  supernovae}}.
\bibinfo{journal}{{\em \mnras}} \bibinfo{volume}{503} (\bibinfo{number}{2}):
  \bibinfo{pages}{2108--2122}. \bibinfo{doi}{\doi{10.1093/mnras/stab614}}.

\bibtype{Article}%
\bibitem[{Qian} and {Woosley}(1996)]{qian_96}
\bibinfo{author}{{Qian} Y} and  \bibinfo{author}{{Woosley} SE}
  (\bibinfo{year}{1996}), \bibinfo{month}{Nov.}
\bibinfo{title}{{Nucleosynthesis in Neutrino-driven Winds. I. The Physical
  Conditions}}.
\bibinfo{journal}{{\em \apj}} \bibinfo{volume}{471}: \bibinfo{pages}{331--+}.
  \bibinfo{doi}{\doi{10.1086/177973}}.

\bibtype{Article}%
\bibitem[{Radice} et al.(2019)]{radice_19}
\bibinfo{author}{{Radice} D}, \bibinfo{author}{{Morozova} V},
  \bibinfo{author}{{Burrows} A}, \bibinfo{author}{{Vartanyan} D} and
  \bibinfo{author}{{Nagakura} H} (\bibinfo{year}{2019}), \bibinfo{month}{May}.
\bibinfo{title}{{Characterizing the Gravitational Wave Signal from
  Core-collapse Supernovae}}.
\bibinfo{journal}{{\em \apjl}} \bibinfo{volume}{876} (\bibinfo{number}{1}):
  \bibinfo{pages}{L9}. \bibinfo{doi}{\doi{10.3847/2041-8213/ab191a}}.

\bibtype{Article}%
\bibitem[{Raffelt} and {Zhou}(2011)]{raffelt_11b}
\bibinfo{author}{{Raffelt} GG} and  \bibinfo{author}{{Zhou} S}
  (\bibinfo{year}{2011}), \bibinfo{month}{May}.
\bibinfo{title}{{Supernova bound on keV-mass sterile neutrinos reexamined}}.
\bibinfo{journal}{{\em \prd}} \bibinfo{volume}{83} (\bibinfo{number}{9}),
  \bibinfo{eid}{093014}. \bibinfo{doi}{\doi{10.1103/PhysRevD.83.093014}}.
\eprint{1102.5124}.

\bibtype{Article}%
\bibitem[{Raffelt} et al.(2025)]{raffelt_26}
\bibinfo{author}{{Raffelt} GG}, \bibinfo{author}{{Janka} HT} and
  \bibinfo{author}{{Fiorillo} DFG} (\bibinfo{year}{2025}),
  \bibinfo{month}{Sep.}
\bibinfo{title}{{Neutrinos from core-collapse supernovae}}.
\bibinfo{journal}{{\em arXiv e-prints}} ,
  \bibinfo{eid}{arXiv:2509.16306}\bibinfo{doi}{\doi{10.48550/arXiv.2509.16306}}.
\eprint{2509.16306}.

\bibtype{Article}%
\bibitem[{Rahman} et al.(2022)]{rahman_22}
\bibinfo{author}{{Rahman} N}, \bibinfo{author}{{Janka} HT},
  \bibinfo{author}{{Stockinger} G} and  \bibinfo{author}{{Woosley} SE}
  (\bibinfo{year}{2022}), \bibinfo{month}{May}.
\bibinfo{title}{{Pulsational pair-instability supernovae: gravitational
  collapse, black hole formation, and beyond}}.
\bibinfo{journal}{{\em \mnras}} \bibinfo{volume}{512} (\bibinfo{number}{3}):
  \bibinfo{pages}{4503--4540}. \bibinfo{doi}{\doi{10.1093/mnras/stac758}}.
\eprint{2112.09707}.

\bibtype{Article}%
\bibitem[{Rampp} and {Janka}(2000)]{rampp_00}
\bibinfo{author}{{Rampp} M} and  \bibinfo{author}{{Janka} HT}
  (\bibinfo{year}{2000}), \bibinfo{month}{Aug.}
\bibinfo{title}{{Spherically Symmetric Simulation with Boltzmann Neutrino
  Transport of Core Collapse and Postbounce Evolution of a 15 $M_\odot$ Star}}.
\bibinfo{journal}{{\em \apjl}} \bibinfo{volume}{539}:
  \bibinfo{pages}{L33--L36}. \bibinfo{doi}{\doi{10.1086/312837}}.

\bibtype{Article}%
\bibitem[{Reddy} et al.(1999)]{reddy_99}
\bibinfo{author}{{Reddy} S}, \bibinfo{author}{{Prakash} M},
  \bibinfo{author}{{Lattimer} JM} and  \bibinfo{author}{{Pons} JA}
  (\bibinfo{year}{1999}), \bibinfo{month}{May}.
\bibinfo{title}{{Effects of strong and electromagnetic correlations on neutrino
  interactions in dense matter}}.
\bibinfo{journal}{{\em \prc}} \bibinfo{volume}{59}:
  \bibinfo{pages}{2888--2918}. \bibinfo{doi}{\doi{10.1103/PhysRevC.59.2888}}.

\bibtype{Article}%
\bibitem[{Roberts} et al.(2012{\natexlab{a}})]{roberts_12c}
\bibinfo{author}{{Roberts} LF}, \bibinfo{author}{{Reddy} S} and
  \bibinfo{author}{{Shen} G} (\bibinfo{year}{2012}{\natexlab{a}}),
  \bibinfo{month}{Dec.}
\bibinfo{title}{{Medium modification of the charged-current neutrino opacity
  and its implications}}.
\bibinfo{journal}{{\em \prc}} \bibinfo{volume}{86} (\bibinfo{number}{6}):
  \bibinfo{pages}{065803}. \bibinfo{doi}{\doi{10.1103/PhysRevC.86.065803}}.

\bibtype{Article}%
\bibitem[{Roberts} et al.(2012{\natexlab{b}})]{roberts_12b}
\bibinfo{author}{{Roberts} LF}, \bibinfo{author}{{Shen} G},
  \bibinfo{author}{{Cirigliano} V}, \bibinfo{author}{{Pons} JA},
  \bibinfo{author}{{Reddy} S} and  \bibinfo{author}{{Woosley} SE}
  (\bibinfo{year}{2012}{\natexlab{b}}), \bibinfo{month}{Feb.}
\bibinfo{title}{{Protoneutron Star Cooling with Convection: The Effect of the
  Symmetry Energy}}.
\bibinfo{journal}{{\em Physical Review Letters}} \bibinfo{volume}{108}
  (\bibinfo{number}{6}): \bibinfo{pages}{061103}.
  \bibinfo{doi}{\doi{10.1103/PhysRevLett.108.061103}}.

\bibtype{Article}%
\bibitem[{Rusakov} et al.(2026)]{rusakov_26}
\bibinfo{author}{{Rusakov} A}, \bibinfo{author}{{Burrows} AS},
  \bibinfo{author}{{Wang} T} and  \bibinfo{author}{{Vartanyan} D}
  (\bibinfo{year}{2026}), \bibinfo{month}{Feb.}
\bibinfo{title}{{An Exploration of the Equation of State Dependence of
  Core-Collapse Supernova Explosion Outcomes and Signatures}}.
\bibinfo{journal}{{\em arXiv e-prints}} ,
  \bibinfo{eid}{arXiv:2602.09025}\bibinfo{doi}{\doi{10.48550/arXiv.2602.09025}}.

\bibtype{Article}%
\bibitem[{Sagert} et al.(2009)]{sagert_09}
\bibinfo{author}{{Sagert} I}, \bibinfo{author}{{Fischer} T},
  \bibinfo{author}{{Hempel} M}, \bibinfo{author}{{Pagliara} G},
  \bibinfo{author}{{Schaffner-Bielich} J}, \bibinfo{author}{{Mezzacappa} A},
  \bibinfo{author}{{Thielemann} FK} and  \bibinfo{author}{{Liebend{\"o}rfer} M}
  (\bibinfo{year}{2009}), \bibinfo{month}{Feb.}
\bibinfo{title}{{Signals of the QCD Phase Transition in Core-Collapse
  Supernovae}}.
\bibinfo{journal}{{\em Physical Review Letters}} \bibinfo{volume}{102}
  (\bibinfo{number}{8}): \bibinfo{pages}{081101:1--4}.
  \bibinfo{doi}{\doi{10.1103/PhysRevLett.102.081101}}.

\bibtype{Article}%
\bibitem[{Schneider} et al.(2019)]{schneider_a_19}
\bibinfo{author}{{Schneider} AS}, \bibinfo{author}{{Roberts} LF},
  \bibinfo{author}{{Ott} CD} and  \bibinfo{author}{{O'Connor} E}
  (\bibinfo{year}{2019}), \bibinfo{month}{Nov.}
\bibinfo{title}{{Equation of state effects in the core collapse of a 20
  -$\mathrm{M}_\odot$ star}}.
\bibinfo{journal}{{\em \prc}} \bibinfo{volume}{100} (\bibinfo{number}{5}),
  \bibinfo{eid}{055802}. \bibinfo{doi}{\doi{10.1103/PhysRevC.100.055802}}.
\eprint{1906.02009}.

\bibtype{Article}%
\bibitem[{Seitenzahl} et al.(2014)]{seitenzahl_14}
\bibinfo{author}{{Seitenzahl} IR}, \bibinfo{author}{{Timmes} FX} and
  \bibinfo{author}{{Magkotsios} G} (\bibinfo{year}{2014}),
  \bibinfo{month}{Sep.}
\bibinfo{title}{{The Light Curve of SN 1987A Revisited: Constraining Production
  Masses of Radioactive Nuclides}}.
\bibinfo{journal}{{\em \apj}} \bibinfo{volume}{792} (\bibinfo{number}{1}),
  \bibinfo{eid}{10}. \bibinfo{doi}{\doi{10.1088/0004-637X/792/1/10}}.
\eprint{1408.5986}.

\bibtype{Article}%
\bibitem[{Serpico} et al.(2012)]{serpico_12}
\bibinfo{author}{{Serpico} PD}, \bibinfo{author}{{Chakraborty} S},
  \bibinfo{author}{{Fischer} T}, \bibinfo{author}{{H{\"u}depohl} L},
  \bibinfo{author}{{Janka} HT} and  \bibinfo{author}{{Mirizzi} A}
  (\bibinfo{year}{2012}), \bibinfo{month}{Apr.}
\bibinfo{title}{{Probing the neutrino mass hierarchy with the rise time of a
  supernova burst}}.
\bibinfo{journal}{{\em \prd}} \bibinfo{volume}{85} (\bibinfo{number}{8}):
  \bibinfo{pages}{085031}. \bibinfo{doi}{\doi{10.1103/PhysRevD.85.085031}}.

\bibtype{Article}%
\bibitem[{Smartt}(2015)]{smartt_15}
\bibinfo{author}{{Smartt} SJ} (\bibinfo{year}{2015}), \bibinfo{month}{Apr.}
\bibinfo{title}{{Observational Constraints on the Progenitors of Core-Collapse
  Supernovae: The Case for Missing High-Mass Stars}}.
\bibinfo{journal}{{\em \pasa}} \bibinfo{volume}{32}: \bibinfo{pages}{16}.
  \bibinfo{doi}{\doi{10.1017/pasa.2015.17}}.

\bibtype{Article}%
\bibitem[{Smartt} et al.(2009)]{smartt_09a}
\bibinfo{author}{{Smartt} SJ}, \bibinfo{author}{{Eldridge} JJ},
  \bibinfo{author}{{Crockett} RM} and  \bibinfo{author}{{Maund} JR}
  (\bibinfo{year}{2009}), \bibinfo{month}{May}.
\bibinfo{title}{{The death of massive stars - I. Observational constraints on
  the progenitors of Type II-P supernovae}}.
\bibinfo{journal}{{\em \mnras}} \bibinfo{volume}{395}:
  \bibinfo{pages}{1409--1437}.
  \bibinfo{doi}{\doi{10.1111/j.1365-2966.2009.14506.x}}.

\bibtype{Article}%
\bibitem[{Smith} et al.(2011)]{smith_11}
\bibinfo{author}{{Smith} N}, \bibinfo{author}{{Li} W},
  \bibinfo{author}{{Filippenko} AV} and  \bibinfo{author}{{Chornock} R}
  (\bibinfo{year}{2011}), \bibinfo{month}{Apr.}
\bibinfo{title}{{Observed fractions of core-collapse supernova types and
  initial masses of their single and binary progenitor stars}}.
\bibinfo{journal}{{\em \mnras}} \bibinfo{volume}{412}:
  \bibinfo{pages}{1522--1538}.
  \bibinfo{doi}{\doi{10.1111/j.1365-2966.2011.17229.x}}.

\bibtype{Article}%
\bibitem[{Staelin} and {Reifenstein}(1968)]{staelin_68}
\bibinfo{author}{{Staelin} DH} and  \bibinfo{author}{{Reifenstein} EC}
  (\bibinfo{year}{1968}), \bibinfo{month}{Dec.}
\bibinfo{title}{{Pulsating radio sources near the Crab Nebula.}}
\bibinfo{journal}{{\em Science}} \bibinfo{volume}{162}:
  \bibinfo{pages}{1481--1483}.
  \bibinfo{doi}{\doi{10.1126/science.162.3861.1481}}.

\bibtype{Article}%
\bibitem[{Steiner} et al.(2010)]{steiner_10}
\bibinfo{author}{{Steiner} AW}, \bibinfo{author}{{Lattimer} JM} and
  \bibinfo{author}{{Brown} EF} (\bibinfo{year}{2010}), \bibinfo{month}{Oct.}
\bibinfo{title}{{The Equation of State from Observed Masses and Radii of
  Neutron Stars}}.
\bibinfo{journal}{{\em \apj}} \bibinfo{volume}{722}: \bibinfo{pages}{33--54}.
  \bibinfo{doi}{\doi{10.1088/0004-637X/722/1/33}}.

\bibtype{Article}%
\bibitem[{Sukhbold} et al.(2016)]{sukhbold_16}
\bibinfo{author}{{Sukhbold} T}, \bibinfo{author}{{Ertl} T},
  \bibinfo{author}{{Woosley} SE}, \bibinfo{author}{{Brown} JM} and
  \bibinfo{author}{{Janka} HT} (\bibinfo{year}{2016}), \bibinfo{month}{Apr.}
\bibinfo{title}{{Core-Collapse Supernovae from 9 to 120 Solar Masses Based on
  Neutrino-powered Explosions}}.
\bibinfo{journal}{{\em \apj}} \bibinfo{volume}{821}: \bibinfo{pages}{38}.
  \bibinfo{doi}{\doi{10.3847/0004-637X/821/1/38}}.

\bibtype{Article}%
\bibitem[{Suwa} et al.(2013)]{suwa_13}
\bibinfo{author}{{Suwa} Y}, \bibinfo{author}{{Takiwaki} T},
  \bibinfo{author}{{Kotake} K}, \bibinfo{author}{{Fischer} T},
  \bibinfo{author}{{Liebend{\"o}rfer} M} and  \bibinfo{author}{{Sato} K}
  (\bibinfo{year}{2013}), \bibinfo{month}{Feb.}
\bibinfo{title}{{On the Importance of the Equation of State for the
  Neutrino-driven Supernova Explosion Mechanism}}.
\bibinfo{journal}{{\em \apj}} \bibinfo{volume}{764}: \bibinfo{pages}{99}.
  \bibinfo{doi}{\doi{10.1088/0004-637X/764/1/99}}.

\bibtype{Article}%
\bibitem[{Sykes} and {M{\"u}ller}(2025{\natexlab{a}})]{sykes_25}
\bibinfo{author}{{Sykes} B} and  \bibinfo{author}{{M{\"u}ller} B}
  (\bibinfo{year}{2025}{\natexlab{a}}), \bibinfo{month}{Mar.}
\bibinfo{title}{{Long-time 2D simulations of fallback supernovae: a systematic
  investigation of explosions dynamics and mass ejection}}.
\bibinfo{journal}{{\em \mnras}} \bibinfo{volume}{538} (\bibinfo{number}{1}):
  \bibinfo{pages}{572--592}. \bibinfo{doi}{\doi{10.1093/mnras/staf317}}.
\eprint{2410.04944}.

\bibtype{Article}%
\bibitem[{Sykes} and {M{\"u}ller}(2025{\natexlab{b}})]{sykes_25b}
\bibinfo{author}{{Sykes} B} and  \bibinfo{author}{{M{\"u}ller} B}
  (\bibinfo{year}{2025}{\natexlab{b}}), \bibinfo{month}{Mar.}
\bibinfo{title}{{Long-time 3D supernova simulations of nonrotating progenitors
  with magnetic fields}}.
\bibinfo{journal}{{\em \prd}} \bibinfo{volume}{111} (\bibinfo{number}{6}),
  \bibinfo{eid}{063042}. \bibinfo{doi}{\doi{10.1103/PhysRevD.111.063042}}.
\eprint{2412.01155}.

\bibtype{Article}%
\bibitem[{Takiwaki} et al.(2014)]{takiwaki_14}
\bibinfo{author}{{Takiwaki} T}, \bibinfo{author}{{Kotake} K} and
  \bibinfo{author}{{Suwa} Y} (\bibinfo{year}{2014}), \bibinfo{month}{May}.
\bibinfo{title}{{A Comparison of Two- and Three-dimensional
  Neutrino-hydrodynamics Simulations of Core-collapse Supernovae}}.
\bibinfo{journal}{{\em \apj}} \bibinfo{volume}{786}: \bibinfo{pages}{83}.
  \bibinfo{doi}{\doi{10.1088/0004-637X/786/2/83}}.

\bibtype{Article}%
\bibitem[{Tamborra} et al.(2012)]{tamborra_12}
\bibinfo{author}{{Tamborra} I}, \bibinfo{author}{{M{\"u}ller} B},
  \bibinfo{author}{{H{\"u}depohl} L}, \bibinfo{author}{{Janka} HT} and
  \bibinfo{author}{{Raffelt} G} (\bibinfo{year}{2012}), \bibinfo{month}{Dec.}
\bibinfo{title}{{High-resolution supernova neutrino spectra represented by a
  simple fit}}.
\bibinfo{journal}{{\em \prd}} \bibinfo{volume}{86} (\bibinfo{number}{12}):
  \bibinfo{pages}{125031}. \bibinfo{doi}{\doi{10.1103/PhysRevD.86.125031}}.

\bibtype{Article}%
\bibitem[{Tamborra} et al.(2014)]{tamborra_14a}
\bibinfo{author}{{Tamborra} I}, \bibinfo{author}{{Hanke} F},
  \bibinfo{author}{{Janka} HT}, \bibinfo{author}{{M{\"u}ller} B},
  \bibinfo{author}{{Raffelt} GG} and  \bibinfo{author}{{Marek} A}
  (\bibinfo{year}{2014}), \bibinfo{month}{Sep.}
\bibinfo{title}{{Self-sustained Asymmetry of Lepton-number Emission: A New
  Phenomenon during the Supernova Shock-accretion Phase in Three Dimensions}}.
\bibinfo{journal}{{\em \apj}} \bibinfo{volume}{792}: \bibinfo{pages}{96}.
  \bibinfo{doi}{\doi{10.1088/0004-637X/792/2/96}}.

\bibtype{Article}%
\bibitem[{Torres-Forn{\'e}} et al.(2018)]{torres_18}
\bibinfo{author}{{Torres-Forn{\'e}} A}, \bibinfo{author}{{Cerd{\'a}-Dur{\'a}n}
  P}, \bibinfo{author}{{Passamonti} A} and  \bibinfo{author}{{Font} JA}
  (\bibinfo{year}{2018}), \bibinfo{month}{Mar.}
\bibinfo{title}{{Towards asteroseismology of core-collapse supernovae with
  gravitational-wave observations - I. Cowling approximation}}.
\bibinfo{journal}{{\em \mnras}} \bibinfo{volume}{474} (\bibinfo{number}{4}):
  \bibinfo{pages}{5272--5286}. \bibinfo{doi}{\doi{10.1093/mnras/stx3067}}.

\bibtype{Article}%
\bibitem[{Tubbs} and {Schramm}(1975)]{tubbs_75}
\bibinfo{author}{{Tubbs} DL} and  \bibinfo{author}{{Schramm} DN}
  (\bibinfo{year}{1975}), \bibinfo{month}{Oct.}
\bibinfo{title}{{Neutrino Opacities at High Temperatures and Densities}}.
\bibinfo{journal}{{\em \apj}} \bibinfo{volume}{201}: \bibinfo{pages}{467--488}.
  \bibinfo{doi}{\doi{10.1086/153909}}.

\bibtype{Article}%
\bibitem[{Ugliano} et al.(2012)]{ugliano_12}
\bibinfo{author}{{Ugliano} M}, \bibinfo{author}{{Janka} HT},
  \bibinfo{author}{{Marek} A} and  \bibinfo{author}{{Arcones} A}
  (\bibinfo{year}{2012}), \bibinfo{month}{Sep.}
\bibinfo{title}{{Progenitor-explosion Connection and Remnant Birth Masses for
  Neutrino-driven Supernovae of Iron-core Progenitors}}.
\bibinfo{journal}{{\em \apj}} \bibinfo{volume}{757}: \bibinfo{pages}{69}.
  \bibinfo{doi}{\doi{10.1088/0004-637X/757/1/69}}.

\bibtype{Article}%
\bibitem[{Varma} et al.(2021)]{varma_21b}
\bibinfo{author}{{Varma} V}, \bibinfo{author}{{M{\"u}ller} B} and
  \bibinfo{author}{{Obergaulinger} M} (\bibinfo{year}{2021}),
  \bibinfo{month}{Dec.}
\bibinfo{title}{{A comparison of 2D Magnetohydrodynamic supernova simulations
  with the COCONUT-FMT and AENUS-ALCAR codes}}.
\bibinfo{journal}{{\em \mnras}} \bibinfo{volume}{508} (\bibinfo{number}{4}):
  \bibinfo{pages}{6033--6048}. \bibinfo{doi}{\doi{10.1093/mnras/stab2983}}.

\bibtype{Article}%
\bibitem[{Vartanyan} et al.(2019)]{vartanyan_19}
\bibinfo{author}{{Vartanyan} D}, \bibinfo{author}{{Burrows} A},
  \bibinfo{author}{{Radice} D}, \bibinfo{author}{{Skinner} MA} and
  \bibinfo{author}{{Dolence} J} (\bibinfo{year}{2019}), \bibinfo{month}{Jan.}
\bibinfo{title}{{A successful 3D core-collapse supernova explosion model}}.
\bibinfo{journal}{{\em \mnras}} \bibinfo{volume}{482}:
  \bibinfo{pages}{351--369}. \bibinfo{doi}{\doi{10.1093/mnras/sty2585}}.

\bibtype{Article}%
\bibitem[{Wang} and {Wheeler}(2008)]{wang_08}
\bibinfo{author}{{Wang} L} and  \bibinfo{author}{{Wheeler} JC}
  (\bibinfo{year}{2008}), \bibinfo{month}{Sep}.
\bibinfo{title}{{Spectropolarimetry of supernovae.}}
\bibinfo{journal}{{\em \araa}} \bibinfo{volume}{46}: \bibinfo{pages}{433--474}.
  \bibinfo{doi}{\doi{10.1146/annurev.astro.46.060407.145139}}.

\bibtype{Article}%
\bibitem[{Wong} et al.(2014)]{wong_14}
\bibinfo{author}{{Wong} TW}, \bibinfo{author}{{Fryer} CL},
  \bibinfo{author}{{Ellinger} CI}, \bibinfo{author}{{Rockefeller} G} and
  \bibinfo{author}{{Kalogera} V} (\bibinfo{year}{2014}), \bibinfo{month}{Jan.}
\bibinfo{title}{{The Fallback Mechanisms in Core-Collapse Supernovae}}.
\bibinfo{journal}{{\em arXiv e-prints}} ,
  \bibinfo{eid}{arXiv:1401.3032}\bibinfo{doi}{\doi{10.48550/arXiv.1401.3032}}.
\eprint{1401.3032}.

\bibtype{Article}%
\bibitem[{Wongwathanarat} et al.(2013)]{wongwathanarat_13}
\bibinfo{author}{{Wongwathanarat} A}, \bibinfo{author}{{Janka} HT} and
  \bibinfo{author}{{M{\"u}ller} E} (\bibinfo{year}{2013}),
  \bibinfo{month}{Apr.}
\bibinfo{title}{{Three-dimensional neutrino-driven supernovae: Neutron star
  kicks, spins, and asymmetric ejection of nucleosynthesis products}}.
\bibinfo{journal}{{\em \aap}} \bibinfo{volume}{552}: \bibinfo{pages}{A126}.
  \bibinfo{doi}{\doi{10.1051/0004-6361/201220636}}.

\bibtype{Article}%
\bibitem[{Woosley}(2017)]{woosley_17}
\bibinfo{author}{{Woosley} SE} (\bibinfo{year}{2017}), \bibinfo{month}{Feb.}
\bibinfo{title}{{Pulsational Pair-instability Supernovae}}.
\bibinfo{journal}{{\em \apj}} \bibinfo{volume}{836} (\bibinfo{number}{2}),
  \bibinfo{eid}{244}. \bibinfo{doi}{\doi{10.3847/1538-4357/836/2/244}}.

\bibtype{Article}%
\bibitem[{Woosley} and {Bloom}(2006)]{woosley_06b}
\bibinfo{author}{{Woosley} SE} and  \bibinfo{author}{{Bloom} JS}
  (\bibinfo{year}{2006}), \bibinfo{month}{Sep}.
\bibinfo{title}{{The Supernova Gamma-Ray Burst Connection}}.
\bibinfo{journal}{{\em \araa}} \bibinfo{volume}{44} (\bibinfo{number}{1}):
  \bibinfo{pages}{507--556}.
  \bibinfo{doi}{\doi{10.1146/annurev.astro.43.072103.150558}}.

\bibtype{Article}%
\bibitem[{Woosley} and {Heger}(2021)]{woosley_21}
\bibinfo{author}{{Woosley} SE} and  \bibinfo{author}{{Heger} A}
  (\bibinfo{year}{2021}), \bibinfo{month}{May}.
\bibinfo{title}{{The Pair-instability Mass Gap for Black Holes}}.
\bibinfo{journal}{{\em \apjl}} \bibinfo{volume}{912} (\bibinfo{number}{2}),
  \bibinfo{eid}{L31}. \bibinfo{doi}{\doi{10.3847/2041-8213/abf2c4}}.
\eprint{2103.07933}.

\bibtype{Article}%
\bibitem[{Woosley} et al.(2002)]{woosley_02}
\bibinfo{author}{{Woosley} SE}, \bibinfo{author}{{Heger} A} and
  \bibinfo{author}{{Weaver} TA} (\bibinfo{year}{2002}), \bibinfo{month}{Nov.}
\bibinfo{title}{{The evolution and explosion of massive stars}}.
\bibinfo{journal}{{\em {\it Rev.~Mod.~Phys.}}} \bibinfo{volume}{74}:
  \bibinfo{pages}{1015--1071}. \bibinfo{doi}{\doi{10.1103/RevModPhys.74.1015}}.

\bibtype{Article}%
\bibitem[{Yasin} et al.(2020)]{yasin_20}
\bibinfo{author}{{Yasin} H}, \bibinfo{author}{{Sch{\"a}fer} S},
  \bibinfo{author}{{Arcones} A} and  \bibinfo{author}{{Schwenk} A}
  (\bibinfo{year}{2020}), \bibinfo{month}{Mar.}
\bibinfo{title}{{Equation of State Effects in Core-Collapse Supernovae}}.
\bibinfo{journal}{{\em \prl}} \bibinfo{volume}{124} (\bibinfo{number}{9}),
  \bibinfo{eid}{092701}. \bibinfo{doi}{\doi{10.1103/PhysRevLett.124.092701}}.

\bibtype{incollection}%
\bibitem[Zampieri(2017)]{zampieri_17}
\bibinfo{author}{Zampieri L} (\bibinfo{year}{2017}), \bibinfo{title}{Light
  curves of type ii supernovae}, \bibinfo{editor}{Alsabti AW} and
  \bibinfo{editor}{Murdin P}, (Eds.), \bibinfo{booktitle}{Handbook of
  Supernovae}, \bibinfo{publisher}{Springer International Publishing},
  \bibinfo{address}{Cham},  \bibinfo{pages}{1--32}.

\bibtype{Article}%
\bibitem[{Zha} et al.(2021)]{zha_21}
\bibinfo{author}{{Zha} S}, \bibinfo{author}{{O'Connor} EP} and
  \bibinfo{author}{{da Silva Schneider} A} (\bibinfo{year}{2021}),
  \bibinfo{month}{Apr.}
\bibinfo{title}{{Progenitor Dependence of Hadron-quark Phase Transition in
  Failing Core-collapse Supernovae}}.
\bibinfo{journal}{{\em \apj}} \bibinfo{volume}{911} (\bibinfo{number}{2}),
  \bibinfo{eid}{74}. \bibinfo{doi}{\doi{10.3847/1538-4357/abec4c}}.

\bibtype{Article}%
\bibitem[{Zhang} and {Burrows}(2013)]{zhang_13b}
\bibinfo{author}{{Zhang} Y} and  \bibinfo{author}{{Burrows} A}
  (\bibinfo{year}{2013}), \bibinfo{month}{Nov.}
\bibinfo{title}{{Transport equations for oscillating neutrinos}}.
\bibinfo{journal}{{\em \prd}} \bibinfo{volume}{88} (\bibinfo{number}{10}),
  \bibinfo{eid}{105009}. \bibinfo{doi}{\doi{10.1103/PhysRevD.88.105009}}.
\eprint{1310.2164}.

\end{thebibliography*}

\end{document}